%% file: master-thesis.tex
\documentclass{scrbook}
\KOMAoptions{
BCOR=10mm,
DIV=12,
draft=false,
fontsize=12pt,
headsepline=true,
numbers=noendperiod,
open=any,
paper=a4,
parskip=half,
toc=listof
}

\usepackage{scrhack}

\usepackage[T1]{fontenc}

\usepackage[utf8]{inputenc}

\usepackage{microtype}

\usepackage[onehalfspacing]{setspace}

\usepackage{mathpazo}

\usepackage[
scale=0.9,
semibold
]{sourcecodepro}

\usepackage[english]{babel}

\usepackage[
backend=biber,
maxbibnames=3,
style=alphabetic,
urldate=long
]{biblatex}
\usepackage{bookmark}

\usepackage{hyperxmp}

\usepackage{hyperref}
\hypersetup{
colorlinks=true,
allcolors=black,
pdfsubject={Master Thesis},
pdftitle={Machine Learning and Evolutionary Computing for GUI-based Regression Testing},
pdfauthor={Daniel Kraus},
pdfcontactemail={daniel.kraus@mailbox.org},
pdfkeywords={Artificial neural network (ANN), genetic algorithm (GA), ML-technique enhanced-EC (MLEC), GUI testing, test generation},
pdflang={en}
}

\usepackage{caption}

\usepackage[newfloat,frozencache]{minted}
\setminted{
baselinestretch=1,
fontsize=\small,
frame=single,
linenos=true,
tabsize=4,
}

\usepackage{csquotes}

\usepackage{siunitx}
\usepackage{amsmath}

\usepackage{booktabs}

\usepackage{tikz}

\usepackage{pgfplots}

\usepackage{tcolorbox}
\tcbset{
boxrule=\heavyrulewidth,
colback=white,
colbacktitle=white,
colframe=black,
coltitle=black,
}
\newtcolorbox[auto counter]{userstory}[2][]{title=\#\thetcbcounter: #2,#1}

\usepackage{chngcntr}

\usepackage{graphicx}
\graphicspath{{resources/img/}}

\usepackage{pdfpages}

\usepackage{glossaries}

\usepackage[
acronym,
toc
]{glossaries-extra}
\setabbreviationstyle[acronym]{long-short}
\makenoidxglossaries
\glssetcategoryattribute{acronym}{glossdesc}{firstuc}
\loadglsentries{resources/acronyms}

\recalctypearea

\setkomafont{disposition}{\bfseries}

\setkomafont{descriptionlabel}{\bfseries}

\newmintinline[inline]{text}{fontsize=\normalsize}

\SetupFloatingEnvironment{listing}{placement=tbp}

\DeclareMathOperator{\dist}{dist}
\DeclareMathOperator{\feas}{feas}
\DeclareMathOperator{\fitness}{fitness}
\DeclareMathOperator{\norm}{norm}
\DeclareMathOperator{\sgn}{sgn}

\newcommand{\dlfj}{\mbox{Deeplearning4j}}

\DeclareSIUnit\inch{''}

\begin{document}

\frontmatter

\includepdf{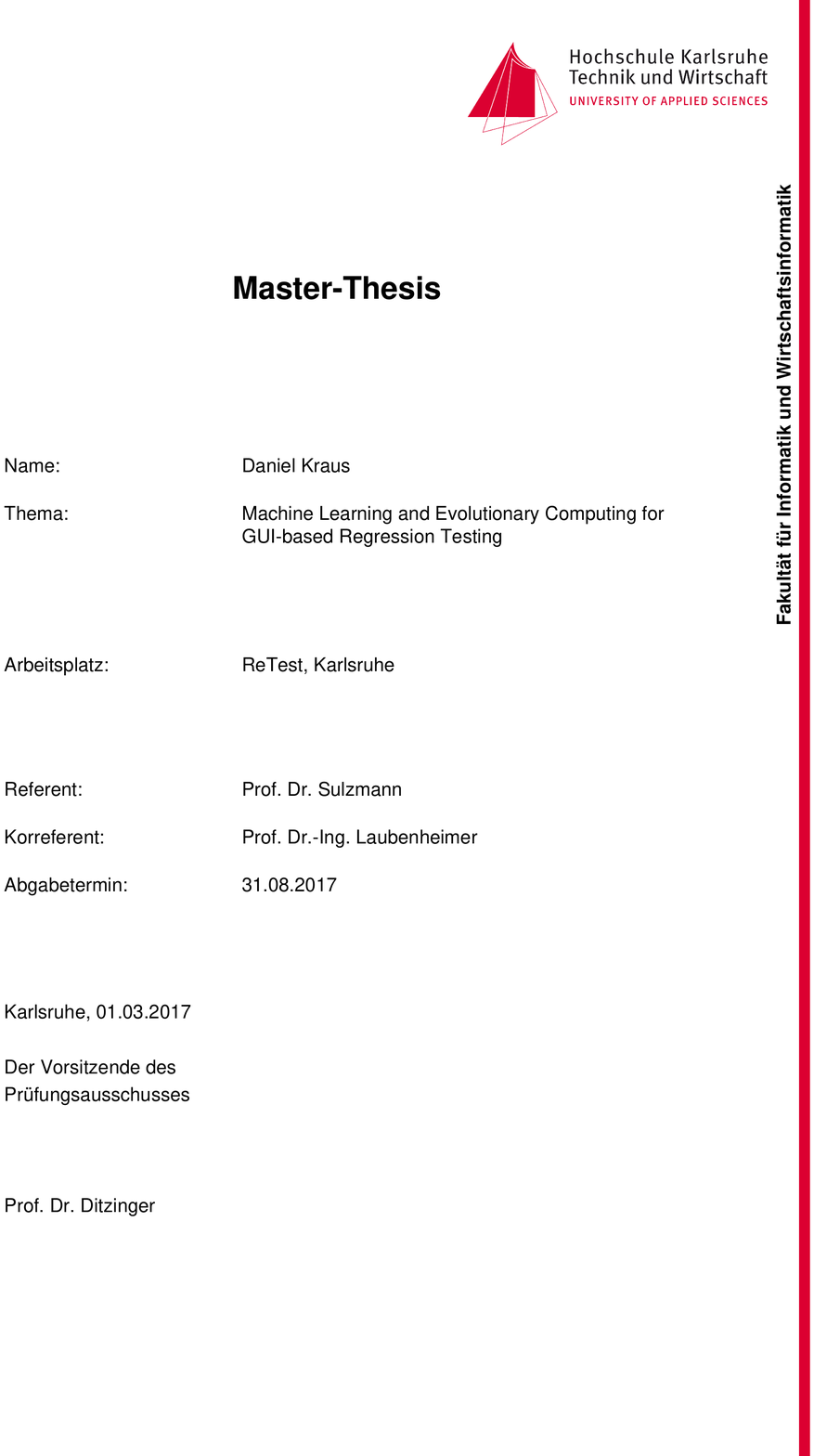}

\thispagestyle{empty}

\textbf{Declaration of Authenticity}

I hereby declare that I have written the present thesis independently and have not used any sources or aids other than those mentioned.\\

\begin{tabular}{@{}l l}
Karlsruhe, October 31, 2017 & \makebox[50mm]{\hrulefill}\\
& Daniel Kraus
\end{tabular}

\vfill

This thesis was created with {\KOMAScript} and {\LaTeX}. Sources available under \url{https://bitbucket.org/beatngu13/master-thesis/}.

\clearpage

\input{chapters/0-preamble}
\newpage
\pdfbookmark[section]{\contentsname}{toc}
\tableofcontents
\listoffigures
\listoftables
\listoflistings
\printnoidxglossary[type=\acronymtype,title=List of Acronyms]


\mainmatter

\input{chapters/1-introduction}
\input{chapters/2-background}
\input{chapters/3-problem}
\input{chapters/4-design}
\input{chapters/5-implementation}
\input{chapters/6-conclusion}


\backmatter

\printbibliography[heading=bibintoc]

\input{chapters/7-appendix}

\end{document}

%% file: resources/acronyms.tex

\newacronym{aec}{AEC}{adaptive EC}

\newacronym{ai}{AI}{artificial intelligence}

\newacronym{ann}{ANN}{artificial neural network}

\newacronym{api}{API}{application programming interface}

\newacronym{awt}{AWT}{Abstract Window Toolkit}

\newacronym{cli}{CLI}{command line interface}

\newacronym{ct}{CT}{continuous testing}

\newacronym{csv}{CSV}{comma-separated values}

\newacronym{dfa}{DFA}{deterministic finite automaton}

\newacronym{dfs}{DFS}{depth-first search}

\newacronym{dom}{DOM}{document object model}

\newacronym{ea}{EA}{evolutionary algorithm}

\newacronym{ec}{EC}{evolutionary computing}

\newacronym{ep}{EP}{evolutionary programming}

\newacronym[
shortplural={ES},
longplural={evolution strategies}
]{es}{ES}{evolution strategy}

\newacronym{etl}{ETL}{extract, transform, load}

\newacronym{exsyst}{EXSYST}{Explorative System Testing}

\newacronym{ga}{GA}{genetic algorithm}

\newacronym{gm}{GM}{golden master}

\newacronym{gpu}{GPU}{graphics processing unit}

\newacronym{gui}{GUI}{graphical user interface}

\newacronym{istqb}{ISTQB}{International Software Testing Qualifications Board}

\newacronym{json}{JSON}{JavaScript object notation}

\newacronym{jvm}{JVM}{Java virtual machine}

\newacronym{ma}{MA}{memetic algoritm}

\newacronym{ml}{ML}{machine learning}

\newacronym{mlec}{MLEC}{ML-technique enhanced-EC}

\newacronym{mosbat}{MoSBaT}{multi-objective search based testing}

\newacronym{nfa}{NFA}{non-deterministic finite automaton}

\newacronym{oop}{OOP}{object-oriented programming}

\newacronym{osi}{OSI}{Open Source Initiative}

\newacronym{rdd}{RDD}{resilient distributed dataset}

\newacronym{rmsprop}{RMSProp}{root mean square propagation}

\newacronym{saas}{SAAS}{software as a service}

\newacronym{sbse}{SBSE}{search-based software engineering}

\newacronym{sbst}{SBST}{search-based software testing}

\newacronym{soa}{SOA}{service-oriented architecture}

\newacronym{sut}{SUT}{system under test}

\newacronym{tap}{TAP}{Test Anything Protocol}

\newacronym{uml}{UML}{Unified Modeling Language}

\newacronym{vcs}{VCS}{version control system}

\newacronym{xp}{XP}{extreme programming}

%% file: chapters/0-preamble.tex


\addchap*{Abstract}

ReTest is a novel testing tool for Java applications with a graphical user interface~(GUI), combining monkey testing and difference testing. Since this combination sidesteps the oracle problem, it enables the generation of GUI-based regression tests. ReTest makes use of evolutionary computing~(EC), particularly a genetic algorithm~(GA), to optimize these tests towards code coverage. While this is indeed a desirable goal in terms of software testing and potentially finds many bugs, it lacks one major ingredient: human behavior. Consequently, human testers often find the results less reasonable and difficult to interpret.

This thesis proposes a new approach to improve the initial population of the GA with the aid of machine learning~(ML), forming an ML-technique enhanced-EC~(MLEC) algorithm. In order to do so, existing tests are exploited to extract information on how human testers use the given GUI. The obtained data is then utilized to train an artificial neural network~(ANN), which ranks the available GUI actions respectively their underlying GUI components at runtime---reducing the gap between manually created and automatically generated regression tests. Although the approach is implemented on top of ReTest, it can be easily used to guide any form of monkey testing.

The results show that with only little training data, the ANN is able to reach an accuracy of \SI{82}{\percent} and the resulting tests represent an improvement without reducing the overall code coverage and performance significantly.



\addchap*{Acknowledgements}

First and foremost, I would like to thank my advisor Prof.~Dr.~Martin Sulzmann for his insightful guidance and his support during this project. I am especially grateful for his enormous help during my independent research proposal, which I did in parallel to this thesis. Furthermore, I thank my colleagues at ReTest for making this possible. In particular, I would like to thank Dr. Jeremias Rößler for pointing me into the right direction.

I would also like to thank Vanessa Fliegauf for her irresistible enthusiasm while proofreading this work to make me become a better writer. Although they did not directly participate in this project, I want to thank Johannes Dillmann and Julian Keppel for being awesome fellow students and making the past six years much easier.

Finally, I would like to thank my family for their endless support, especially when I became a father during my studies. My grandmother Gertrud Bart, my parents Brigitte and Jürgen Kraus, my brother Dennis Kraus, and particularly both my lovely partner and my adorable daughter, Melanie and Marie-Louise Fliegauf. Without these people, I would not have been able to do this.

%% file: chapters/1-introduction.tex


\chapter{Introduction}

The present thesis was created in cooperation with ReTest\footnote{\url{https://retest.de/}.} in Karlsruhe, Germany. ReTest is a small-sized business that develops a test automation tool of the same name and offers various support, training, and consulting services, mostly in the area of software testing. The company was founded in 2014 by Jeremias Rößler as a one-man business to make the former research project~\cite{gfz12b} of the Saarland University become a reliable and user-friendly product. Today, ReTest employs several software development, marketing, and sales specialists. The tool itself matured as well, helping various national and international organizations to implement test automation.

Besides standard functionalities for creating, executing, and maintaining tests, ReTest exhibits two special properties. First, it supports \emph{difference testing}, which captures the whole state of the \emph{\gls{sut}} that is visible through the \emph{\gls{gui}}. If a change is detected, it can be either accepted or ignored with a single action---just like a \gls{vcs} would do. Consequently, no assertions need to be defined since the entire state is consulted, which leads to a faster test creation. These tests are usually also more stable because additional information is available for identifying \gls{gui} components, a common issue in \gls{gui}-based testing. Second, ReTest offers \emph{monkey testing} to test the \gls{sut} fully automatic, supported by a simple form of \emph{\gls{ai}}. In doing so, a \emph{\gls{ga}} optimizes towards code coverage in order to test as many parts of the \gls{sut} as possible. Since difference testing in combination with monkey testing sidesteps the \emph{oracle problem}, it also enables the generation of \gls{gui}-based \emph{regression tests}. Hence, these tests aim to avoid inconsistencies~(the regressions) between the different versions of the \gls{sut}.

\section{Motivation}

Today, companies are constantly exposed to changing market conditions. In the course of this and in the context of the so-called \enquote{digital change}, more and more businesses use software to keep up with this fast-moving and volatile environment. According to the German Federal Ministry of Economics and Energy, \SI{27}{\percent} of companies in Germany are already highly digitalized~\cite{fed17}. This affects 20 billion devices that are connected via the Internet---even half a trillion by 2020.

Software must adapt to these conditions, too, which leads to the fact that the underlying code base is changed almost on a daily basis. These changes must not adversely affect the correctness of the \gls{sut}. \emph{Software testing} can reduce this risk based on observations about the runtime behavior of the software~\cite[6]{sls14}, which is why consistent testing is an indispensable activity that decisively determines the success or failure of software. Within this context, \emph{test automation} describes the automatic execution of otherwise manual tests~\cite[7]{bucp15}. This has many advantages, most importantly it improves the efficiency of software testing, which in turn enables companies to implement \gls{ct}. \cite[1]{hm16} predicts that about \SI{50}{\percent} of all companies will implement \gls{ct} as a result of the DevOps movement. Also \cite[1]{giu16} is of the opinion that the market for test automation is prospering because more and more businesses see this as an essential building block to deliver better software faster.

Nonetheless, according to \cite[41]{csh16}, only \SI{29}{\percent} of the surveyed businesses use test automation today. \SI{45}{\percent} justify this by the fact that no suitable tool is available. Even worse are the numbers when it comes to \emph{test generation} since the term cannot be found at all; although the automatic generation of tests can further boost software testing efficiency by reducing the required amount of manual intervention to a minimum. One of the biggest issues with test generation is that human testers have the ability to construe \emph{meaningful} tests from an interface like a \gls{gui}, whereas machines can hardly do that. As a result, the generated tests typically tend to be missing the link to human behavior, which is why human testers often find the results less reasonable and difficult to interpret.

Nowadays, the question arises how \gls{ai} can help here. But although the World Quality Report states that \gls{ai} will be an inherent part of the future of software testing~\cite[27]{csh16}, none of the test automation tool vendors mentioned in the report as yet leverage it in production. Apart from the leading tool manufacturers, the number of small-sized businesses that offer new, innovative, and \gls{ai}-based software testing products slowly increases. A notable example is Appdiff\footnote{\url{https://appdiff.com/}.} which employs \emph{\gls{ml}} in order to automatically generate regression tests on the \gls{gui} level for mobile applications, although the tool is currently not available for external use. Also the research community is increasingly using \gls{ai} for real-world problems such as software testing. For instance, \cite{ep16} uses a combination of random \gls{gui} inputs and the imitation of user behavior---inferred via \gls{ml}---resulting in a significantly improved test generation. These examples show that the application of \gls{ai}, especially \gls{ml}, for test generation is not just a promising research direction, but also seems to be mature enough for production use.

\section{Goals}

The goal of this thesis is to investigate how ReTest's code coverage-optimizing \gls{ga} can be improved with the aid of \gls{ml}. In particular, the use of \emph{\glspl{ann}} shall be evaluated, in order to see how they can be utilized to support the given \gls{ai}-based monkey testing mechanism for generating regression tests via the \gls{gui}. This includes taking into account existing advances in enhancing \emph{\gls{ec}} with \gls{ml}, namely \emph{\gls{mlec}} algorithms~\cite[69]{zhap11}. In terms of improvement, the main objective is to~(optionally) enrich the generated tests with human behavior to change their characteristics. That is, having the ability to move from non-functional testing towards functional testing. But rather than on a highly-optimized \gls{ml} model, the focus shall be on a robust prototype, including an \gls{etl} pipeline, that can be easily extended for a later use in production.

\section{Contribution and Outline}
\label{sec:contribution-outline}

The main contributions of this thesis can be summarized as follows:
\begin{description}
\item[C.1] In-depth description of the current state of ReTest's \gls{ai}-based monkey testing mechanism.
\item[C.2] Design and implementation of a simple \gls{ann} for ranking \gls{gui} actions respectively their underlying \gls{gui} components at runtime, using production-ready libraries.
\item[C.3] Identification and extraction of relevant features for training the described \gls{ann} based on existing tests including a corresponding \gls{etl} pipeline.
\item[C.4] Presentation of a general framework for enhancing monkey testing based on the aforementioned methods.
\item[C.5] Prototypical implementation of a corresponding \gls{mlec} algorithm on top of ReTest as well as the experimental evaluation of this prototype.
\end{description}

The remainder of this document is structured as follows: Chapter~\ref{ch:background} presents the required background knowledge this thesis is based on along with C.1. In chapter~\ref{ch:problem}, the problems concerning code coverage in software testing and the current monkey testing approach in ReTest are analyzed---including the specification of requirements for the planned prototype---to better understand the given task. Chapter~\ref{ch:design} presents a concrete design that addresses the previously identified limitations as well as the design part of C.2, C.3, and C.4. In chapter~\ref{ch:implementation}, this design is evaluated through a prototypical implementation on top of ReTest, representing C.5. Finally, chapter~\ref{ch:conclusion} summarizes and reflects the findings of this thesis, which is followed by a discussion of possible future work. It should be noted that the work the individual chapters are based upon was performed in a different chronological order than the chapters suggest. The entire project followed an iterative-incremental approach, but for the sake of readability, the chapters follow a traditional waterfall model.

\section{Related Work}

Several relatively new contributions exist when it comes to the use of \gls{ml} in the context of \gls{gui}-based testing. As mentioned before, \cite{ep16} combines monkey testing and the imitation of human behavior for client-side web applications. Instead of using \gls{ec} to explore the \gls{sut}, the prototypical implementation employs \gls{ml} to identify so-called \enquote{macro events} within event traces. A macro event can be interpreted as an atomic sequence of low-level \gls{gui} events that represents a logical step from a user perspective. For instance, selecting a menu item might first trigger a mouse over event on the menu header, followed by a mouse over event, the actual click, and a mouse out event on the corresponding menu item. The event traces are splitted such that only per-page sequences remain, which are then used to perceive recurring patterns. Similar sequences are subsequently grouped into macro event clusters, where each cluster is converted into a \gls{dfa}. According to the results, reusing these \glspl{dfa} during test generation leads to a higher branch coverage and a greater number of covered use cases compared to pure monkey testing. However, the test generation outcome strongly relies on the quality of the event traces. If these traces are poorly chosen, it may not be able to explore much of the \gls{sut}. Because the approach of the present thesis combines \gls{ml} and \gls{ec}, it can still fall back to only use the \gls{ga} if the \gls{ann} yields  suboptimal results.

Another notable example that was mentioned before is \cite{arb17}. The company behind, Appdiff\footnote{\url{https://appdiff.com/}.}, offers a tool for \gls{gui}-based testing of mobile applications as \gls{saas}, but so far no customer has access to the tool itself. It uses supervised learning to classify the current \gls{sut} state (e.g. a login dialog or a privacy policy information) and to generate reasonable \gls{gui} actions. To achieve the first part, an \gls{ann} is trained on large amounts of screenshots and \gls{dom} information that are labeled accordingly. A similar approach is used to train the network on actions: the input here is the set of all \gls{gui} elements within a window and the output is a recommendation of a human-like action for each element. Unfortunately, not many details are available because it is a commercial tool, but Arbon states that once the network is trained, it is capable of generating tests for almost any app. Although this is quite fascinating, the approach comes with a burden: since the training is based on vast amounts of image data, it might become very time-consuming. Moreover, the training data itself has to be labeled manually, which in turn is rather inefficient. The prototype in this thesis does this automatically by extracting the mandatory knowledge from existing tests.

\cite{espp16} is using \gls{ml} for \gls{gui}-based testing as well. The implementation is based on TESTAR\footnote{\url{https://testar.org/}.}, an open source tool for generating tests for desktop, mobile, and web applications. It uses a model-free reinforcement learning technique to guide the selection of \gls{gui} actions. The \gls{ml} model learns which action is optimal for each state and is given a reward if it explores new states of the \gls{sut}, where a state is defined as a separate window. A chosen discount establishes how this reward decreases when actions are being repeated in order to bias the test generation mechanism towards unexplored states. With relatively little knowledge~(e.g. no code coverage information) about the \gls{sut}, the approach is able to outperform pure monkey testing. But the results also reveal that the configuration of the \gls{ml} model heavily depends on the \gls{sut}. That is, parameters which yield good results with a particular \gls{sut} may lead to worse results---compared to monkey testing---if they are used with a different application. It is also important to note that the proposed technique only optimizes towards the exploration of the \gls{sut}, which is measured in terms of visited windows. Consequently, the approach probably does not create human-like sequences of \gls{gui} actions such as filling out a form. Since the \gls{ml} model of the present thesis is trained with features that are~(partially) relative to the previously selected \gls{gui} component, it provides a finer granularity when it comes to the execution of multiple actions within a single state.

Regarding the use of \gls{ec} for \gls{gui}-based testing, especially the field of \gls{sbst} offers a rich variety of tools. For example, \cite{mhj16} generates tests for Android applications with the aid of a \gls{mosbat} algorithm. Such algorithms are basically \glspl{ea} that are able to address multiple objectives during the search; in the context of test generation, this could be code coverage in combination with test length and past bug detection. The tool itself, Sapienz, integrates monkey testing with a systematic exploration of the \gls{sut}. When Sapienz tested over \num{1000} apps from the Google Play Store, it revealed 558 crashes. This is a great advantage that tools developed for mobile or web applications inherent because a large number of \glspl{sut} are easily available, whereas tools that focus on desktop applications---such as ReTest---can hardly do that. Even though Sapienz leads to impressive results, the tool itself does not use \gls{ml} or any other technique in order to include human behavior in its test generation mechanism. More importantly, the author of this thesis is not aware of any solution that employs an \gls{mlec} algorithm in the context of \gls{gui}-based regression testing, although the \textcquote[74]{zhap11}{\textelp{} good results of MLEC algorithms on numerical benchmark functions also encourage the research of applying the MLEC algorithms to numerous real-world applications.}

\glsresetall

%% file: chapters/2-background.tex


\chapter{Background}
\label{ch:background}

This chapter communicates the fundamental knowledge that is necessary for the further understanding of the present thesis. Section~\ref{sec:software-testing} starts with the description of several core concepts in software testing with an emphasis on GUI-based test automation. This also includes ReTest's underlying techniques, namely monkey testing and difference testing. In section~\ref{sec:machine-learning}, a basic introduction to machine learning can be found which focuses on artificial neural networks as they serve a special role in the latter chapters. Section~\ref{sec:evolutionary-computing} does the same towards evolutionary computing respectively genetic algorithms and explains how ReTest leverages them to improve the generation of regression tests.

\section{Software Testing}
\label{sec:software-testing}

Software systems are created to address the required tasks of their stakeholders---also know as \emph{requirements}\footnote{Unless explicitly stated otherwise, all definitions in this section are taken from~\cite{sls14}.}. A \emph{failure} occurs when such a requirement is not fulfilled, but it is important to distinguish between the occurrence of failures and their actual causes. A failure is caused by a \emph{bug}~(or defect or fault) in the corresponding software, whereas a bug is usually caused by an \emph{error}~(or mistake) made by person.

Since most software systems are dynamic by nature, the underlying code base changes almost on a daily basis. These changes must not adversely affect the correctness of the given software. \emph{Software testing} can reduce the risk of bugs based on observations about the runtime behavior of the software~\cite[6]{sls14}, which is why consistent testing is an indispensable activity that decisively determines the success or failure of software. According to Martin, the test code is even more important than the production code itself: \textcquote{mar13}{You can~(and do) create the system from the tests, \emph{but you can't create the tests from the system.}} Although this is not necessarily true, as shown in section~\ref{subsec:difference-testing}, not only agile software development methodologies such as \gls{xp}\footnote{\url{https://en.wikipedia.org/wiki/Extreme_programming}.} give software testing a central role. Within the V-model\footnote{\url{https://en.wikipedia.org/wiki/V-Model_(software_development)}.}, both activities---development and testing---are equally important. In the field of software testing, the V-model is quite special because it further defines different \emph{test levels} that are widely adopted:
\begin{displaycquote}[41]{sls14}
\begin{description}
\item[Component test] verifies whether each software component correctly fulfills its specification.
\item[Integration test] checks if groups of components interact in the way that is specified by the technical system design.
\item[System test] verifies whether the system as a whole meets the specified requirements.
\item[Acceptance test] checks if the system meets the customer requirements, as specified in the contract and/or if the system meets user needs and expectations.
\end{description}
\end{displaycquote}
Henceforth, the term \emph{unit test} will be used instead of component test as it is highly popular in \gls{oop}.

Regardless of these test levels, a \emph{test case} usually defines various conditions~(e.g. inputs and expected outputs) for its \emph{test object}. If the test object refers to a whole software system, this thesis will explicitly identify it as \emph{\gls{sut}}. Finally, a set of test cases is often combined to a \emph{test suite} where the postconditions of a previous test can be used as the preconditions for a following test.

\subsection{Test Automation}

\emph{Test automation} can be loosely defined as \textcquote[7]{bucp15}{\textelp{} the execution of otherwise manual tests by machines.} On the one hand, this broad definition shows that basically a vast amount of software testing tasks can be automated~(although this does not take the actual costs of automation into account). On the other hand, test automation has its limits. Bucsics et~al. state that this is where testers use their \enquote{\textelp{} intellectual, creative, and intuitive dimension \textelp{}} That is, for instance, exploratively creating new test cases. When test automation takes care of simple and recurring tasks, testers can spend more time doing tasks like this.

\begin{figure}[ht]
\centering
\includegraphics{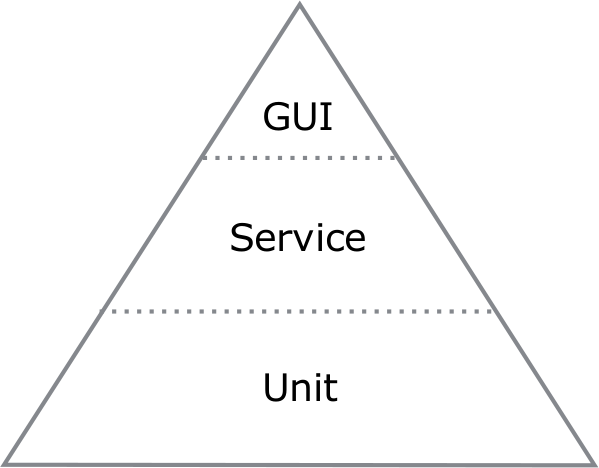}
\captionbelow[Test automation pyramid.]{Test automation pyramid~\cite[312]{coh09}.}
\label{fig:test-pyramid}
\end{figure}

In general, test automation can be applied on any test level---from unit tests to acceptance tests---but the corresponding procedures and tools vary. A widely-used strategy for test automation at different levels is the \emph{test automation pyramid}~(see figure~\ref{fig:test-pyramid}). Accordingly, the foundation should be a large number of unit tests. The reasoning behind this is that these tests are usually faster to execute and easier to write. Furthermore, since unit tests are directly related to code, it is often less complex to locate the cause of a failing test. The service level borrows its name from \gls{soa}, but it is not restricted to these kinds of systems. Without taking the many different service definitions into account, one can map this layer to integration tests. The basic idea is that a group of components, which offers a dedicated service, should be tested in isolation to ensure they work together the intended way. The reason for this \enquote{extra} layer is that tests via the \emph{\gls{gui}}, the tip of the test automation pyramid representing the system tests, tend to be \textcquote{fow12}{\textelp{} brittle, expensive to write, and time consuming to run.} Although the \gls{gui} does not necessarily has to serve as the test interface for the system tests.

However, a common approach for \gls{gui}-based system testing are test robots. Such a test robot uses a script to perform several actions~(e.g. mouse clicks or keyboard strokes) on the \gls{gui}, which usually describes a test case. These scripts are often generated with the aid of \emph{capture and replay}. That is, a test case is performed once by a tester while being recorded~(capture). Afterwards, the recorded test case can be executed arbitrarily often~(replay). Since capture and replay does not require programming skills, it is frequently used by domain experts. Theoretically, these types of tests are good candidates for \emph{regression testing}, i.e. tests \textcquote[75]{sls14}{\textelp{} of a previously tested program following modification to ensure that faults have not been introduced or uncovered \textelp{}} But in practice testers are confronted with different problems:
\begin{displaycquote}[1]{mm09}
\textelp{} these test cases often cause difficulty during software maintenance and regression testing, because relatively minor changes to the GUI can cause a test case to break, or, cease to be executable against an updated version of the software. When such a situation occurs, a large manual effort is often required to repair some subset of the cases in a test suite, or worse yet;
\end{displaycquote}
McMaster and Memon further formalized this as the GUI element identification problem, which is basically the reason why \gls{gui}-based testing has the reputation of being fragile. To overcome this issue, patterns such as page objects~\cite{fow13} have been introduced. This pattern has its origin in browser-based regression testing, where a class~(in terms of \gls{oop}) acts as an interface to a web page. The usually manually created test scripts then use this interface to interact with the \gls{gui}. If the \gls{gui} changes afterwards, only the corresponding page objects have to be adapted instead of all affected tests. Compared with capture and replay, then again these tests become expensive to create. In both cases, the actual test execution is rather slow as the tests have to use the actual \gls{gui}. This is especially painful in regression testing as a complete run of all tests is usually too time consuming. Therefore, there is normally a selection of regression test cases to balance risks and costs~\cite[76]{sls14}. It is also worth mentioning that there are tools, such as PhantomJS\footnote{\url{http://phantomjs.org/}.}, which are capable of running tests \enquote{headlessly}, i.e. without the need of rendering the \gls{gui}. This normally improves execution performance, but comes with various trade-offs. For instance, if there is no \gls{gui}, one can also take no screenshots during test execution for later examination.

\subsection{Monkey Testing}
\label{subsec:monkey-testing}

Although \emph{monkey testing} is often used synonymously for fuzz testing and random testing, the \gls{istqb}\footnote{\url{http://istqb.org/}.} defines these terms as follows:
\begin{displaycquote}{ist16}
\begin{description}
\item[Fuzz Testing] A software testing technique used to discover security vulnerabilities by inputting massive amounts of random data, called fuzz, to the component or system.
\item[Monkey Testing] Testing by means of a random selection from a large range of inputs and by randomly pushing buttons, ignorant of how the product is being used.
\item[Random Testing] A black-box test design technique where test cases are selected, possibly using a pseudo-random generation algorithm, to match an operational profile. This technique can be used for testing non-functional attributes such as reliability and performance.
\end{description}
\end{displaycquote}
Regardless of this, the present thesis will use \textcquote[18]{nym00}{\textelp{} the term \enquote{monkey} to refer broadly to any form of automated testing done randomly and without any \enquote{typical user} bias.} It is believed that the name is derived from the infinite monkey theorem which \textcquote{wik17b}{\textelp{} states that a monkey hitting keys at random on a typewriter keyboard for an infinite amount of time will almost surely type a given text, such as the complete works of William Shakespeare.} In terms of software testing, the idea is that the monkey will cover many---not all, as there is no infinite amount of time---test cases which have not been considered by developers respectively testers.

\begin{listing}
\begin{minted}{java}
Robot robot = new Robot();
Random rand = new Random();
Dimension screenSize = Toolkit.getDefaultToolkit().getScreenSize();
int maxX = screenSize.width;
int maxY = screenSize.height;
int maxLength = 42;

while (true) {
	robot.mouseMove(rand.nextInt(maxX), rand.nextInt(maxY));
	robot.mousePress(InputEvent.BUTTON1_DOWN_MASK);
	robot.mouseRelease(InputEvent.BUTTON1_DOWN_MASK);
	robot.delay(200);

	String inputString = RandomStringUtils // Apache Commons Lang.
			.random(rand.nextInt(maxLength));

	for (char inputChar : inputString.toCharArray()) {
		robot.keyPress(inputChar);
		robot.keyRelease(inputChar);
		robot.delay(10);
	}

	robot.keyPress(KeyEvent.VK_ENTER);
	robot.keyRelease(KeyEvent.VK_ENTER);
}
\end{minted}
\caption[Dumb monkey implementation.]{Dumb monkey implementation~\cite{roe17}.}
\label{lst:dumb-monkey}
\end{listing}

\begin{listing}[hb!]
\begin{minted}{java}
WebDriver driver = new FirefoxDriver();
driver.get("https://retest.de/");
Random rand = new Random();
int maxLength = 42;

while (true) {
	List<WebElement> links = driver.findElements(By.tagName("a"));
	links.get(rand.nextInt(links.size())).click();

	Thread.sleep(500L);

	List<WebElement> fields = driver
			.findElements(By.xpath("//input[@type='text']"));
	WebElement field = fields.get(rand.nextInt(fields.size()));

	String inputString = RandomStringUtils // Apache Commons Lang.
			.random(rand.nextInt(maxLength));
	field.sendKeys(inputString);
	
	Thread.sleep(500L);
}
\end{minted}
\caption[Smart monkey implementation.]{Smart monkey implementation~\cite{roe17}.}
\label{lst:smart-monkey}
\end{listing}

In general, two types of monkeys can be considered: \emph{dumb} and \emph{smart} ones. Smart monkeys usually know their past and current location, where they can go, and are sometimes capable to recognize if a given result conforms to the expected result~\cite{exf11}. They normally retrieve this knowledge from some sort of state table or model of the \gls{sut}~\cite[19]{nym00}. This enables smart monkeys to~(randomly) choose from a set of legal actions, whereas dumb monkeys often waste resources trying to do something illegal in the current state. Furthermore, this knowledge can be used to respect behavior, which dumb monkeys are not able to. For example, when testing an \gls{api}, smart monkeys may know that for a given method \inline{null} parameters are not allowed by design. Therefore, they have the ability to ignore the response or simply skip \inline{null} as an input. Dumb monkeys could cause a \inline{NullPointerException} in such a situation, i.e. they yield a false positive since they are not aware of the preconditions. Nonetheless, dumb monkeys are a good tool to unveil crashes and hangs~\cite[21]{nym00}. Listing~\ref{lst:dumb-monkey} shows a rudimentary implementation of such a monkey. It uses Java's default \inline{Robot} to dispatch mouse clicks and key strokes. Obviously, the monkey is relatively dumb as it simply fires random events without any knowledge about the \gls{sut}. If, for instance, the monkey is testing a website, it may click on an external link from time to time---the monkey then starts to test the internet, which is definitely not the intention. By giving the monkey \emph{context}, it is fairly easy to make it a bit smarter; in order to do so, listing~\ref{lst:smart-monkey} leverages the Selenium\footnote{\url{http://seleniumhq.org/}.} \gls{api}. Although it is still possible that clicks on external links happen, the implementation now can be easily extended to address this issue, e.g. by filtering \inline{links}.

Both types have their advantages and disadvantages as further illustrated by table~\ref{tab:dumb-vs-smart-monkeys}. Nowadays, various industrial tools, e.g. UI/Application Exerciser Monkey\footnote{\url{https://developer.android.com/studio/test/monkey.html}.} for mobile~(Android) applications or \inline{gremlin.js}\footnote{\url{https://github.com/marmelab/gremlins.js/}.} for web applications, are freely available and usually work out of the box. The majority of these tools can be categorized as dumb monkeys because they mostly focus on firing random \gls{gui} events at the \gls{sut} for reliability testing.

\begin{table}[ht]
\small
\centering
\begin{tabular}{l l l} \toprule
&
Dumb Monkeys &
Smart Monkeys \\ \midrule
Applicability &
Early stage &
Later stage \\
Capability &
Limited and basic tests &
Depends on state model \\
Costs &
Low &
Medium to high \\
Number of bugs &
Less &
More \\
Type of bugs &
Crashes and hangs &
Non-functional \\
Good for &
Reliability testing &
Load and stress testing \\ \bottomrule
\end{tabular}
\captionbelow[Comparison of dumb and smart monkey testing.]{Comparison of dumb and smart monkey testing~\cite{exf11}.}
\label{tab:dumb-vs-smart-monkeys}
\end{table}

\subsection{Difference Testing}
\label{subsec:difference-testing}

Almost every developer has to work with legacy code at some point in their career. Strictly speaking, this means that the corresponding code is inherited from someone else. But most people use the adjective \enquote{legacy} when they face code that is difficult to change, because a part of the software system is deprecated and there is no support available anymore. Feathers comes up with a different definition that says that legacy code is simply code with no tests:
\begin{displaycquote}[xvi]{fea04}
Code without tests is bad code. It doesn't matter how well written it is; it doesn't matter how pretty or object-oriented or well-encapsulated it is. With tests, we can change the behavior of our code quickly and verifiably. Without them, we really don't know if our code is getting better or worse.
\end{displaycquote}
This is a dilemma because if there are no tests, how is one supposed to change the code? And if there is no specification, how is one supposed to write a test? One could try to create tests based on old specification documents or do trial-and-error, but this can be time consuming and dangerous. Moreover, Feathers points out that in \enquote{\textelp{} nearly every legacy system, what the system does is more important than what it is supposed to do.} Hence, bug finding is not the actual intention. This is also known as the \emph{oracle problem}, which says that testers rely on~(partial) oracles to decide whether a piece of software behaves correctly~\cite[465]{wey82}. Regression testing sidesteps this problem by using the software itself as the oracle~\cite[521]{barp15}. That is, the results of a previous~(typically stable) version serve as the oracle for the tests. Feathers applies this idea to the situation described above and calls it \emph{characterization testing}~\cite[186--188]{fea04}. As the name suggests, such tests aim to characterize the behavior of the test object in order to document the current behavior---regardless of its correctness. This sort of oracle is named \emph{consistency oracle} as it compares the consistency between two versions~\cite[57]{hof98}.

One usually starts with an assertion that will fail, captures the given output, and uses the result to adapt the expected value of the assertion to make it pass. \cite{man12} also suggests to \enquote{bombard} the test object with a sufficiently large amount of random inputs to increase the total number of test cases while using a constant seed to make the tests repeatable. The captured results are often referred to as the \emph{\gls{gm}}\footnote{The term usually describes a build within the software release lifecycle that is ready to be delivered~\cite{wik17c}.} since they represent a temporary oracle; this is also why the technique is further known as golden master testing. As soon as the characterization tests helped to form enough understanding of the test object, one can modify the corresponding code and replay the tests during that process to see if the previous behavior has changed in an unintended way. But rather than doing this all by hand, libraries such as Approval Tests\footnote{\url{http://approvaltests.com/}.} embrace characterization testing by providing facilities to store and compare the \gls{gm}. Other implementations make use of similar techniques: TextTest\footnote{\url{http://texttest.sourceforge.net/}.} utilizes log files to serve as the \gls{gm}~(called approval testing), whereas Depicted\footnote{\url{https://github.com/bslatkin/dpxdt/}.} is based on screenshots~(called perceptual diff testing). The problem is that characterization testing, including its derivatives, has two downsides:
\begin{enumerate}
\item It is not possible to~(semantically) compare unknown formats such as PDFs or \glspl{gui}.
\item Ignoring volatile and unimportant elements~(e.g. a time/date string) often requires a lot of work.
\end{enumerate}
This is due to the way the \gls{gm} and a divergent result are typically compared, which is basically a text- or pixel-oriented diff between two files. Instead of checking for differences on this level, one actually wants to see differences in terms of \emph{behavior}---especially in \gls{gui}-based regression testing, where the~(dynamic) runtime behavior is not directly related to the~(static) underlying source code.

\begin{figure}
\centering
\includegraphics[width=\textwidth]{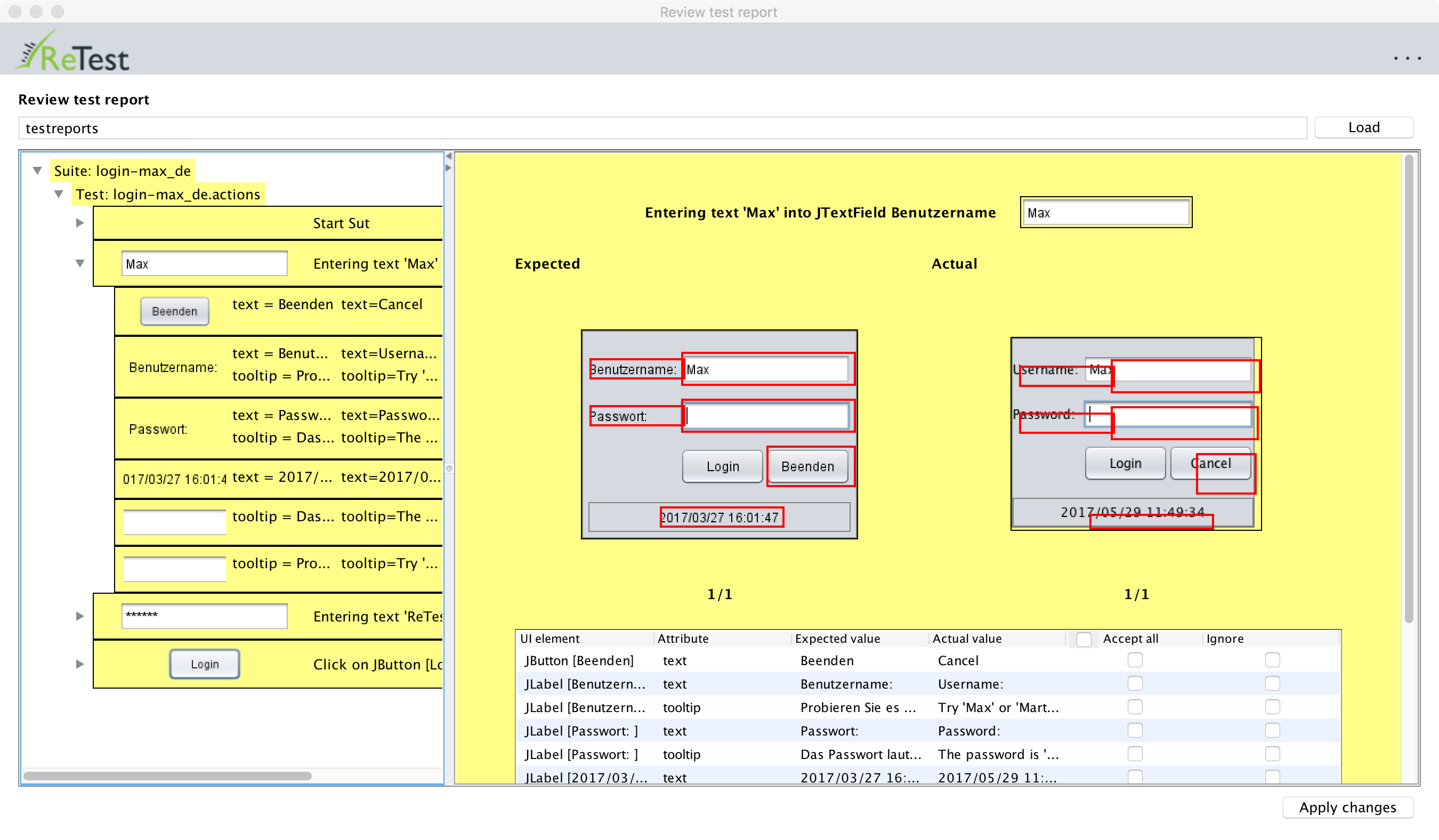}
\captionbelow{Difference testing report in ReTest.}
\label{fig:difference-testing-report}
\end{figure}

\begin{figure}[h]
\centering
\includegraphics{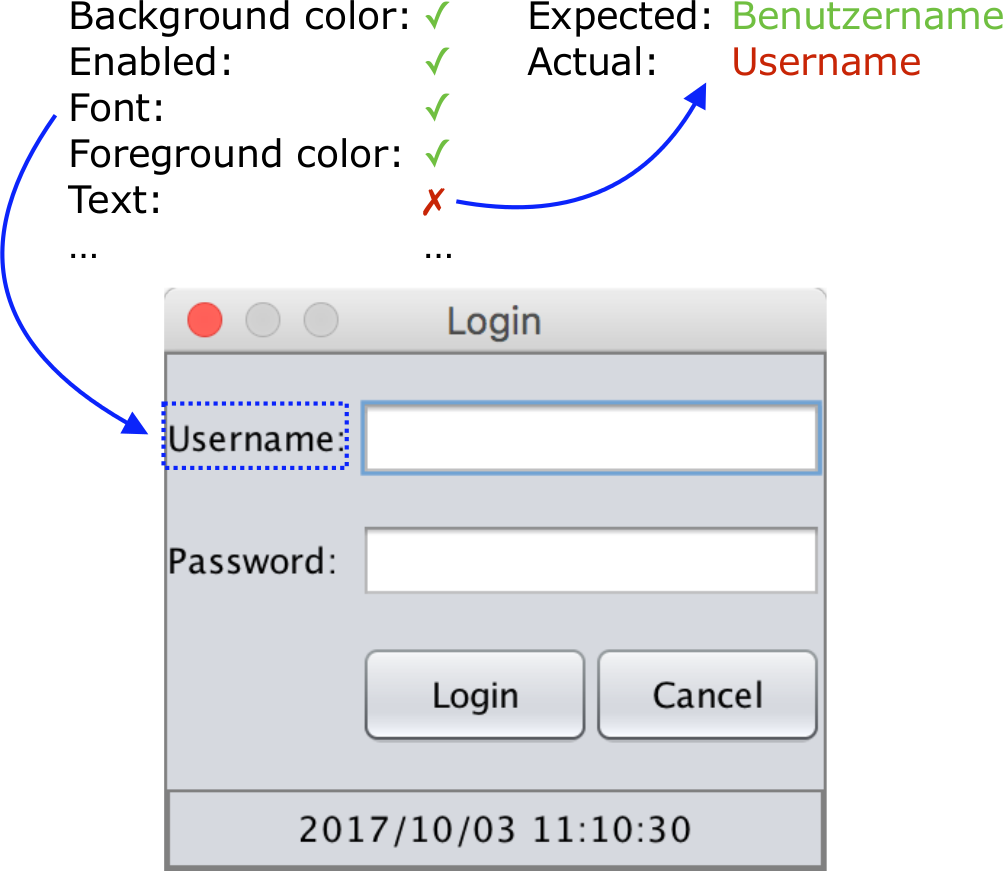}
\captionbelow{Native comparison with difference testing.}
\label{fig:difference-testing-comparison}
\end{figure}

Back in 2016, ReTest extended characterization testing for this purpose and named it \emph{difference testing}~\cite[4]{ret17}. If, for instance, \glspl{gui} are being compared, difference testing aims to only show differences that belong to the behavior by capturing the whole visible state of the \gls{sut}. Figure~\ref{fig:difference-testing-report} shows an example of a ReTest test report, in which a \gls{gui} was translated from German~(expected) to English~(actual). As can be seen, the given format---Swing-based \glspl{gui}---is compared natively. That is, individual \gls{gui} components and their attributes are used instead of text or pixels as illustrated in figure~\ref{fig:difference-testing-comparison}. In the given example, the text attribute caused a difference because it changed from \enquote{Benutzername} to \enquote{Username}. Such a difference can be either accepted or ignored to update the \gls{gm}, just like a \gls{vcs} would do. In the case of a regression, the test report can be used to document and reproduce the unintended change. As a result, multiple advantages can be achieved:
\begin{itemize}
\item Because more information is available for component identification, the GUI element identification problem becomes less of an issue.
\item Less visible properties, such as tool tips or enabled states, are taken into account, too.
\item No assertions need to be defined, which leads to a much faster test creation.
\item The \gls{gm} can be used as a lightweight documentation of the \gls{sut}.
\item Less maintenance effort since differences can be simply adopted or ignored---similar to a \gls{vcs}---which can reduce test flakiness\footnote{A flaky test is a test that fails intermittently. This can happen if the test~(inadvertently) depends on timing or environmental conditions.}.
\item Test cases can be automatically generated because there is no oracle problem.
\end{itemize}
Nonetheless, difference testing comes with the same drawback as other consistency oracles: historic bugs may remain undiscovered~\cite[57]{hof98}. But, as already mentioned, when working with legacy code or during regression testing, this problem can be neglected. Although difference testing is currently only implemented by ReTest for \gls{gui}-based system testing of Java applications, the technique can be adapted for other test levels and formats as well. For example, integration testing of service layer responses in \gls{json}.

\section{Machine Learning}
\label{sec:machine-learning}

\emph{\Gls{ml}} is a subfield of \gls{ai}, which basically studies algorithms that give software the ability to learn from data. These software systems may improve their performance over time to, for instance, predict future outcomes based on learning of historical data. According to Mitchell, \gls{ml} can be described more precisely as follows:
\begin{displaycquote}[2]{mit97}
A computer program is said to \textbf{learn} from experience $E$ with respect to some class of tasks $T$ and performance measure $P$, if its performance at tasks in $T$, as measured by $P$, improves with experience $E$.
\end{displaycquote}
Nowadays, the application of \gls{ml} ranges from weather forecasting over stock trading to autonomous driving and many other domains, where it is used to perform classification, regression, and clustering among other things. For illustration, one can consider the task $T$ of learning the board game checkers\footnote{\url{https://en.wikipedia.org/wiki/Draughts}.}, where the performance $P$ is being measured as the ratio of games won against opponents. $E$ could be attained via different types of training experience; in general, the training experience can be classified into two independent properties:
\begin{enumerate}
\item Direct or indirect?
\begin{description}
\item[Direct learning] Learning through direct training examples of concrete board states and the optimal move for each of these states.
\item[Indirect learning] Learning through indirect feedback of move sequences and their final result, i.e. won, lost, or draw.
\end{description}
\item Teacher or not?
\begin{description}
\item[Supervised learning] A teacher provides board state examples and rates the moves of the learner based on the correct solution.
\item[Unsupervised learning] The learner has no knowledge about the training data and has to discover patterns in it.
\item[Reinforcement learning] The learner is punished or rewarded for games he loses respectively wins against himself or others.
\end{description}
\end{enumerate}

Throughout this thesis, the problem of improving $P$ in regards to $T$ will be reduced to learning a particular \emph{target function} $f$~\cite[7]{mit97}. In the selected example of learning checkers, this could be $f: B \rightarrow M$, where $B$ and $M$ are the sets of legal board states and legal moves. It is important to note that $f$ denotes the optimal solution for the given task, which is usually unknown and very difficult to learn. Therefore, most of the time \gls{ml} algorithms are only expected to deliver an approximation of $f$---the \emph{hypothesis} $h$. In order to do so, a learning algorithm searches through a \emph{hypothesis space} $H$~(e.g. linear functions or logical descriptions) and adjusts its internal parameters until it finds a hypothesis $h \approx f$. The composition of a specific learning algorithm combined with a specific hypothesis space is normally referred to as a \emph{model}. Picking an appropriate hypothesis space can be difficult since there is an inevitable trade-off between the expressiveness of $H$ and the complexity of finding $h$ in it~\cite[652--653]{rn03}. This is also true when choosing among multiple consistent hypotheses, which requires another trade-off between the computational complexity of $h$ and its accuracy.

Besides the type of training experience, Mitchell further mentions that it is important how well the available data \textcquote[6]{mit97}{\textelp{} represents the distribution of examples over which the final system performance $P$ must be measured.} If, for example, the learner only gains experience by playing checkers against himself, he probably misses crucial moves typically played by humans. Consequently, learning is most effective when the training experience follows the distribution of future inputs. This assumption is crucial as the majority of \gls{ml} algorithms are based on induction. Accordingly, if $h$ is derived from a statistically significant amount of training examples, it will approximate $f$ over unknown examples, too~\cite[23]{mit97}. In order to evaluate this for a particular model, the data is commonly split into two sets: \emph{training and test}~(or validation). As the names suggest, the training set is used to learn a hypothesis, whereas the test set is used to validate the accuracy of this hypothesis. A popular heuristic is to use $\frac{2}{3}$ for training and $\frac{1}{3}$ for testing without affecting the overall distribution. When both sets are strongly dissimilar, then the trained model tends to \emph{overfit}. That is, a hypothesis $h \in H$ is said to overfit if another hypothesis $h' \in H$ exists which achieves a higher accuracy over the entire distribution. One way to reduce the risk of overfitting---and to deal with small data sets as well---is to use $k$-fold cross-validation, where statistical cross-validation\footnote{\url{https://en.wikipedia.org/wiki/Cross-validation_(statistics)}.} is applied $k$ times with different partitions for training and testing in each run.

\begin{listing}
\begin{minted}{text}
wheels,chassis,pax,vtype
4,2,4,Car
9,20,25,Bus
5,14,18,Bus
5,2,1,Car
9,17,25,Bus
1,1,1,Bike
4,4,2,Car
9,15,36,Bus
1,1,1,Bike
5,1,4,Car
4,2,1,Car
\end{minted}
\caption[Training data as CSV.]{Training data as CSV~\cite[102]{bel15}.}
\label{lst:csv-data-example}
\end{listing}

Listing~\ref{lst:csv-data-example} shows an excerpt of \gls{csv} including a header. Each column stands for a selected \emph{feature}, which is used as the input for the learner. Features are usually constructed from raw input variables and may require further processing to be interpretable by the learning algorithm\footnote{An introduction to variable and feature selection can be found in \cite{ge03}.}. Each row is a $n$-dimensional \emph{feature vector}, representing some sort of object instance. In the given example, the classification of vehicles should be learned in a supervised manner. The training experience is supervised because the last column consists of \emph{labels}, specifying the desired output. Hence, in this case, $n$ equals the number of features plus the label. In general, two basic feature types do exist:
\begin{enumerate}
\item Quantitative~(numerical).
\item Qualitative~(categorical).
\end{enumerate}
Quantitative features can be further categorized into discrete~(finite) and continuous~(infinite) features, whereas qualitative features are either ordinal~(ordered) or nominal~(unordered). Accordingly, the example consists of the following types:
\begin{itemize}
\item $\text{wheels} \in \mathbb{N}$: number of wheels $\Rightarrow$ quantitative, discrete.
\item $\text{chassis} \in \mathbb{N}$: chassis length in meters $\Rightarrow$ quantitative, discrete.
\item $\text{pax} \in \mathbb{N}$: number of passengers $\Rightarrow$ quantitative, discrete.
\item $\text{vtype} \in \{\text{Bike}, \text{Bus}, \text{Car}\}$: vehicle type $\Rightarrow$ qualitative, unordered.
\end{itemize}
The way the data is finally normalized, heavily depends on the present learning algorithm as well as the given task. Being objective by standardizing all features is considered to be a good starting point, but if some features are intrinsically more important, one should assign the weights of these features based on domain knowledge~\cite[11]{kr90}.

\subsection{Artificial Neural Networks}
\label{subsec:anns}

\begin{figure}[ht]
\centering
\includegraphics{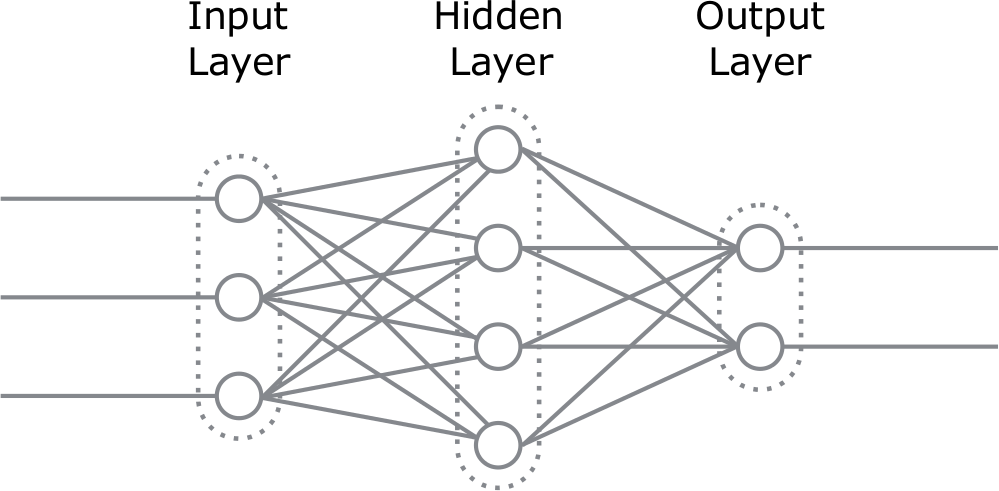}
\captionbelow{Basic structure of multilayer networks.}
\label{fig:multilayer-networks}
\end{figure}

\emph{\Glspl{ann}} are an \gls{ml} model based on the parallel architecture of animal brains, which form a system of highly interconnected \emph{neurons}~\cite[91]{bel15}. A neuron is a cell that is able to transmit and process chemical as well as electrical signals. Analogously, \glspl{ann} consist of densely interconnected units, each taking an arbitrary number of inputs $x_1, x_2, \ldots, x_n \in \mathbb{R}$ and producing an output $o \in \mathbb{R}$. The structure of an \gls{ann} can be loosely described as a graph. Most of the time this graph constitutes an acyclic, feed-forward network, in which the units are grouped into interconnected layers. Although these layers can come in many different forms, using three distinct layers is a widely-used variant. In this case, the units of an input, a hidden, and an output layer are sequentially connected to each other~(see figure~\ref{fig:multilayer-networks}). When \glspl{ann} are involved that use more than one hidden layer, it is often referred to as \emph{deep learning}~\cite{wik17a}.

\begin{figure}
\centering
\includegraphics{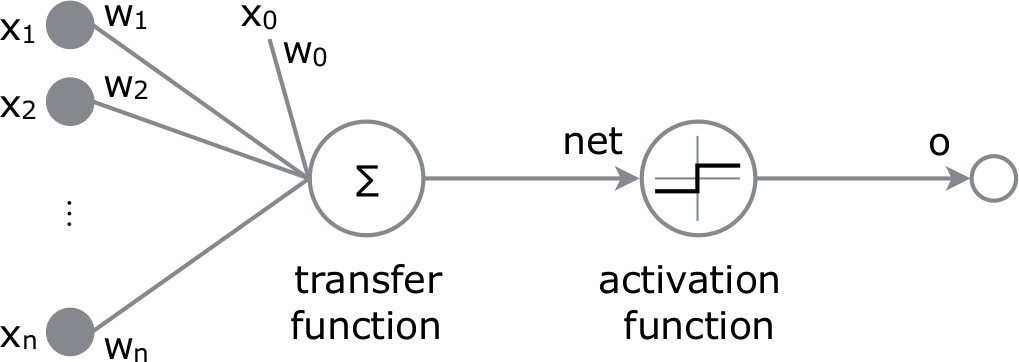}
\captionbelow[Perceptron unit.]{Perceptron unit~\cite[87]{mit97}.}
\label{fig:perceptron}
\end{figure}

There exist several designs for the simple units of an \gls{ann}, one well-known type is the \emph{perceptron} as illustrated in figure~\ref{fig:perceptron}. Each incoming edge corresponds to an input $x_i$, where $w_i$ describes the \emph{weight} of that input. The perceptron first computes the linear combination of the inputs and their weights~(the \emph{transfer function} computing $net$); afterwards, it outputs a 1 or a $-1$ depending on whether the result is greater or smaller than a certain threshold~(the \emph{activation function} computing $o$). Just like Mitchell, this thesis will denote that threshold as the negative weight~$-w_0$, combined with the additional constant input $x_0 = 1$. All weights are enclosed in the vector $\vec{w} = (w_0, w_1, \ldots, w_n)$, whereas the inputs are in $\vec{x} = (x_0, x_1, \ldots, x_n)$. The perceptron is now formally defined by\footnote{Unless explicitly stated otherwise, all formulas in this section are taken from~\cite{mit97}.}:
\begin{equation*}
o(\vec{x}) = \sgn(\vec{w}\vec{x}) 
\end{equation*}
Where:
\begin{equation*}
\sgn(net) =
\begin{cases}
1 & net > 0 \\
-1 & \text{else}
\end{cases}
\end{equation*}
Since learning a perceptron---or an \gls{ann} in general---essentially means picking the weights $w_0, w_1, \ldots, w_n$, the hypothesis space $H$ equals the set of all possible weight vectors $\vec{w} \in \mathbb{R}^{n + 1}$. Accordingly, a hypothesis $h$ spans a hyperplane in the $n$-dimensional space, separating the instances with respect to their output. Of course, this is only possible if the data is \emph{linearly separable}. A set of instances is said to be linearly separable if there exists at least one straight line that can separate the data.

To understand how multilayer networks learn, it is helpful to first understand how weights for single units are determined. In case of the perceptron, one way to do this is by starting with random weights that are iteratively adapted until all instances are correctly classified. In each iteration, the current weights are updated by adding the following delta according to the \emph{perceptron training rule}:
\begin{equation*}
\Delta w_i = \eta (t - o) x_i
\end{equation*}
The term $(t - o)$ quantifies the delta between the target output $t$ and the current output $o$ for the input $x_i$. When both values are equal, i.e. the desired target output is reached, the difference between $t$ and $o$ is 0---and so is $\Delta w_i$. Hence, the current weight is not updated anymore. $\eta$ is a positive constant named the \emph{learning rate}. The smaller it is, the smaller is also the change to the weight in each iteration. If the learning rate is too small, the convergence may become very slow; if it is too large, the algorithm may overleap the global minimum.

A problem with the perceptron training rule is that it is only guaranteed to converge when the data is linearly separable, whereas the \emph{delta rule} is capable to handle non-linearly separable data as well. The delta rule uses \emph{gradient descent}\footnote{\url{https://en.wikipedia.org/wiki/Gradient_descent}.} to search the hypothesis space, which is the basis to learn networks with many layers respectively units. Here, $\Delta w_i$ is defined as follows:
\begin{equation}
\label{eq:single-delta-rule}
\Delta w_i = -\eta \frac{\partial E}{\partial w_i}
\end{equation}
The learning rate $\eta$ now determines the gradient descent step size, whereas $\frac{\partial E}{\partial w_i}$ means that, in each step, the weights in $\vec{w}$ are changed so that they follow the direction that produces the steepest decrease---which is why $\eta$ is preceded by a negative sign---in terms of the \emph{training error} $E$. A common way to measure this in relation to the training set is:
\begin{equation}
\label{eq:single-error-function}
E(\vec{w}) \equiv \frac{1}{2} \sum_{d \in D} (t_d - o_d)^2
\end{equation}
$d$ denotes an instance in the set of training examples $D$, where $t_d$ and $o_d$ are the corresponding target and current output. Although $D$ influences the error as well, the assumption is that the set does not change during training, which makes $E$ a function exclusively of $\vec{w}$. Thus, the training error is simply defined as half the squared difference between $t_d$ and $o_d$. By differentiating $E$ from equation~\ref{eq:single-error-function} and substituting the result into the equation~\ref{eq:single-delta-rule}, one can retrieve the final gradient descent update rule:
\begin{equation*}
\Delta w_i = \eta \sum_{d \in D} (t_d - o_d) x_{id}
\end{equation*}

Although it is now possible to handle non-linearly separable data, one must consider the fact that even multiple layers of linear units, such as the perceptron, are still only capable to produce linear functions. Therefore, a unit with a non-linear output is needed in order to express highly non-linear decision surfaces. Very popular is the \emph{sigmoid} unit, which uses the same transfer function as the perceptron unit, but the activation function is based on the sigmoid function:
\begin{equation*}
o(\vec{x}) = \sigma(\vec{w}\vec{x})
\end{equation*}
Where:
\begin{equation*}
\sigma(net) = \frac{1}{1 + e^{-net}}
\end{equation*}
The plot of the sigmoid function shows a smooth \enquote{S}-curve that monotonically increases with its input, producing an output between 0 and 1. Since it maps all inputs into this interval, it is sometimes also called the squashing function.

To now learn the weights for multilayer networks with a fixed number of units and interconnections, the \emph{backpropagation}\footnote{\url{https://en.wikipedia.org/wiki/Backpropagation}.} algorithm leverages gradient descent to minimize the network error. First of all, $E$ must be redefined to sum the errors of the network's output units:
\begin{equation*}
E(\vec{w}) \equiv \frac{1}{2} \sum_{d \in D} \sum_{k \in K} (t_{kd} - o_{kd})^2
\end{equation*}
The difference here is that $K$ is the set of units in the output layer, where $t_{kd}$ and $o_{kd}$ correspond to the target respectively current output of the $k$-th output unit for the training example $d$. It is important to note that the error surface now may have multiple local minima, which is why gradient descent can get stuck locally instead of finding the global minimum. Added to this are the following definitions:
\begin{itemize}
\item A \emph{node} is an input of the network or an output of a unit within the network and is assigned with an index.
\item $x_{ij}$ and $w_{ij}$ is the input respectively the weight between node $i$ and $j$.
\item $\delta_i$ is the error term for unit $i$, defined as $\delta_i = -\frac{\partial E}{\partial net_i}$.
\end{itemize}
Especially the extension of the error term is necessary because the training examples only provide the target output $t_{kd}$ for the output units, not for the hidden units. For this reason, the error is propagated backwards---which is where the name derives from. Consequently, in case of an output unit $k$, the error term is:
\begin{equation*}
\delta_k = o_k (1 - o_k) (t_k - o_k)
\end{equation*}
$(t_k - o_k)$ is just the same as in the delta rule, but it is additionally multiplied by the derivative of the sigmoid function $o_k (1 - o_k)$. For a hidden unit $i$ in layer $m$, the error is defined as follows:
\begin{equation*}
\delta_i = o_i (1 - o_i) \sum_{j \in m + 1} w_{ji} \delta_j
\end{equation*}
The term describes the summed errors of the next deeper layer $m + 1$ influenced by unit $i$, where each error is multiplied by the weight $w_{ji}$ between these two units. That means, the weight update rule is:
\begin{equation*}
\Delta w_{ji} = \eta \delta_j x_{ji}
\end{equation*}
This is also known as the \emph{stochastic gradient descent} version of the backpropagation algorithm, which comes in many flavors\footnote{\url{https://en.wikipedia.org/wiki/Stochastic_gradient_descent\#Extensions_and_variants}.}. Generally, each propagation is immediately followed by a weight update. This procedure is repeated until one or more termination criteria are met~(e.g. an error threshold combined with a global timeout). Picking an adequate criterion is important as to few iterations may lead to a low accuracy, whereas to many can cause overfitting.

\section{Evolutionary Computing}
\label{sec:evolutionary-computing}

Just as \gls{ml}, \emph{\gls{ec}} is another subfield of \gls{ai}, \textcquote[70]{zhap11}{\textelp{} inspired by the mechanisms of biological evolution and behaviors of living organisms.} \gls{ec} is traditionally applied to problems such as optimization, modeling, and simulation, but over the past years more and more real-world problems have been addressed. For instance, \gls{sbse} applies \gls{ec} to software engineering problems such as performance optimization or automatic maintenance~\cite{harp12, hmz12}. \Gls{sbst} is a subfield of \gls{sbse}, which focuses on the use of \gls{ec} within the context of software testing~\cite{hjz15, mcm11}; an example is ReTest itself, section~\ref{subsec:retest-ga} explains in detail how \gls{ec} is used here for the generation of regression tests.

\begin{table}[ht!]
\small
\centering
\begin{tabular}{r c l} \toprule
Evolution &
&
Problem solving \\ \midrule
Environment &
$\Leftrightarrow$ &
Problem \\
\makebox[\widthof{Candidate solution}][r]{Individual} &
$\Leftrightarrow$ &
Candidate solution \\
Fitness &
$\Leftrightarrow$ &
Quality \\ \bottomrule
\end{tabular}
\captionbelow[EC metaphor linking evolution to problem solving.]{EC metaphor linking evolution to problem solving.~\cite[14]{es03}.}
\label{tab:ec-metaphor}
\end{table}

Fundamentally, \gls{ec} relates natural evolution to trial-and-error problem solving as illustrated by table~\ref{tab:ec-metaphor}. Eiben and Smith describe this as follows\footnote{Words are not emphasized in the original source.}:
\begin{displaycquote}[13]{es03}
A given \emph{environment} is filled with a population of \emph{individuals} that strive for survival and reproduction. The \emph{fitness} of these individuals is determined by the environment, and relates to how well they succeed in achieving their goals. In other words, it represents their chances of survival and of multiplying. Meanwhile, in the context of a stochastic trial-and-error~(also known as generate-and-test) style problem solving process, we have a collection of \emph{candidate solutions}. Their \emph{quality}~(that is, how well they solve the \emph{problem}) determines the chance that they will be kept and used as seeds for constructing further candidate solutions.
\end{displaycquote}
While \gls{ec} includes many different implementations of the same basic idea, like \gls{ep}\footnote{\url{https://en.wikipedia.org/wiki/Evolutionary_programming}.} or \glspl{es}\footnote{\url{https://en.wikipedia.org/wiki/Evolution_strategy}.}, this section will focus particularly on \emph{\glspl{ga}} since they serve as the foundation for ReTest's code coverage-optimized test generation mechanism. However, regardless of the concrete implementation, all \glspl{ea} are based on the Darwinian theory of evolution~\cite{dar59}. Eiben and Smith further point out that for the purpose of \gls{ec}, Darwin's theory as well as genetics can be simplified as follows: Each individual represents a dual entity with an invisible code~(its \emph{genotype}) and observable traits~(its \emph{phenotype}). The phenotypical properties~(e.g. strong muscles or attractive sent) define the success in terms of survival and reproduction. Hence, natural and sexual selection act on the phenotype level. New individuals can have one single parent~(\emph{asexual reproduction}) or two parents~(\emph{sexual reproduction}), but in both cases their genomes are not identical in comparison with their parent genomes. This is due to small reproductive variants and, additionally, in the case of sexual reproduction, because of the combination of two parents genotypes. Consequently, genotypical variations translate to phenotypical variants and, therefore, are a subject to natural and sexual selection, too. This is also why some scientists argue that one should think about \enquote{gene pools} with competing genes, rather than populations with individuals~\cite{daw76}.

\begin{figure}[ht]
\centering
\includegraphics{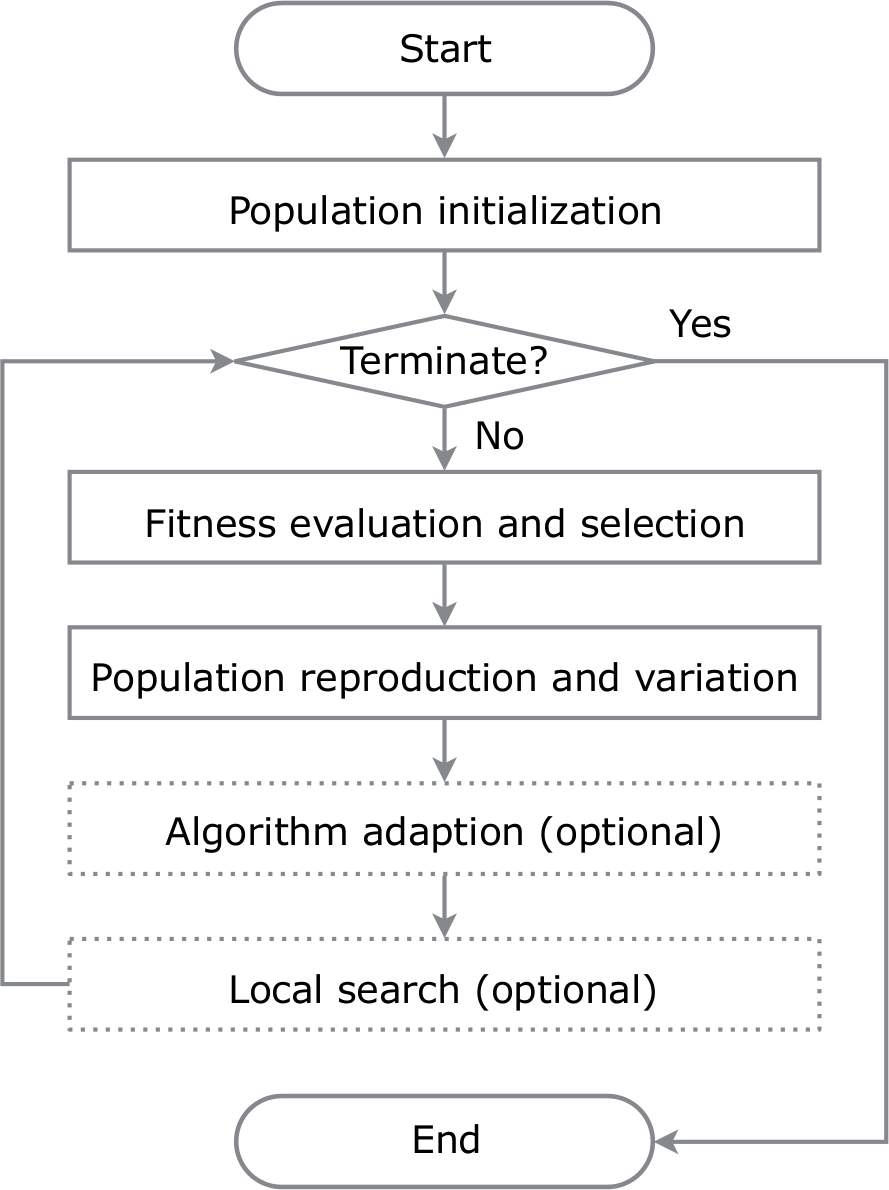}
\captionbelow[General EC framework.]{General EC framework~\cite[68]{zhap11}.}
\label{fig:ec-framework}
\end{figure}

From an algorithmic perspective, all \gls{ec} implementations also share a similar framework with three fundamental and two optional operations~(see figure~\ref{fig:ec-framework}). First of all, the algorithm starts with the initialization of the population, which is usually seeded with random individuals. Sometimes problem-specific heuristics are used, although this means a trade-off between the fitness of the initial population and the computational effort~\cite[34]{es03}. Afterwards, the \emph{fitness function} evaluates the quality of the individuals and the best candidate solutions are selected. This is one of the most crucial steps as the calculated fitness defines the requirements the population should adapt to over time. This also requires an abstraction of the real-world problem, bridging to the problem-solving space of the \gls{ea}. Yet, individuals with a low fitness are often given a small chance to survive to, for instance, avoid local minima. The third step takes the previously selected individuals and uses them for reproduction and variation. For this purpose, operators are grouped into two types based on their arity: \emph{mutation}~(unary) and \emph{crossover}~($n$-ary). Mutation generates a new, slightly modified child from a given genotype~(i.e. asexual reproduction), whereas crossover recombines the genotypes of usually two parents into a single child~(i.e. sexual reproduction). In addition, \gls{aec} performs \emph{algorithm adaption}~\cite{zhap12}, where the configuration of the algorithm is controlled dynamically. For instance, an \gls{aec} algorithm may adapt its mutation probability over time. \Glspl{ma} also use \emph{local search}~\cite{olc10} techniques such as hill climbing\footnote{\url{https://en.wikipedia.org/wiki/Hill_climbing}.} to improve their converge performance. All these operations---apart from the population initialization---are repeated until a particular termination condition is met. This can be a time limit, an upper bound for the number of generations, or something domain-specific~(e.g. a code coverage criterion when generating test cases). One must note that \glspl{ea} are stochastic, hence, there is no guarantee that an optimum is ever satisfied, which is why it is important to include stopping criteria that can be met eventually.

\Glspl{ga} are the most popular type of \gls{ec}, driven by various factors. For example, \glspl{ga} \textcquote[250]{mit97}{\textelp{} are easily parallelized and can take advantage of the decreasing costs of powerful computer hardware.} According to Eiben and Smith, a \gls{ga} traditionally has a fixed workflow~\cite[99--100]{es03}: given a population of $\mu$ individuals, parent selection creates an intermediary population of the same size with possible duplicates. Afterwards, this intermediary population is shuffled in order to get random pairs. Consecutive pairs are then used for crossover with a probability of $p_c$, in which children replace their parents immediately. The result undergoes mutation, where each mutable part is mutated with a probability of $p_m$. The new intermediary population represents the next generation, which replaces the previous one entirely. Depending on the configuration of $\mu$, $p_c$, and $p_m$, there is a small chance that individuals remain unchanged between generations.

\begin{figure}
\centering
\includegraphics{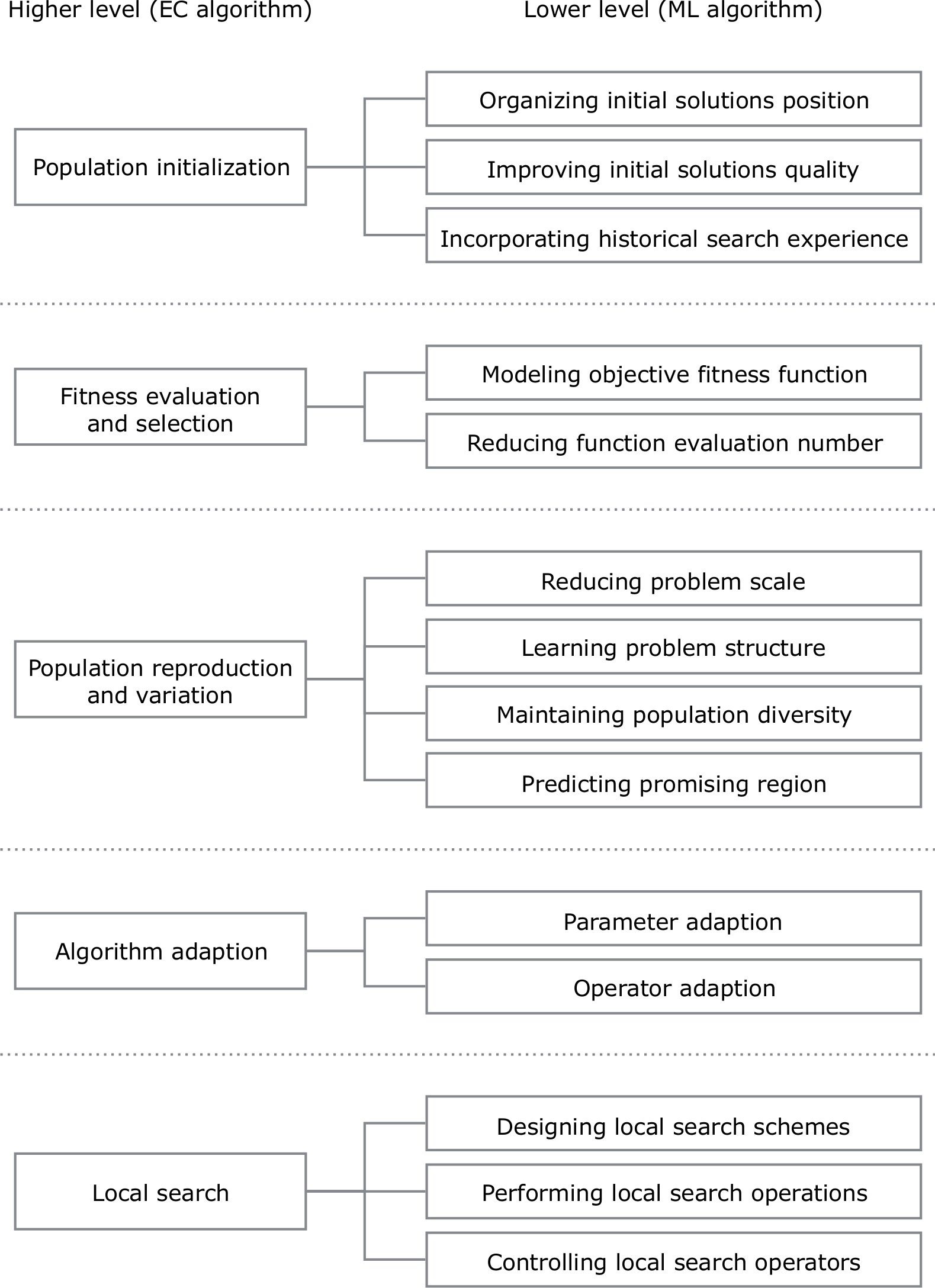}
\captionbelow[MLEC taxonomy.]{MLEC taxonomy~\cite[70]{zhap11}.}	
\label{fig:mlec-taxonomy}
\end{figure}

Due to the recent rise of \gls{ml}, more and more approaches aim to improve \gls{ec} with the aid of \gls{ml}---namely \gls{mlec}. According to Zhang et al., \gls{mlec} algorithms \textcquote[69]{zhap11}{\textelp{} have been proven to be advantageous in both convergence speed and solution quality.} Such algorithms usually extract historical information to understand the given search behavior, which is then used to assist future searches for the global optimum. The authors also provide a taxonomy~(see figure~\ref{fig:mlec-taxonomy}) that classifies the existing research spectrum and which illustrates how \gls{ml} can be used to enhance \glspl{ea} based on the previously described operations. For example, an \gls{mlec} algorithm may use historical search experience to improve the quality of the initial population~\cite{lm04, wy09}. Although \gls{mlec} algorithms come a with computational burden and, therefore, require a trade-off between their benefits and the additional costs, they appeal for a wider range of complex real-world applications because of their improved search speed and accuracy~\cite[74]{zhap11}.

\subsection{ReTest's Genetic Algorithm}
\label{subsec:retest-ga}

As mentioned before, ReTest uses a \gls{ga} to generate code coverage-optimized regression tests during monkey testing. This becomes possible by difference testing, which circumvents the oracle problem. This section describes in detail how the underlying \gls{ga} works.

ReTest started as an extension of EvoSuite~\cite{fa11} for \gls{gui}-based Java applications. EvoSuite itselfs generates unit tests with the aid of a \gls{ga} and mutation testing\footnote{\url{https://en.wikipedia.org/wiki/Mutation_testing}.}. The \gls{ga} optimizes the test suite generation for code coverage, whereas mutation testing is used to rate the importance of assertions. Since its first release back in 2011, a vast amount of optimizations have been introduced and EvoSuite has won several tool competitions\footnote{\url{http://evosuite.org/publications/}.}. However, generating tests at the unit level often results in non-sensical test cases; they often cause false failures, \textcquote[1423]{gfz12a}{\textelp{} created through violations of implicit preconditions, but that never occur in the actual application.} The predecessor of ReTest---namely \gls{exsyst}---sidestepped this issue by only using the \gls{gui}, hence, a system interface:
\begin{displaycquote}[68]{gfz12b}
As system input is controlled by third parties, the program must cope with every conceivable input. If the program fails, it always is the program's fault: At the system level, every failing test is a true positive. The system interface thus acts as a precondition for the system behavior.
\end{displaycquote}

While the EvoSuite team discontinued working on \gls{exsyst}, ReTest adopted its approach and further enhanced it. When regression tests are generated, the \gls{ga} optimizes on whole test suites, rather than individual test cases.  This not just leads to better results on average~\cite{rojp16}, but also avoids excessive test lengths. Moreover, a granularity of test suites minimizes the problems that would arise when mutation or crossover is being applied to a test case. To be more specific, a \emph{test suite} $S$ consists of several \emph{test cases} $C_0, C_1, \ldots, C_n$, each containing a sequence of \gls{gui} \emph{actions}~$(a_0, a_1, \ldots, a_m$). When such a sequence is modified, it may not be executable anymore as an action might become infeasible within a certain state.

\begin{figure}[hb]
\centering
\includegraphics{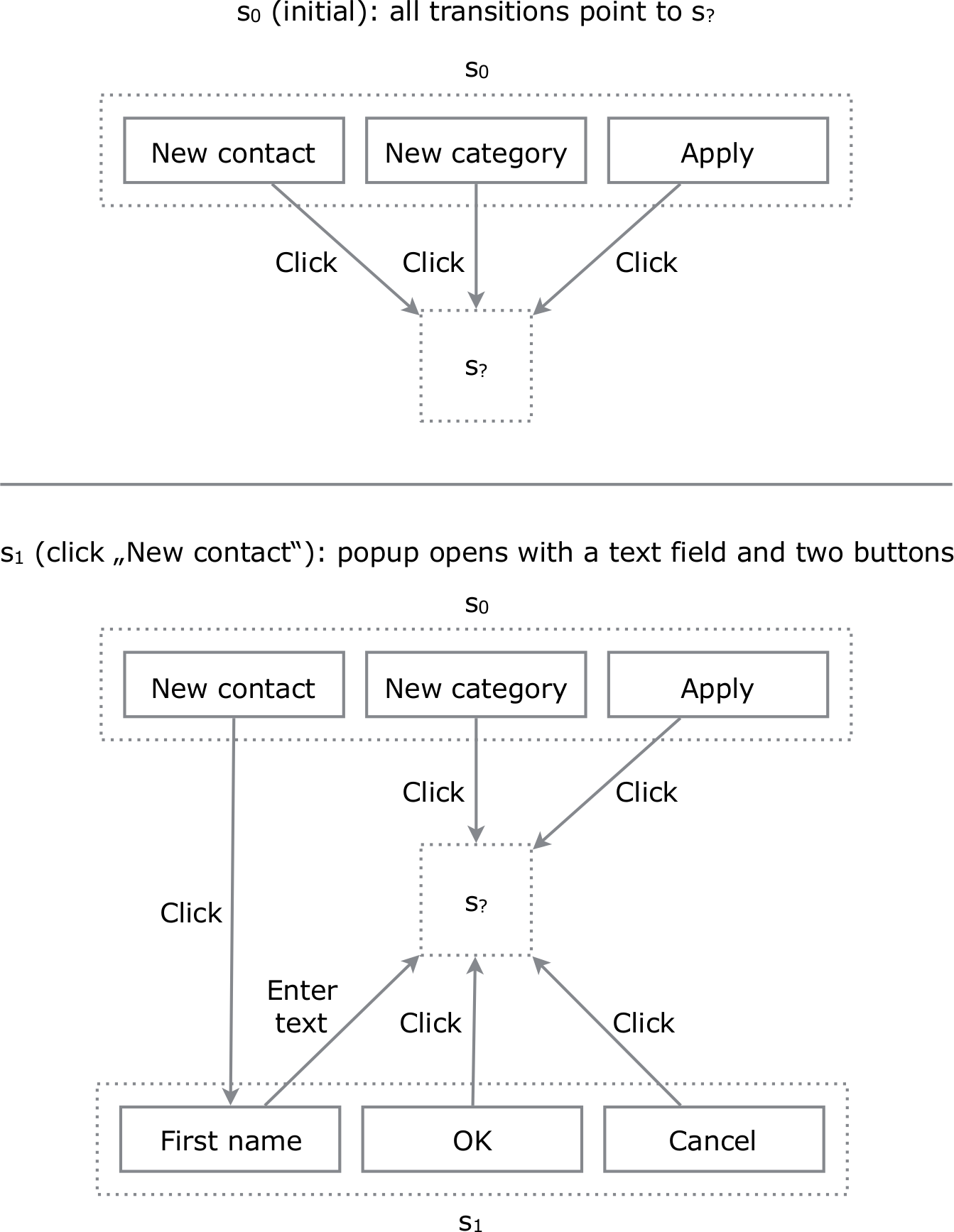}
\captionbelow[State graph model in ReTest.]{State graph model in ReTest~\cite[71]{gfz12b}.}	
\label{fig:state-graph}
\end{figure}

The initial population is composed of randomly generated test suites, in which each contains $k$ test cases, where $k$ is randomly chosen from the interval $[1,\; 10]$. The actions are selected with the aid of monkey testing, hence, again randomly. But in each step, previously unexplored states are given precedence. This is achieved with the help of a \gls{nfa} of the \gls{sut}---the \emph{state graph}. It represents a behavioral model of the \gls{sut}, observed via the \gls{gui}, and is updated iteratively. The model is said to be non-deterministic because this observation may not completely include all variables that determine the overall state of the \gls{sut}~(e.g. time or the state of an external system such as a database). Consequently, a transition does not necessarily lead to the same resulting state when executed later on. A state tells which actions, represented by transitions, on which \gls{gui} components are available in a certain state. Like any \gls{nfa}, the state graph can be formally defined using a 5-tuple $(S, \Sigma, \Delta, s_0, T)$:
\begin{itemize}
\item $S$: the finite set of states, where each state $s_i$ represents a distinct \gls{sut} state.
\item $\Sigma$: the finite set of actions, in which each action $a_k$ represents a possible \gls{gui} action.
\item $\Delta$: the partially defined transition function $\Delta : S \times \Sigma \rightarrow P(S)$, where $P(S)$ denotes the power set of $S$.
\item $s_0$: the initial \gls{sut} state.
\item $T$: the set of states $T \subseteq S$ that terminate the \gls{sut}.
\end{itemize}
As illustrated in figure~\ref{fig:state-graph}, beginning from the initial state $s_0$, all transitions point to the unknown state $s_?$. In this case, the tuple looks as follows:
\begin{itemize}
\item $S = \{s_0, s_?\}$
\item $\Sigma = \{a_0, a_1, a_2\}$
\item $\Delta(s_0, a_0) = \Delta(s_0, a_1) = \Delta(s_0, a_2) = s_?$
\item $s_0$
\item $T = \emptyset$
\end{itemize}
Where $a_0$ represents a click on \enquote{New contact}, $a_1$ a click on \enquote{New category}, and $a_2$ a click on \enquote{Apply}. After $a_0$ is being executed, the tuple changes accordingly. For any given state $s_i$, the set of actions $A_{s_i}$ contains all possible actions for any available \gls{gui} component in $s_i$, unless $s_i = s_?$; then, the set of actions of the last known state in the sequence is taken into account. Furthermore, the feasibility of an action $a_k$ in state $s_i$---which is needed for the mutation operation as described below---is defined by:
\begin{equation*}
\feas(a_k, s_i) =
\begin{cases}
\text{true} & a_k \in A_{s_i} \\
\text{false} & \text{else}
\end{cases}
\end{equation*}
If $s_i = s_?$, then the assumption is always that $a_k$ is potentially feasible:
\begin{equation*}
\feas(a_k, s_?) = \text{true}
\end{equation*}
When the execution of a randomly selected action leads to a new state, it will become part of the model and the prior transition to $s_?$ will now point to that new state. This whole procedure is repeated until the sequence respectively the test case has the desired length, which is chosen randomly out of the interval $[1,\; l]$, where $l$ is a fixed upper bound.

After the initial population was created, the \gls{ga} takes over control. The crossover operator produces two offspring test suites $O_0$ and $O_1$ from two parents $P_0$ and $P_1$. The first half of $O_0$ contains the first $\alpha |P_0|$ test cases from $P_0$, followed by the last $(1 - \alpha) |P_1|$ test cases from $P_1$. For $O_1$ it is simply vice versa, i.e. it contains the first $\alpha |P_1|$ test cases from $P_1$ and the the last $(1 - \alpha) |P_0|$ test cases from $P_0$. As $\alpha$ is either 0 or 1, $|O_i| \leq \max(|P_{i_1}|, |P_{i_2}|)$ is always true and, therefore, the number of test cases does not grow unproportionally. Test cases with a length of 0 are deleted.

Although the \gls{ga} only optimizes towards whole test suites, the mutation operator represents a special case as it is applied to test cases, too. For a test suite $S$, mutation may add new test cases or change existing test cases. A new test case is added with a probability of $\sigma = 0.1$, the $n$-th insertion happens with a probability of $\sigma^n$ and stops when no more test cases are added. An existing test case is changed with a probability of $\frac{1}{|S|}$. Mutating a test case $C$ can result in up to three operations being applied to the given sequence of actions, each with a probability of $\frac{1}{3}$:
\begin{itemize}
\item Change: The parameters of each action may be changed randomly with a probability of $\frac{1}{|C|}$.
\item Deletion: Each action might be removed with a probability of $\frac{1}{|C|}$ as well.
\item Insertion: With a decreasing probability of $2^{1 - n}$, a new action is being inserted at a random position $p$.
\end{itemize}
When a new action is inserted, such as:
\begin{equation*}
C = (a_0, a_1, a_2) \Rightarrow C' = (a_0, a_1, a_p, a_2)
\end{equation*}
Then, the state $s_p$ is searched by following the sequence of actions from $s_0$ up to $p$~(i.e. $a_0$ and $a_1$ in the given example). If a subsequent action of $a_p$ is identified as being infeasible, then the whole sequence is said to be infeasible. In order to repair the test case, all infeasible actions are removed so that only feasible ones remain. Repair procedures do not need to execute the enclosing test case---which drastically reduces the required amount of time. This is because the state graph can be leveraged to create feasible action sequences from existing states and transitions. However, if a later execution of the repaired test case reveals that it is still infeasible, the execution is suspended and the state graph is updated accordingly. This may happen if a subsequent version of \gls{sut} changes its behavior and one or more actions now lead to different states. These \gls{sut} versions are not necessarily stable releases; every change that is built can cause an update of the state graph.

As said before, the optimization towards code coverage focuses on entire test suites with respect to all branches~(see section~\ref{sec:code-coverage-software-testing} for different code coverage criteria types), based on the work of~\cite{rojp16}. During runtime, each branch can be mapped to a \emph{branch distance}~\cite{wbs01}, which describes how close an input was to fulfill a particular condition, guarding a given branch. To determine the fitness of a given test suite $S$, the minimum branch distance $\dist_\text{min}(b, S)$ is calculated for every branch $b \in B$, where $B$ is the set of all branches of the \gls{sut}. Therefore, each condition must be executed at least twice in order to cover each branch:
\begin{equation*}
\dist(b, S) =
\begin{cases}
0 & \text{if $b$ is covered} \\
\norm(\dist_\text{min}(b, S)) & \text{if the condition is executed at least twice} \\
1 & \text{else}
\end{cases}
\end{equation*}
Where $\norm : \mathbb{N} \rightarrow [0, 1]$ is the normalization function; it is used to prevent the domination of individual branches. Consequently, the fitness function is defined as follows, whereupon $M$ is the set of all methods and $M_S$ denotes the set of methods executed by a test suite $S$\footnote{ReTest uses the same fitness function as EvoSuite, but with its own representation and search operators.}:
\begin{equation*}
\fitness(S) = |M| - |M_S| + \sum_{b_k \in B} \dist(b_k, S)
\end{equation*}

Although the primary objective is to improve on code coverage, the \gls{ga} also implicitly optimizes towards test length. This is achieved during selection when individuals are being ranked. If two test suites have an identical fitness, then the shorter one is selected, which leads to less execution time and less maintenance effort. Moreover, in order to address elitism~\cite[89]{es03}, each new generation preserves the $e$ best solutions of the previous generation, in case all new individuals have a worse fitness, where $e$ is a positive constant.

%% file: chapters/3-problem.tex


\chapter{Problem Analysis}
\label{ch:problem}

The goal of this chapter is to understand the acute problem to provide a basis for a concrete design. Initially, section~\ref{sec:code-coverage-software-testing} describes control flow-based code coverage criteria in general and outlines their weaknesses when used in software testing. Section~\ref{sec:missing-link} further illustrates this based on a specific example with ReTest's test generation mechanism. Section~\ref{sec:requirements} specifies various functional and non-functional requirements for the aspired prototype in the form of user stories, incorporating the previously identified limitations.

\section{Code Coverage and Software Testing}
\label{sec:code-coverage-software-testing}

In terms of control flow-based code coverage criteria, generally three distinct types can be derived:
\begin{displaycquote}[146]{gj08}
\begin{itemize}
\item[--]\emph{Block coverage}: A block is a set of sequential statements not having any in-between flow of control, both inward and outward. Complete block coverage requires that every such block in the program be exercised at least once in the test executions.
\item[--]\emph{Branch coverage}: An evaluation point in the code may result in one of the two outcomes---true or false, each of which represents a branch. Complete branch coverage requires that every such branch be exercised at least once in the test executions.
\item[--]\emph{Predicate coverage~(or Condition coverage)}: A predicate is a simple atomic condition in a logical expression. Complete predicate coverage requires that every such simple condition must evaluate to TRUE as well as FALSE at least once in the test executions.
\end{itemize}
\end{displaycquote}
As the fitness function suggests, ReTest itself uses branch coverage. While this is indeed a desirable goal in software testing~\cite{gop16} and---as stated by Gupta and Jalote---offers a good trade-off between effectiveness and efficiency, it is not a good measure for the overall quality of a test suite per se~\cite[195]{wmo12}. One can consider the following simple programming task:
\begin{displayquote}
Given $n \in \mathbb{N}$, implement a method which computes the factorial of $n$.
\end{displayquote}
Where the factorial function is formally defined by:
\begin{equation*}
n! =
\begin{cases}
1 & n = 0 \\
n \cdot (n - 1)! & n > 0
\end{cases}
\end{equation*}
Respectively:
\begin{equation*}
n! = \prod \limits_{i = 1}^{n} = 1 \cdot 2 \cdot \ldots \cdot n
\end{equation*}
According to this description, a naive implementation might look like listing~\ref{lst:factorial-naive}. When striving for branch coverage, a corresponding test using any number greater than zero is sufficient because all branches will be covered.

\begin{listing}[h]
\begin{minted}{java}
public static int of(int n) {
    return n > 0 ? n * of(--n) : 1;
}
\end{minted}
\caption{Naive factorial implementation.}
\label{lst:factorial-naive}
\end{listing}

However, this is not enough. Boundary conditions such as negative numbers, zero, or a value causing an~(integer) overflow should always be considered since they are a reasonable addition~\cite[121--128]{sls14}. Taking this into account, a rudimentary~(JUnit 5-based\footnote{\url{http://junit.org/junit5/}.}) test class could be listing~\ref{lst:factorial-test}.

\begin{listing}[ht!]
\begin{minted}{java}
public class FactorialTest {

    @Test
    void factorial_should_handle_negative_numbers() {
        assertEquals(1, Factorial.of(-1));
    }

    @Test
    void factorial_should_handle_zero() {
        assertEquals(1, Factorial.of(0));
    }

    @Test
    void factorial_should_handle_small_numbers() {
        assertAll(
                () -> assertEquals(1, Factorial.of(1)),
                () -> assertEquals(2, Factorial.of(2)),
                () -> assertEquals(6, Factorial.of(3)),
                () -> assertEquals(24, Factorial.of(4)),
                () -> assertEquals(120, Factorial.of(5))
        );
    }

    @Test
    void factorial_should_handle_int_overflow() {
        // 13! > Integer.MAX_VALUE.
        assertThrows(ArithmeticException.class,
        		() -> Factorial.of(13));
    }

}
\end{minted}
\caption{Factorial test class.}
\label{lst:factorial-test}
\end{listing}

Running these test cases shows that the implementation returns \num{1932053504} when $n = 13$ is given, whereas it should be $13! = 12! \cdot 13 = \num{479001600} \cdot 13 = \num{6227020800}$. Java's \inline{int} data type has a size of \SI{32}{\bit}, which equals the interval of $[{-2^{31}},\; {2^{31} - 1}]$. Thus, $13!$ exceeds \inline{Integer.MAX_VALUE} and causes an overflow. These types of bugs are generally hard to debug and can cause huge problems~\cite{gle96}. With that in mind, a more sophisticated implementation should throw an \inline{ArithmeticException} when an overflow is being detected. Listing~\ref{lst:factorial-advanced} is not just aware of that: As it is based on a stream\footnote{Streams---introduced in Java~8---are an abstraction to process data in a declarative way. For more information see~\cite{urm14}.} instead of recursion, it also avoids a \inline{StackOverflowError} if, for instance, an arbitrary-precision data type like \inline{BigInteger} would have been used and the implementation recurses too deeply.

\begin{listing}[h]
\begin{minted}{java}
public static int of(int n) {
    return IntStream.rangeClosed(1, n).reduce(1, Math::multiplyExact);
}
\end{minted}
\caption{Advanced factorial implementation.}
\label{lst:factorial-advanced}
\end{listing}

This basic example shows that, even on unit level, solely focusing on branch coverage may leave out important test cases. Although ReTest's approach---combining difference testing as well as monkey testing on \gls{gui} level---is not just unique, but also effective and efficient, it is by far no silver bullet. Especially the branch coverage-optimizing \gls{ga} lacks one major ingredient: human behavior. The next section further illustrates this based on a specific example with ReTest.

\section{The Missing Link}
\label{sec:missing-link}

Human testers have the ability to construe \emph{meaningful} tests from an interface~(e.g. a class, an \gls{api}, or the actual \gls{gui}) or an informal specification, whereas machines can hardly do that. Even though monkey testing potentially finds many bugs---as already mentioned in section~\ref{subsec:monkey-testing}---the generated test cases tend to be missing the link to actual human behavior. This is obvious since ReTest's test generation mechanism heavily depends on random decisions; the consequence is that human testers often find the results less reasonable and difficult to interpret. This observation also corresponds to the results of Ciupa~et~al., which say that random testing unveils bugs that humans miss and vice versa~\cite[165]{ciup08}. But rather than serving as a complement to manually created tests, ReTest aims to~(optionally) enrich monkey testing with the behavior of human testers in order to change the characteristics of the generated regression tests. That is, having the ability to move from non-functional testing towards functional testing.

To give an example, listing~\ref{lst:test-generation-example} shows some generated tests using the \gls{tap}\footnote{\url{https://testanything.org/}.}. This is actually not the result of the test generation step, but of the replay step. When tests respectively suites are generated during monkey testing, then the result is a so-called \enquote{execsuite}, which stands for executable suite. To understand what an execsuite is and what it does, one must first understand ReTest's basic building blocks:
\begin{description}
\item[Action sequence] A sequence of~(coherent) \gls{gui} actions.
\item[Test] A test case composed of one or more action sequences.
\item[Suite] A test suite composed of one or more tests.
\item[Executable suite] A suite also containing the corresponding \gls{gui} state.
\end{description}
This means that a generated execsuite is not just a test suite with test cases; because it additionally contains the \gls{gui} state of the \gls{sut} that was captured during monkey testing, it represents a branch coverage-optimized regression test suite. When such an execsuite is being replayed, various reports including a \gls{tap} record are created as well.

\begin{listing}
\begin{minted}[breaklines=true]{text}
1..1
ok 1 generated-suite_20170921-1808
    1..15
    ok 1 generated
        1..4
        ok 1 Click on JLabel [Password: ]
        ok 2 Click on JButton [Login]
        ok 3 Click on JButton [OK]
        ok 4 Click on JButton [Cancel]
    ok 2 generated
        1..5
        ok 1 Entering text 'licaletacerabarn' into JPasswordField Password
        ok 2 Click on JLabel [Username: ]
        ok 3 Entering text 'mbrodysinablenute' into JTextField Username
        ok 4 Entering text 'javax.swing.plaf.nimbus.NimbusIcon' into JTextField Username
        ok 5 Click on JButton [Cancel]
    ok 3 generated
        1..5
        ok 1 Entering text 'Max' into JTextField Username
        ok 2 Entering text 'ReTest' into JPasswordField Password
        ok 3 Click on JButton [Login]
        ok 4 DoubleClick on TableCell [76137] (6/2) of JTable[Address book]
        ok 5 Click on JButton [Close]
    ok 4 generated
        1..8
        ok 1 Entering text 'Max' into JTextField Username
        ok 2 Entering text 'ReTest' into JPasswordField Username
        ok 3 Click on JButton [Login]
        ok 4 DoubleClick on TableCell [Musterweg] (4/2) of JTable[Address book]
        ok 5 DoubleClick on TableCell [Schneider] (1/4) of JTable[Address book]
        ok 6 Click on JButton [Add address]
        ok 7 Click on Tab [Calculator]
        ok 8 Click on JMenuItem [Close]
    ok 5 generated
		# …
\end{minted}
\caption{Test generation example as TAP report.}
\label{lst:test-generation-example}
\end{listing}

Throughout this thesis, ReTest's default demo application will be used as the \gls{sut}, which also happened in the case of listing~\ref{lst:test-generation-example}. The demo is a simple Java Swing application that has three main windows: a login dialog, an address book tab, and a calculator tab. Figure~\ref{fig:demo-login} shows that login dialog, whereas the other windows can be found in appendix~\ref{app:demo-address-book} and \ref{app:demo-calculator}. Before the example was generated, ReTest was given a simple execsuite that logged into the \gls{sut}. ReTest uses the given execsuite(s) to create the corresponding state graph beforehand, which serves two main purposes:
\begin{enumerate}
\item The created state graph can be used to reach specific states of the \gls{sut}. In the present case, ReTest is able to log in to expose the main functionality since it now knows the credentials.
\item Additional \gls{sut}-specific information can be extracted. For instance, seeding of numerical and string constants from the analyzed Java bytecode~\cite{rfa16} to generate text input actions, which may help to reach unexplored states.
\end{enumerate}

\begin{figure}[h]
\centering
\includegraphics[scale=0.7]{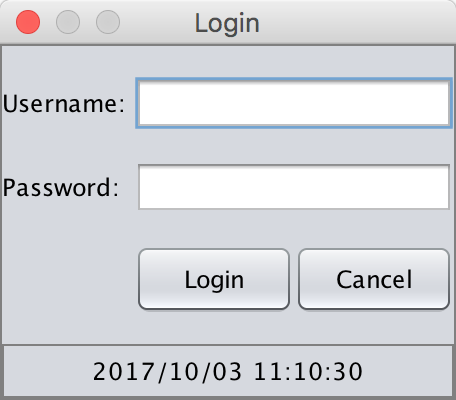}
\captionbelow{Login dialog of ReTest's demo SUT.}
\label{fig:demo-login}
\end{figure}

The problem is that although all this data is available, the generated tests are still quite randomized as the \gls{tap} report shows: it took three attempts to log in, and then the only action that has been executed was a click on a table cell in the address book tab. As mentioned before, this definitely makes sense from a monkey testing perspective because these tests can be a good complement to manually created tests. For example, the first test examines what happens without credentials, the second test checks for invalid credentials, and the third test finally logs in. But there are also less reasonable actions, such as clicks on labels without an effect or frequent use of the close button which terminates the \gls{sut}. This disadvantageous randomness is mostly caused by the way the initial population of the \gls{ga} is created. As described in section~\ref{subsec:retest-ga}, it simply consists of randomly generated test suites, whose test cases are generated via random walks on the \gls{sut}. In addition, the crossover and mutation operators further mix things up.

\section{Requirements}
\label{sec:requirements}

Since the focus of this project is on a robust prototype, including an \gls{etl}\footnote{\url{https://en.wikipedia.org/wiki/Extract,_transform,_load}.} pipeline, that can be easily extended for a later use in production, it is important to formulate a minimum set of functional and non-functional requirements that should be addressed. In agile software development methodologies, this is typically done with \emph{user stories}\footnote{\url{https://en.wikipedia.org/wiki/User_story}.}. A user story is \textcquote[83]{rub12}{\textelp{} a convenient format for expressing the desired business value for many types of product backlog items, especially features.} The product backlog represents an ordered list of requirements for the product and is initially only filled with epics, which denote comprehensive user stories that depict the high-level activities of a~(future) user. In general, user stories typically have the following format:
\begin{tcolorbox}[title=\#0: User Story Title]
As a \texttt{<user role>} I want to \texttt{<goal>} so that \texttt{<benefit>}.
\end{tcolorbox}
However, this is not mandatory and often replaced by free text, in particular, when non-functional requirements are being described. Section~\ref{subsec:user-stories} will list the user stories that were specified by various stakeholders at ReTest. Henceforth, these user stories will be referenced via their unique key~(e.g. \#0 in case of the previous example). An overview of all user stories can be found in table~\ref{tab:user-stories}. It is important to note that the only user role that has been identified is \enquote{advanced user}. This is due to the mere fact that the intended prototype focuses on a robust and extensible implementation, rather than user experience, which is beyond the scope of this thesis. Moreover, the author would like to point out that the use of words such as \enquote{must} or \enquote{should} is based on RFC 2119\footnote{\url{https://tools.ietf.org/html/rfc2119/}.}.

\begin{table}
\small
\centering
\begin{tabular}{l l} \toprule
Key & Title \\\midrule
\#\ref{story:data-extraction} & Training Data Extraction \\
\#\ref{story:training-evaluation} & Monkey Training and Evaluation \\
\#\ref{story:enhanced-monkey} & Enhanced Monkey Testing \\
\#\ref{story:monkey-performance} & Monkey Testing Performance \\
\#\ref{story:open-source} & Open Source Libraries \\
\#\ref{story:java-compatibility} & Java Compatibility \\
\#\ref{story:robustness-extensibility} & Robustness and Extensibility \\\bottomrule
\end{tabular}
\captionbelow{User stories overview.}
\label{tab:user-stories}
\end{table}

\subsection{User Stories}
\label{subsec:user-stories}

\begin{userstory}[label={story:data-extraction}]{Training Data Extraction}
As an advanced user I want to extract training data from my existing tests so that I can reuse this knowledge for test generation.
\end{userstory}

\begin{userstory}[label={story:training-evaluation}]{Monkey Training and Evaluation}
As an advanced user I want to train and evaluate the monkey based on the aforementioned training data.
\end{userstory}

\begin{userstory}[label={story:enhanced-monkey}]{Enhanced Monkey Testing}
As an advanced user I want to leverage the trained monkey so that I can improve the generated tests in terms of human behavior.
\end{userstory}

\begin{userstory}[label={story:monkey-performance}]{Monkey Testing Performance}
Retrieving information from the trained monkey at runtime must not affect the overall performance in terms of execution time and branch coverage.
\end{userstory}

\begin{userstory}[label={story:open-source}]{Open Source Libraries}
Only open source libraries must be used that are professionally maintained and have an active community.
\end{userstory}

\begin{userstory}[label={story:java-compatibility}]{Java Compatibility}
The used libraries should be compatible with Java or at least operate on the \gls{jvm}.
\end{userstory}

\begin{userstory}[label={story:robustness-extensibility}]{Robustness and Extensibility}
The prototypical implementation should focus on robustness and extensibility instead of optimization to simplify a later transfer to production.
\end{userstory}

%% file: chapters/4-design.tex


\chapter{Design Choices}
\label{ch:design}

This chapter presents a concrete design which addresses the limitations that were identified in the problem analysis. Section~\ref{sec:overall-concept} starts by giving an overview of the overall concept, forming a general framework for enhancing monkey testing based on the aforementioned methods. Section~\ref{sec:feature-engineering} describes the feature engineering process in detail, followed by an explanation of the chosen \gls{ml} model in section~\ref{sec:ml-model}. Finally, section~\ref{sec:prototype-architecture} specifies the architecture of the planned prototype.

\section{Overall Concept}
\label{sec:overall-concept}

Based on \cite{zhap11}, figure~\ref{fig:mlec-taxonomy}~(see section~\ref{sec:evolutionary-computing}) provides a taxonomy that classifies the existing \gls{mlec} research spectrum. As can be seen in the figure, there are essentially five possible connecting factors from an \gls{ec} perspective:
\begin{enumerate}
\item Population initialization.
\item Fitness evaluation and selection.
\item Population reproduction and variation.
\item Algorithm adaption.
\item Local search.
\end{enumerate}
According to Eiben and Smith, the \textcquote[172]{es03}{\textelp{} most obvious way in which existing knowledge about the structure of a problem or potential solutions can be incorporated into an EA is in the initialisation phase.} Also, as mentioned in section~\ref{sec:missing-link}, the bulk of the disadvantageous randomness is caused by the initial population of ReTest's \gls{ga}. Consequently, this leaves plenty of room for optimization to reduce the gap between manually created and automatically generated regression tests. Although the subsequent crossover and mutation operators make it difficult to create contiguous and human-like sequences of \gls{gui} actions, the individual actions can still be improved.

\begin{figure}
\small
\centering
\begin{tikzpicture}
\begin{axis}[
axis x line=bottom,
axis y line=left,
domain=0:150,
no markers,
samples=150,
ticks=none,
xlabel=Time,
ylabel=Best value in population,
ymin=-0.5,
ymax=2.5]
\addplot{log10(x)}; 
\end{axis}
\end{tikzpicture}
\captionbelow[Typical progress of an EC algorithm.]{Typical progress of an EA~\cite[42]{es03}.}
\label{fig:ec-progress}
\end{figure}

Traditionally, the initialization of the population is kept simple in most \glspl{ea}; this is also true in the case of ReTest, where the initial population is merely seeded by randomly generated individuals. The reason for this is that \glspl{ea} usually make rapid progress in the beginning, but start flattening out later on~(see figure~\ref{fig:ec-progress}). This quick improvement---which typically only takes a few generations---makes it questionable whether the additional complexity and the extra computational effort that come with a subtle initialization are reasonable. Furthermore, Eiben and Smith mention \cite{sd96} to highlight the importance of providing the \gls{ea} with sufficient diversity for evolution:
\begin{displaycquote}[174]{es03}
They~\textins{Surry and Radcliffe} concluded that the use of a small proportion of derived solutions in the initial population aided genetic search, and as the proportion was increased, the \emph{average} performance improved. However, the \emph{best} performance came about from a more random initial population. In other words, as the proportion of solutions derived from heuristics used increased, so did the mean performance, but the variance in performance decreased. This meant that there were fewer really bad runs, but also fewer really good runs.
\end{displaycquote}

Besides common optimization methods such as seeding or selective initialization~\cite[172--174]{es03}, especially existing advances in the area of \gls{mlec} algorithms that improve the initial solutions quality~(e.g. \cite{pas09}, \cite{rts08}, or \cite{ya01}) look promising---which is the method of choice in the present thesis. In case of the given problem domain and ReTest's \gls{ga} implementation, there are essentially two possibilities to choose between in order to optimize towards human behavior:
\begin{enumerate}
\item \gls{gui} actions.
\item \gls{gui} components.
\end{enumerate}
It is important to note that a smart selection, be it \gls{gui} actions or their underlying \gls{gui} components, does not aim to improve the fitness of the initial population since this is only measured in branch coverage. But rather the goal is to create a more human-like usage of the \gls{sut} respectively its \gls{gui}. Thus, any improvement is already considerably better than a random initial population, which is a crucial step towards the automation of functional testing. However, with respect to user story~\#\ref{story:monkey-performance}~(monkey testing performance) and \cite{sd96}, the design must not affect the overall performance and should still address the importance of diversity.

Without going into too much detail regarding the current implementation, ReTest's \gls{ga} invokes \inline{MonkeyExecutor#getNextAction(NormalState)} while creating the initial population. It returns a \gls{gui} action~(\inline{Action} class) within the current state~(\inline{NormalState} class) that gets executed next. The method itself contains various \inline{if} statements which choose the next action according to the following set of rules in descending order of priority:
\begin{enumerate}
\item Terminate the \gls{sut} if the current state is an exit state.
\item Follow the current road map~(see below).
\item Execute an unexplored action.
\item Create and follow a road map to a state with unexplored actions.
\item Execute a random action.
\end{enumerate}
This whole routine is already very effective, leading to good branch coverage results when generating tests during monkey testing. Additionally, the surrounding architecture is rather monolithic due to historical reasons, which makes changing the \gls{gui} action selection mechanism quite difficult. Yet, it is possible to integrate a smart selection without changing the fundamental mode of operation: When, for instance, a set of unexplored actions is available in the current state, then the model can rank these actions based on observations of human behavior. In principle, this knowledge is already available---the mandatory data can be extracted from existing tests as they basically \enquote{document} how humans use the present \gls{sut} respectively its \gls{gui}.

In regards to the decision between \gls{gui} actions or their underlying \gls{gui} components, picking the latter has two advantages. First, recommending components only says what to do, not how to do it. When the model selects a component, the \gls{ga} is still free to choose an arbitrary action on top of it~(e.g. a left, right, or double mouse click), which promotes diversity for evolution. Second, actions potentially require more variables as they consist of the underlying component and the action itself. Consequently, components possibly lead to a simpler model and enable faster prototyping. Therefore, the overall concept can be summarized as follows:
\begin{itemize}
\item Use an \gls{mlec} algorithm that improves the quality of the initial population towards human behavior.
\item Exploit existing tests to extract information on how human testers use the given \gls{sut} respectively its \gls{gui}.
\item Rank the available \gls{gui} components at runtime based on the extracted knowledge.
\end{itemize}
These methods form a general framework that can be easily used to guide any form of monkey testing with the aid of \gls{ml}---regardless of point 1, the use of an \gls{ea} and, therefore, the creation of an \gls{mlec} algorithm. The following sections describe the relevant details.

\section{Feature Engineering}
\label{sec:feature-engineering}

Since the learning task has been defined, it is now also possible to create the input for the learner---the features. For each generated \gls{gui} action during the population initialization, the question is: Which \gls{gui} component is most likely to be used by a human in the current \gls{sut} state? The proposed approach models this task as a binary classification problem that is learned in a supervised manner. The corresponding target function is defined as $f: S \rightarrow G$, where $S$ and $G$ are the sets of \gls{sut} states and \gls{gui} components. To extract the training data, existing tests are exploited so that for each \gls{gui} action $a_k$ within a test case $C$, every \emph{possible target component} in the given state $s_i$ is compared to the \emph{previous target component} of the previous action $a_{k - 1}$, where the \emph{correct target component} of the current action is labeled accordingly. One advantage of this method is that with only a few tests, many feature vectors can be extracted to train the model. For example, if 10 test cases are available, each containing 10 actions, and the average number of possible target components in each state is 25, then this already leads to $10 \cdot 9 \cdot 25 = \num{2250}$ feature vectors. Here, it is important to note two things:
\begin{enumerate}
\item The first action is always skipped because there is no previous action for comparison, which is why the number of actions that can be used for data extraction is effectively $|C| - 1$.
\item The data may become very imbalanced as every action only has one correct target component~(labeled \inline{true}), whereas the number of possible target components in each state is unbounded~(labeled \inline{false}).
\end{enumerate}

The main objective when selecting features is actually three-fold: \textcquote[1157]{ge03}{\textelp{} improving the prediction performance of the predictors, providing faster and more cost-effective predictors, and providing a better understanding of the underlying process that generated the data.} This usually requires deep domain knowledge and the use of sophisticated heuristics. However, Guyon and Elisseeff also point out that trying the simplest things first is almost always a good starting point. In the case of the given learning task, the chosen features should abstract the structure of the present \gls{gui} in such a way that the aforementioned objectives are being addressed.

The following two sections describe the selected features~(see table~\ref{tab:features-overview}) based on their relation~(absolute or relative) to the corresponding \gls{gui} component. It should be noted that the selection is designed for the targeted Swing platform respectively the underlying \gls{awt}. But, in general, the features should be applicable for other toolkits and platforms as well.

\begin{table}[h]
\small
\centering
\begin{tabular}{l l l l} \toprule
& Relation & Category & Data type \\\midrule
Enabled & Absolute & Qualitative, unordered & Boolean \\
Preferred type & Absolute & Qualitative, unordered & Boolean \\
Focus distance & Relative & Quantitative, discrete & Integer \\
Path distance & Relative & Quantitative, discrete & Integer \\
Point distance & Relative & Quantitative, continuous & Floating point \\\bottomrule
\end{tabular}
\captionbelow{Features overview.}
\label{tab:features-overview}
\end{table}

\subsection{Absolute Features}

Absolute features are properties of a possible target component that are \emph{absolute} to the component itself, i.e. they do not depend on the previous target component.

\minisec{Enabled}

\begin{figure}[h]
\centering
\includegraphics[scale=0.7]{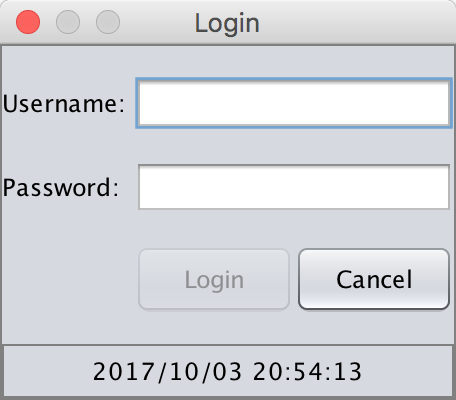}
\captionbelow{Disabled login button compared to enabled cancel button.}
\label{fig:demo-login-enabled}
\end{figure}

The enabled state of a \inline{JComponent}~(Swing) respectively a \inline{Component}~(\gls{awt}) determines whether the component can respond to user input and generate events. Components are enabled by default and may alter their visual representation when they are disabled~(see figure~\ref{fig:demo-login-enabled}). If a component is disabled, it is usually not reasonable to apply an action because it is not able to respond.

\minisec{Preferred Type}

\begin{figure}[h]
\centering
\includegraphics{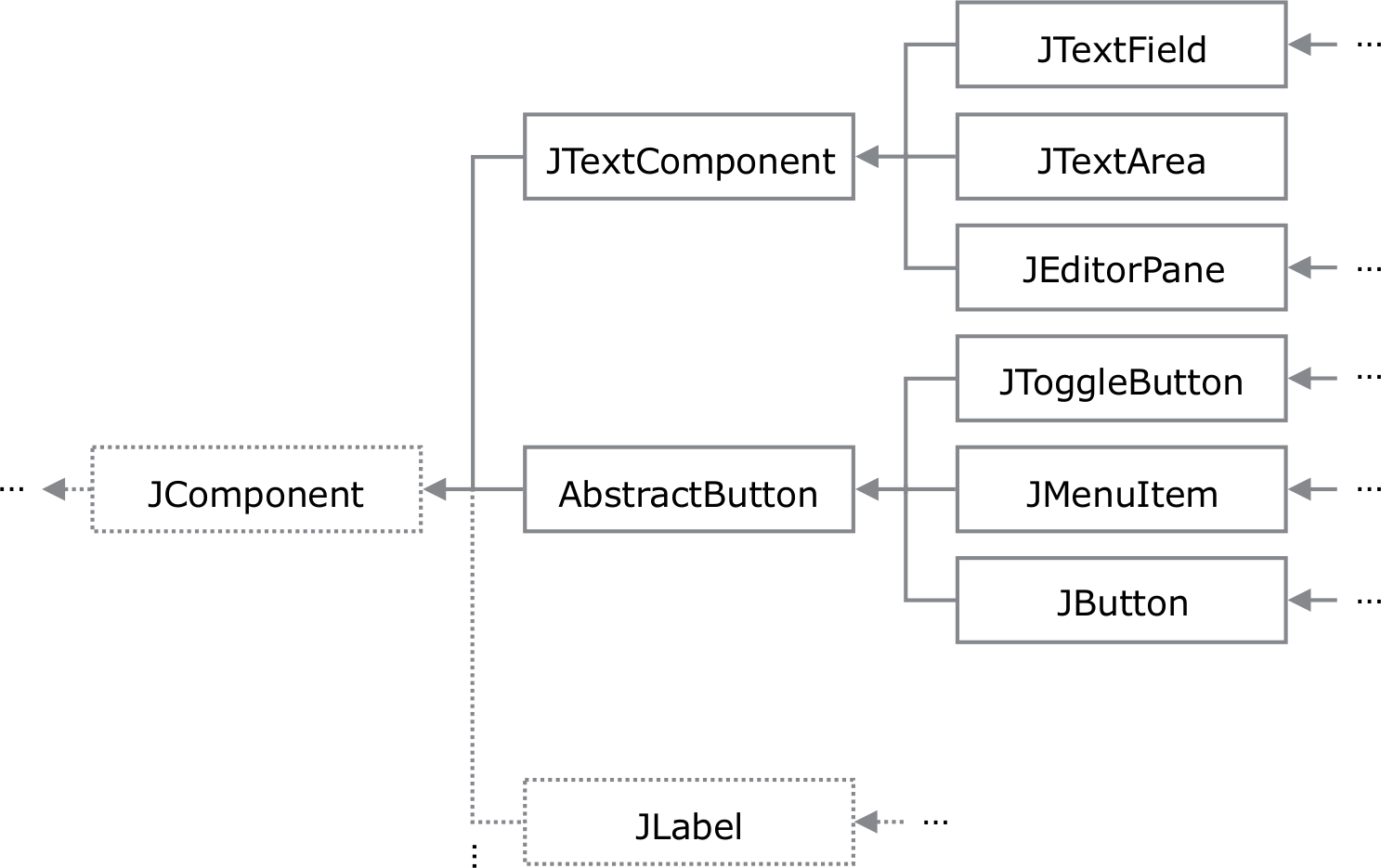}
\captionbelow{Preferred types within the Swing component hierarchy.}
\label{fig:swing-hierarchy}
\end{figure}

The preferred type feature categorizes all Swing types to provide an indicator whether a possible target component is generally a desirable target. For instance, a \inline{JLabel} typically does not respond to clicks or similar events; it may display a tool tip on a mouse over event, but it is usually not helpful when exploring the \gls{sut}. Preferred types are defined as any instance of \inline{AbstractButton} or \inline{JTextComponent}~(see figure~\ref{fig:swing-hierarchy}).

\subsection{Relative Features}

Relative features are properties of a possible target component that are \emph{relative} to the previous target component.

\minisec{Focus Distance}

\begin{figure}[h]
\centering
\includegraphics{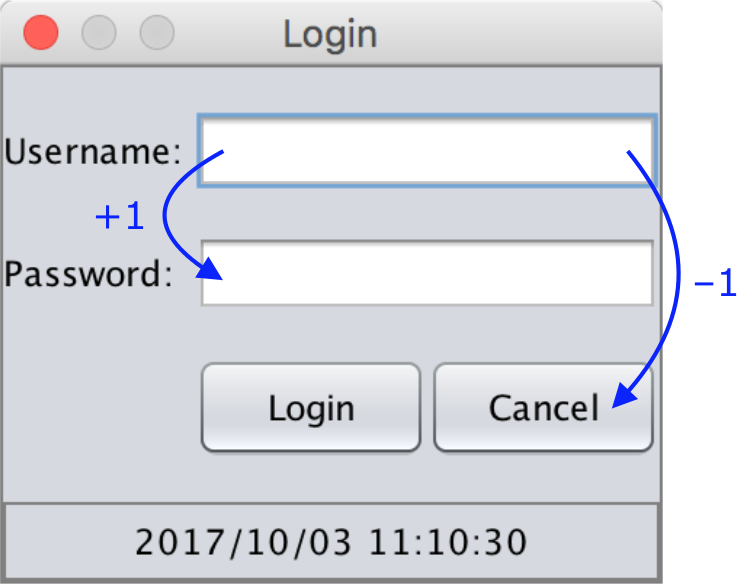}
\captionbelow{Focus distance calculation.}
\label{fig:demo-login-focus-distance}
\end{figure}

The focus distance describes the distance when traversing forward~($+1$) or backward~($-1$) in the \gls{gui} with respect to the previous target component, which owns the focus. Traversing usually happens with key strokes~(tab or arrow keys) and is generally a strong indicator for picking the next target component, especially in the case of forms. In Swing, the order is defined within a \inline{FocusTraversalPolicy} that can be retrieved via the component's focus cycle root. As the name suggests, the policy represents a cycle as illustrated in figure~\ref{fig:demo-login-focus-distance}. Going from the username text field forward to the password text field equates a distance of 1, whereas going backward to the cancel button means $-1$.

\minisec{Path Distance}

\begin{figure}[h]
\centering
\includegraphics{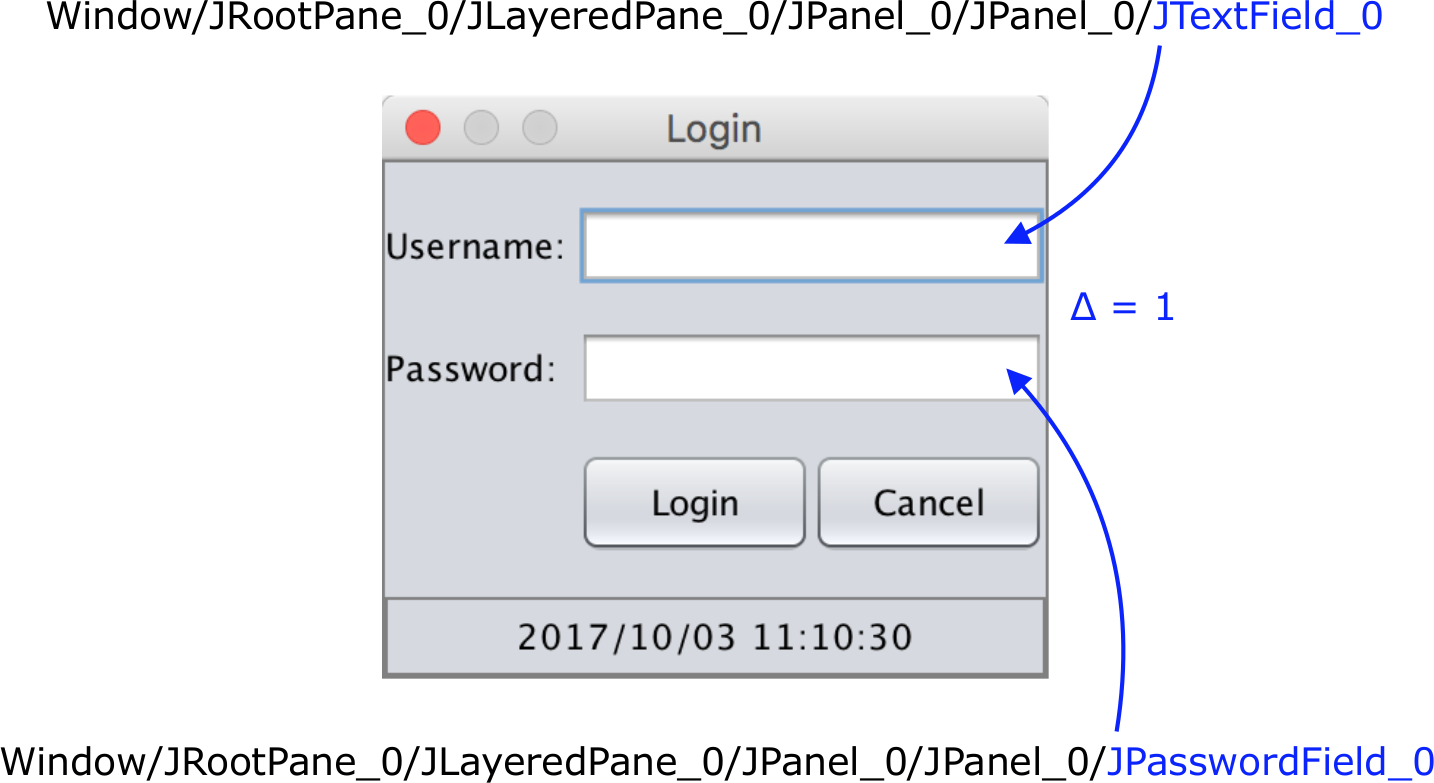}
\captionbelow{Path distance calculation.}
\label{fig:demo-login-path-distance}
\end{figure}

Each component has a unique path, which is similar to e.g. a XPath\footnote{\url{https://en.wikipedia.org/wiki/XPath}.} locator in Selenium. The path distance determines the distance between two of such paths, revealing the lowest common parent component. Figure~\ref{fig:demo-login-path-distance} gives an example based on the username text field and the password text field. For each different path element, a distance of 1 is added; the path distance is only 0 if both components are the same.

\minisec{Point Distance}

\begin{figure}
\centering
\includegraphics{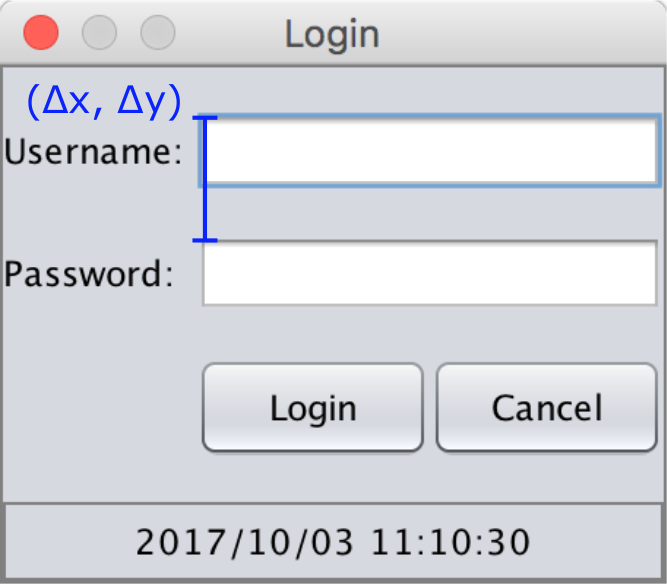}
\captionbelow{Point distance calculation.}
\label{fig:demo-login-point-distance}
\end{figure}

The point distance simply equates the absolute distance between the upper-left corner of the previous target component and a possible target component, measured in the coordinate system of the enclosing window~(see figure~\ref{fig:demo-login-point-distance}). Usually, the closer two components are, the more they belong together.

\section{Machine Learning Model}
\label{sec:ml-model}

As implied in the previous sections of this thesis, the \gls{ml} algorithm that is going to be used is an \gls{ann}. This decision is based on the following reasons:
\begin{itemize}
\item \gls{ann} learning is well-suited for training which may contain errors~\cite[85]{mit97}. Depending on the available tests, some might also include less reasonable actions to cover boundary conditions. This untypical behavior could lead the model in the wrong direction.
\item Due to the recent rise of \glspl{ann} and especially deep learning, several pro\-duc\-tion-ready libraries are already available. This helps to create a robust and extensible prototype.
\item \glspl{ann} are very efficient when it comes to handling vast amounts of data, which is why they are often used in~(near) real-time scenarios~\cite[92]{bel15}. Although real time is not an issue here, this ability might help to not decelerate the existing test generation mechanism.
\item The model implies that one is not exactly sure how the input and output nodes relate to each other~\cite[94]{bel15}. This is also the case with the given set of features, whose impact on the results is uncertain.
\item There is already some in-house knowledge regarding the use of \glspl{ann}. This not just promotes the acceptance of \glspl{ann} at ReTest, but may also help with potential issues during the implementation phase.
\end{itemize}

The next two sections describe the chosen topology of the network and the settings of its hyperparameters. It should be noted that---as already mentioned in section~\ref{sec:contribution-outline}---the entire project followed an iterative-incremental approach. That is, the content presents the configuration that yielded the best results~(see section~\ref{sec:evaluation}).

\subsection{Network Topology}
\label{subsec:network-topology}

The network uses a standard structure of a feed-forward network, consisting of three layers: an input, a hidden, and an output layer. The input and the hidden layers are composed of sigmoid units, whereas the units of the output layer use the \emph{softmax}\footnote{\url{https://en.wikipedia.org/wiki/Softmax_function}.} function as their activation function. The output of the softmax function represents a categorical distribution, which is commonly used for~(binary) classification problems like the given task. Although it is possible to add more layers, the two sigmoid layers can already express many target functions and keep the training times relatively short~\cite[115]{mit97}.

\begin{figure}[h]
\centering
\includegraphics{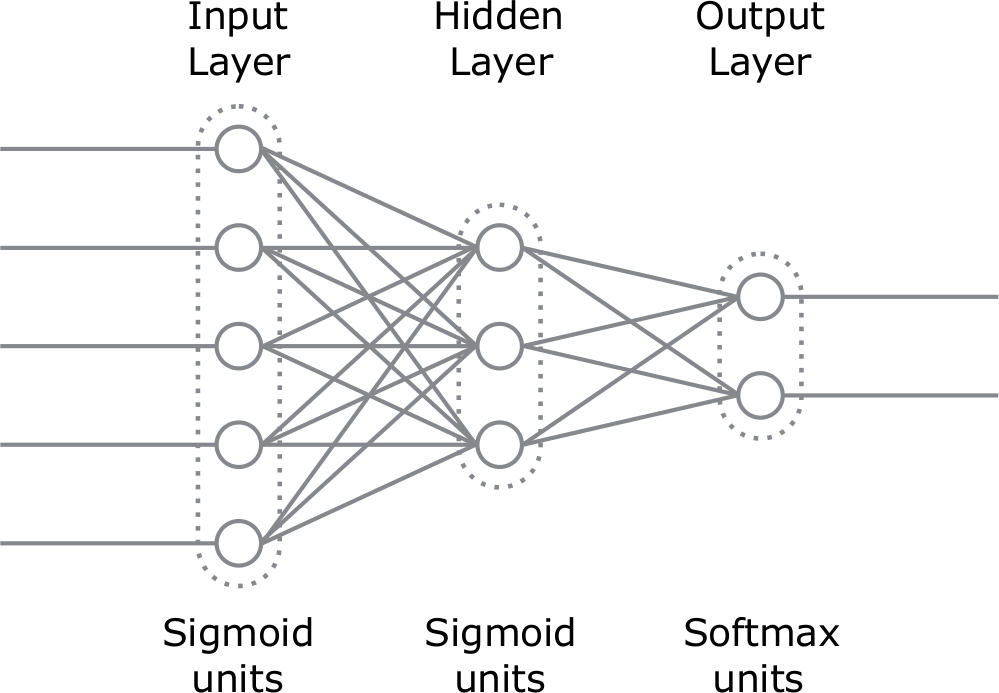}
\captionbelow{Chosen network topology.}
\label{fig:network-topology}
\end{figure}

The number of units in the input layer is uniquely determined by the shape of the training data, which consists of the five previously mentioned features. The quantitative features~(focus distance, path distance, and point distance) are already suitable as inputs since they are represented by numerical values. The qualitative features~(enabled and preferred type) use a simple label encoding that maps each boolean to a scalar, where \inline{false} becomes~0 and \inline{true} becomes~1. Alternative approaches are, for instance, one-hot encoding\footnote{\url{https://en.wikipedia.org/wiki/One-hot}.}. It ensures that the (Hamming) distance between all encodings is equal, but because there are only two categorical values, label encoding suffices. Since the \gls{ann} uses softmax in the output layer, the number of units here is determined by the number of labels, which is two~(\inline{false} and \inline{true} respectively 0 and 1). Various empirically-derived heuristics exist to choose the number of units in the hidden layer; the most commonly-used rule of thumb is that \textcquote[157]{hea08}{\textelp{} the optimal size of the hidden layer is usually between the size of the input and size of the output layers.} Therefore, three sigmoid units are used within the hidden layer. Figure~\ref{fig:network-topology} illustrates the resulting network topology.

\subsection{Hyperparameter Settings}
\label{subsec:hyperparameters}

\glspl{ann} have many hyperparameters that can be tuned. The network might learn very slow, or maybe not at all, if these parameters are poorly chosen. The following settings are used to optimize the network for the task at hand:
\begin{description}
\item[Normalization] The given data is comprised of features with varying scales. On the one hand, boolean features such as the enabled state only have numerical values between 0 and 1. On the other hand, the point distance feature is represented by a floating-point value, retrieved via the Swing environment. Even within this environment different scales may occur since they are measured in the coordinate system of the enclosing window. To overcome this issue, the training and test data is normalized so that it has a mean of 0 and a standard derivation of 1.
\item[Weight initialization] As outlined in section~\ref{subsec:anns}, one way to initialize the weights is by starting with random values. However, there are also sophisticated algorithms to ensure that the weights are neither too big nor too small, which could make the input signal grow respectively shrink disproportionately. The initial weights of the network are calculated with the \emph{Xavier initialization}~\cite{gb10}, resulting in a Gaussian distribution with mean 0 and a variance of $\frac{2}{n_m + n_{m + 1}}$, where $n$ is the number of units in layer $m$.
\item[Epochs and iterations] In general, an epoch is a full pass through the given data. Hence, the \gls{ann} has seen every example of the training set after an epoch. An iteration is an update of the network's parameters, which can happen many times within an epoch. For training the \gls{ann}, the data is split into \emph{minibatches}. A minibatch refers to the number of feature vectors that is used when computing the parameters. If, for example, the training data is split into two minibatches $A$ and $B$, then two iterations create the sequence $(A, A, B, B)$, whereas two epochs result in the sequence $(A, B, A, B)$. Besides long training times, the model might overfit if too many epochs and iterations are being used. If the values are too low, the model may not have enough time to learn. One technique to avoid this is \emph{early stopping}\footnote{\url{https://en.wikipedia.org/wiki/Early_stopping}.}, which is utilized to train the network for 400 epochs on \SI{70}{\percent} of the available data, using a single iteration and a minibatch size of 128. At the end of each epoch, the current network is evaluated via the test set and replaces the previous network if it yields a better accuracy. After all epochs are over, the model with the best performance survives.
\item[Learning rate] The learning rate $\eta$ is said to be one of, if not the most important hyperparameter and usually lies within the range of $10^{-1}$ to $10^{-6}$~\cite{sky17a}. To have a good trade-off between convergence speed and overall accuracy, it is set to a medium value of $\eta = 10^{-3}$.
\item[Loss function] In order to measure the training error $E$, a \emph{loss function}~(or cost function or error function) is needed. Most-commonly used for binary classification problems is the \emph{cross entropy}\footnote{\url{https://en.wikipedia.org/wiki/Cross_entropy}.} between two probability distributions~(the current output $o$ and the target output $t$), which is also used for the softmax units of the network's output layer. To additionally address the imbalance of the data\footnote{The \gls{etl} pipeline also uses undersampling to increase the label distribution of correct target components within the data set. For more details see section~\ref{sec:data-transformation}.}, the weight vector $(0.8 \; 1.2)$ is attached to the labels \inline{false} and \inline{true} respectively 0 and 1. This way, the weight of the loss for the rare case \inline{true} is increased, whereas it is decreased for examples that are labeled \inline{false}.
\item[Regularization] Regularization methods are another technique to avoid overfitting by basically adding a penalty as the model complexity increases, which helps to better generalize. The network uses the \emph{Tikhonov regularization}\footnote{\url{https://en.wikipedia.org/wiki/Tikhonov_regularization}.}~(or $L_2$ or ridge regression) with a coefficient of $10^{-3}$ in order to add a portion of the squared weights as a penalty on top of the loss function.
\item[Weight update] The stochastic gradient descent version of the backpropagation algorithm is used as the weight update rule $\Delta w_{ji}$ of the network, which was already presented in section~\ref{subsec:anns}. In particular, the \emph{\gls{rmsprop}}\footnote{\url{https://en.wikipedia.org/wiki/Stochastic_gradient_descent\#RMSProp}.} is utilized to update the weights. It essentially divides the learning rate for a particular weight by an average of the magnitudes of recent gradients for that weight. Compared to a standard stochastic gradient descent, the \gls{rmsprop} optimization often leads to much faster training times~\cite{sky17a}.
\end{description}

\section{Prototype Architecture}
\label{sec:prototype-architecture}

The prototypical implementation is split into the following three packages, each reflecting a coherent set of features~(also known as package-by-feature~\cite{oha08}):
\begin{itemize}
\item \inline{de.retest.ml.extract}: Contains everything related to the extraction of the actual data. This includes a hook into the existing system, the knowledge to create feature vectors via the ReTest and Swing \glspl{api}, and a way to persist that data in a convenient format.
\item \inline{de.retest.ml.transform}: This package is responsible to transform the raw extracted data into a format that is readable for the \gls{ann} along with an undersampling mechanism.
\item \inline{de.retest.ml.model}: Encloses the construction of the \gls{ann} and the actual training process, including an \gls{api} for the existing monkey testing mechanism to use the resulting model.
\end{itemize}

\begin{figure}[h]
\centering
\includegraphics[width=\textwidth]{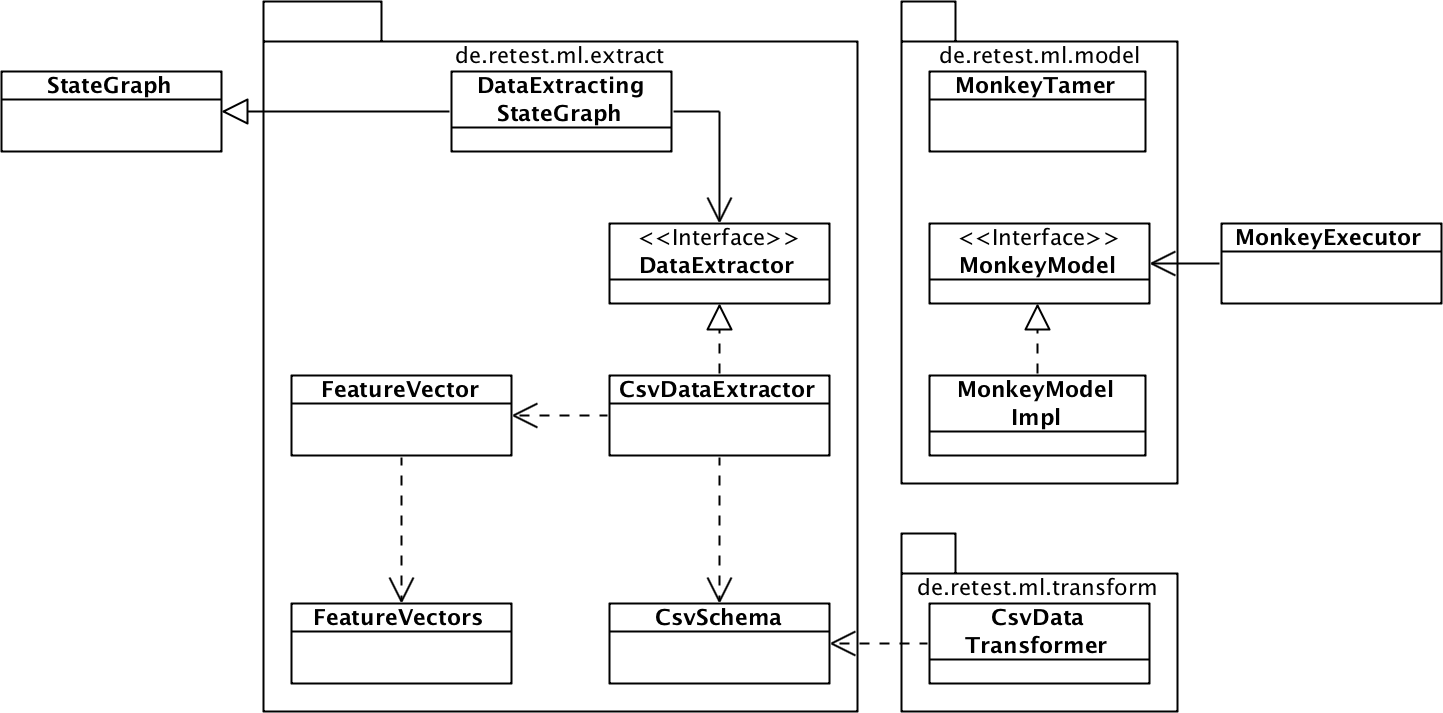}
\captionbelow{Simplified UML class diagram of the prototype architecture.}
\label{fig:uml-class}
\end{figure}

Figure~\ref{fig:uml-class} shows a simplified~(no attributes and no methods) \gls{uml} class diagram, giving an overview of the created classes and their relationships with each other as well as to the existing ReTest system. The entire architecture of the prototype is independent from the use of an \gls{ea} and basically can be used by any monkey testing mechanism~(although the actual \glspl{api} would require slight changes to be more generic).

\inline{DataExtractingStateGraph} extends \inline{StateGraph}, as introduced in section~\ref{subsec:retest-ga}, to hook into ReTest and extract the training data while the state graph is being created from the existing tests. \inline{DataExtractor} is a generic interface to persist that data in various formats, which is initially implemented by \inline{CsvDataExtractor} to generate \gls{csv} files; a convenient and well-know format that is supported by many \gls{etl} and \gls{ml} libraries. \inline{FeatureVector} is a simple data class that encloses the features for a single training example, whereas \inline{FeatureVectors} contains the actual knowledge to create these feature vectors via the ReTest and Swing \glspl{api}. \inline{CsvSchema} is a helper class that offers an enumeration and several constants regarding the concrete \gls{csv} schema.

\inline{CsvDataTransformer} is the only class in the \inline{de.retest.ml.transform} package. As previously described, it is responsible for transforming the raw extracted data into a format such that the \gls{ann} is able to learn. This is the part where it is reasonable to integrate an external library, which offers the respective functionality. \emph{\dlfj{}}\footnote{\url{https://deeplearning4j.org/}.} is used for this purpose as it covers multiple aspects. Backed by Skymind\footnote{\url{https://skymind.ai/}.}, it is open source according to the \gls{osi} definition~\cite{osi07} and written in Java, which addresses both user story~\#\ref{story:open-source} as well as user story~\#\ref{story:java-compatibility}. Furthermore, it can accelerate \gls{ann} training via native and distributed \glspl{gpu} and it is also capable of performing \gls{etl} operations that optionally run on Spark\footnote{\url{https://spark.apache.org/}.}. Hence, it is suitable for processing large amounts of data.

\dlfj{} is also used in \inline{de.retest.ml.model}, mainly by \inline{MonkeyTamer} to construct, train, and persist the model. Afterwards, this model is loaded via \inline{MonkeyModelImpl} and ready for use. \inline{MonkeyModel} provides the \gls{api} for ReTest---used by \inline{MonkeyExecutor}---which returns the predicted correct target component based on the passed state~(i.e. the list of all possible target components) and the previous target component.

%% file: chapters/5-implementation.tex


\chapter{Prototypical Implementation}
\label{ch:implementation}

The purpose of this chapter is to cover the most important aspects of the actual implementation as well as the evaluation of it. First of all, section~\ref{sec:feature-extraction} shows how the features are extracted via the ReTest and Swing \glspl{api} and provides a description on how the data is persisted. Section~\ref{sec:data-transformation} examines the transformation process of the raw data into a format that allows the \gls{ann} to learn from it. In section~\ref{sec:ann-construction-training}, it is shown how the network is constructed and trained with the aid of \dlfj{}. Section~\ref{sec:retest-integration} presents the integration with ReTest, followed by section~\ref{sec:evaluation}, which evaluates both the accuracy of the model and the results of the extended test generation mechanism.

\section{Feature Extraction}
\label{sec:feature-extraction}

In order the extract the specified features, it is necessary to first hook into ReTest's test execution mechanism so that one is able to capture the required data for every executed \gls{gui} action. This is achieved by extending \inline{StateGraph} with \inline{DataExtractingStateGraph}. As mentioned in section~\ref{sec:missing-link}, ReTest replays the given set of execsuites to create the corresponding state graph before tests are being generated. The modified state graph has a property \inline{de.retest.ml.extractTrue} which, if set to \inline{true}, tells ReTest a) to use \inline{DataExtractingStateGraph} instead of \inline{StateGraph} and b) to stop after the state graph has been created, i.e. to not generate tests afterwards. This way, one can launch the feature extraction via the existing \gls{cli}. The actual implementation of the adapted state graph is rather simple. The class holds a reference to \inline{CsvDataExtractor} and invokes its \inline{extract} method~(see listing~\ref{lst:csvdataextractor-extract}) every time a transition is added, which is done for each action within an execsuite. \inline{ComponentDescriptor} denotes an abstraction for components in a \gls{gui} toolkit-independent way and contains information such as the component's state (\inline{StateCriteria} class) and how to identify it (\inline{UniqueCompIdentCriteria} class). The remaining code behaves just like its \inline{super} implementation, making the extraction process transparent to the rest of the system.

\begin{listing}[h]
\begin{minted}{java}
@Override
public void extract(NormalState fromState, Action currentAction) {
	if (previousAction == null) {
		writeHeader();
	} else {
		ComponentDescriptor previousTarget =
				previousAction.getTargetComponentDescriptor();
		List<ComponentDescriptor> possibleTargets =
				new ArrayList<>(fromState.getWindowDescriptors());
		ComponentDescriptor correctTarget =
				currentAction.getTargetComponentDescriptor();

		writeFeatureVectors(previousTarget, possibleTargets,
				correctTarget);
	}

	previousAction = currentAction;
}
\end{minted}
\caption{Core domain objects in \inline{CsvDataExtractor}.}
\label{lst:csvdataextractor-extract}
\end{listing}

\inline{CsvDataExtractor} is responsible for persisting the extracted features in a \gls{csv} file. The class' field \inline{previousAction} is used as a reference to the previous target component. If this reference is \inline{null}, the current action represents the first action. In this case, the \gls{csv} file header~(which names each column) is added; otherwise, the three core domain objects are extracted and handed over to the method \inline{writeFeatureVectors}. As mentioned before, the basic idea is to compare the previous target component with any possible target component in the current state, where the correct target component is labeled \inline{true}. To do so, the \inline{writeFeatureVector} method uses \gls{dfs} to walk the component hierarchy that is contained in \inline{possibleTargets} and adds a row to the \gls{csv} file for each possible target component, which can be found in listing~\ref{lst:csvdataextractor-createcsvrow}. It is important to note that components respectively their descriptors are compared for equality via their paths, not \inline{equals(Object)}. This is because a component might change its state after an action has been applied, which would result in a failure for every comparison since the possible target components are retrieved via \inline{fromState}~(e.g. an arbitrary state $s_i$), whereas the correct target component is part of \inline{currentAction}~(possibly pointing to another state $s_j$).

\begin{listing}
\begin{minted}{java}
private String createCsvRow(ComponentDescriptor previousTarget,
		ComponentDescriptor possibleTarget,
		ComponentDescriptor correctTarget) {
	String component = possibleTarget.toString()
			.replaceAll(CSV_DELIMITER, StringUtils.EMPTY);
	FeatureVector featureVector = FeatureVector.of(previousTarget,
			possibleTarget, environment);
	boolean label = ComponentDescriptors.pathEquals(possibleTarget,
			correctTarget);

	return component + CSV_DELIMITER
			+ featureVector.enabled + CSV_DELIMITER
			+ featureVector.preferredType + CSV_DELIMITER
			+ featureVector.focusDistance + CSV_DELIMITER
			+ featureVector.pathDistance + CSV_DELIMITER
			+ featureVector.pointDistance + CSV_DELIMITER
			+ label + "\n";
}
\end{minted}
\caption{Creation of CSV rows.}
\label{lst:csvdataextractor-createcsvrow}
\end{listing}

\begin{listing}
\begin{minted}{java}
static int getFocusDistance(Component from, Component to) {
	if (from.equals(to)) {
		return 0;
	}
	Container root = from.getFocusCycleRootAncestor();
	if (root == null) {
		logger.debug("No focus cycle root for {}.", from);
		return MIN_FOCUS_DISTANCE;
	}
	FocusTraversalPolicy policy = root.getFocusTraversalPolicy();
	HashSet<Component> visited = new HashSet<>();
	Component after = from;
	Component before = from;
	for (int dist = 1; dist <= Math.abs(MIN_FOCUS_DISTANCE); dist++) {
		if (after != null) {
			visited.add(after);
			after = policy.getComponentAfter(
					after.getFocusCycleRootAncestor(), after);
			if (to.equals(after)) {
				return dist;
			}
		}
		if (before != null) {
			visited.add(before);
			before = policy.getComponentBefore(
					before.getFocusCycleRootAncestor(), before);
			if (to.equals(before)) {
				return -dist;
			}
		}
		if (after == null && before == null
				|| visited.contains(after)
				&& visited.contains(before)) {
			break;
		}
	}
	return MIN_FOCUS_DISTANCE;
}
\end{minted}
\caption{Focus distance feature extraction.}
\label{lst:featurevectors-getfocustraversaldistance}
\end{listing}

The data class \inline{FeatureVector} internally uses \inline{FeatureVectors} to compute its fields. Listing~\ref{lst:featurevectors-getfocustraversaldistance} exemplifies this for the extraction of the focus distance feature, which is the most complicated to obtain~(the other methods can be found in the appendices~\ref{lst:extract-enabled-feature}, \ref{lst:extract-preferredtype-feature}, \ref{lst:extract-pathdistance-feature}, and \ref{lst:extract-pointdistance-feature}). The shown method \inline{getFocusTraversalDistance} is actually not part of the public interface of \inline{FeatureVectors} as indicated by the missing access level modifier. The public variant expects component descriptors, hence, it differs in terms of the method signature. This is due to the fact that the implementation always depends on the toolkit. In case of the Swing/\gls{awt} focus subsystem, the implementation first checks for various boundary conditions such as component equality or a missing root. Each focus traversal cycle has only one root and each component belongs to exactly one focus traversal cycle, whereas containers belong to two cycles: one rooted at the container itself as well as one rooted at the nearest ancestor. The constant \inline{MIN_FOCUS_DISTANCE} is an arbitrarily chosen lower bound that is set to $-10$. In general, it can be considered impractical if a user has to do 10 or more key strokes in a row to get to a specific \gls{gui} component. Since it is especially unusual to go back to previously completed components~(e.g. in case of a typing error within a text field), the algorithm prefers positive values over negative values and stops after 10 iterations of the \inline{for} loop. It searches before and after the previous target component~(\inline{from}) to find the given possible target component~(\inline{to}) and returns the corresponding distance. It only returns \inline{MIN_FOCUS_DISTANCE} in the following situations:
\begin{itemize}
\item Both \inline{after} and \inline{before} are \inline{null}.
\item All components of the focus cycle already have been visited.
\item The focus distance is $\geq$ \inline{Math.abs(MIN_FOCUS_DISTANCE)}.
\end{itemize}
It should be noted that although this approach works, practically non-focusable components may produce strange results. Regarding \inline{Component#isFocusable()}, the specification of the \gls{awt} focus subsystem~\cite{ora00}---which is inherited by Swing---says that \enquote{\textelp{} all Components return true from this method.} Even components such as \inline{JLabel} yield \inline{true}, although \inline{Component#hasFocus()} is always \inline{false} and they cannot gain focus from a user perspective.

\begin{listing}[ht!]
\begin{minted}{java}
public class FeatureVectorsIntTest extends AssertJSwingJUnitTestCase {

	Sut frame;
	FrameFixture window;

	@Override
	protected void onSetUp() {
		frame = GuiActionRunner.execute(() -> new Sut());
		window = new FrameFixture(robot(), frame);
		window.show();
	}
	
	@Test
	public void getFocusDistance_should_handle_different_components()
			throws Exception {
		assertThat(FeatureVectors.getFocusDistance(
					frame.tfUsername, frame.pfPassword))
				.as("tfUsername to pfPassword").isEqualTo(1);
		assertThat(FeatureVectors.getFocusDistance(
					frame.tfUsername, frame.btnLogin))
				.as("tfUsername to btnLogin").isEqualTo(2);
		assertThat(FeatureVectors.getFocusDistance(
					frame.tfUsername, frame.btnCancel))
				.as("tfUsername to btnCancel").isEqualTo(-1);
	}
	
	// …

}
\end{minted}
\caption{Focus distance integration test.}
\label{lst:featurevectorsinttest}
\end{listing}

To address user story~\#\ref{story:robustness-extensibility}~(robustness and extensibility), critical functionalities like this is verified with the aid of unit and integration tests, which is also why the internal \inline{getFocusTraversalDistance} method is package-private~(no explicit modifier) to allow access from the test class. Listing~\ref{lst:featurevectorsinttest} shows an example of a~(JUnit 4-based\footnote{\url{http://junit.org/junit4/}.}) integration test that uses AssertJ Swing\footnote{\url{https://joel-costigliola.github.io/assertj/assertj-swing.html}.} to operate on a real \gls{gui}. The extended base class \inline{AssertJSwingJUnitTestCase} invokes \inline{onSetUp()} before each \inline{@Test} annotated method to set up the \gls{sut}~(\inline{frame}). Fixtures such as \inline{window} also offer methods to simulate user interaction and provide assertions to verify the state of components. However, the shown example simply starts the \gls{sut} and uses its components to the verify results of the \inline{getFocusDistance} method for different components. Other tests verify the computed focus distance for the same component~(i.e. \inline{from} equals \inline{to}), for non-focusable components, and for the corner case where the distance after equals the distance before, in which the former should be preferred.

\section{Data Transformation}
\label{sec:data-transformation}

Once the features have been extracted, \inline{CsvDataTransformer} can be used to transform the raw data into a format such that the \gls{ann} is able to learn. As mentioned in section~\ref{subsec:anns}, \glspl{ann} expect inputs that are real numbers~(section~\ref{subsec:network-topology} already discussed how this is to be done conceptually). \inline{CsvDataTransformer} is equipped with its own \inline{main} method, but also offers the public class method \inline{transform}. Both expect the absolute path to the raw \gls{csv} file, the former as a program argument and the latter as a \inline{java.nio.file.Path}. The data transformation essentially consists of the following three steps:
\begin{enumerate}
\item Define the schema of the given \gls{csv} file.
\item Determine the transform process for this schema.
\item Execute this transform process on Spark.
\end{enumerate}
The implementation uses DataVec\footnote{\url{https://deeplearning4j.org/datavec}.}, the \dlfj{} vectorization and \gls{etl} library. Listing~\ref{lst:csvdatatransformer-createschema} shows the first transformation step that defines the \gls{csv} schema. Generally, the \inline{Schema} class describes the layout of tabular data by naming the contained columns and describing their data types~(which optionally include restrictions). The shown schema corresponds to the created \gls{csv} row in listing~\ref{lst:csvdataextractor-createcsvrow}, but compared to the originally defined features in table~\ref{tab:features-overview}---which constitute the actual input for the \gls{ann}---the transformation process must be aware of the identifier in the first column that is only used for debugging purposes. Constants such as \inline{COMPONENT} or \inline{FOCUS_DISTANCE} are statically imported from the previously introduced \inline{CsvSchema} helper class. \inline{BOOLEAN_CATEGORICALS} belongs to \inline{CsvDataTransformer} and is only used locally to define a fixed order for the boolean categoricals, where \inline{false} is at index~0 and \inline{true} at index~1.

\begin{listing}[h]
\begin{minted}{java}
private static Schema createSchema() {
	return new Schema.Builder()
			.addColumnsString(COMPONENT)
			.addColumnCategorical(ENABLED, BOOLEAN_CATEGORICALS)
			.addColumnCategorical(PREFERRED_TYPE, BOOLEAN_CATEGORICALS)
			.addColumnInteger(FOCUS_DISTANCE)
			.addColumnInteger(PATH_DISTANCE)
			.addColumnDouble(POINT_DISTANCE)
			.addColumnCategorical(LABEL, BOOLEAN_CATEGORICALS)
			.build();
}
\end{minted}
\caption{CSV schema definition.}
\label{lst:csvdatatransformer-createschema}
\end{listing}

Now that the \gls{csv} schema has been defined with the \inline{Schema} class, it is possible to determine the data transformation on top of the schema via \inline{TransformProcess}. An instance of the class defines a sequential list of data transformations that can be executed on Spark. As can be seen in listing~\ref{lst:csvdatatransformer-createtransformprocess}, the transform process for the given schema is relatively simple: the first step removes the component identifier as it is useless for the \gls{ann}, the second step performs the label encoding for the boolean categoricals.

\begin{listing}
\begin{minted}{java}
private static TransformProcess createTransformProcess(Schema schema) {
	return new TransformProcess.Builder(schema)
			.removeColumns(COMPONENT)
			.categoricalToInteger(ENABLED, PREFERRED_TYPE, LABEL)
			.build();
}
\end{minted}
\caption{Transform process definition.}
\label{lst:csvdatatransformer-createtransformprocess}
\end{listing}

Section~\ref{sec:feature-engineering} already pointed out that the data may become very imbalanced because every action only has one correct target component~(labeled \inline{true}), whereas the number of possible target components in each state is unbounded~(labeled \inline{false}). One way to cope with this issue is to simply collect more data of the rare examples, i.e. more correct target components. Unfortunately, this is not effective in this case, which can be illustrated by revisting the example given in section~\ref{sec:feature-engineering}: If 10 test cases are available, each containing 10 actions, and the average number of possible target components in each state is 25, then this creates $10 \cdot 9 \cdot 25 = \num{2250}$ feature vectors. But only \SI{4}{\percent}~($10 \cdot 9 = 90$ out of \num{2250}) of these examples are labeled \inline{true}. Increasing the number of test cases or actions obviously also increases the number of feature vectors that are labeled \inline{false}. Therefore, the ratio effectively remains the same. Another way to address label imbalance is by adapting the labeling itself. For instance, one might be able to merge rare labels into a single, more frequent label. However, this is not an option because the given task represents a binary classification problem. A third possibility is a technique called sampling\footnote{\url{https://en.wikipedia.org/wiki/Oversampling_and_undersampling_in_data_analysis}.}, which basically adjusts the label distribution within the data set. \emph{Oversampling} means that examples of the rare label are shown to the algorithm with a higher frequency, whereas \emph{undersampling} shows the more frequent label less often; the effect of both is essentially the same. Since DataVec does not support a mechanism to perform sampling, a custom solution for Spark is implemented.

\begin{listing}[ht!]
\begin{minted}{java}
private static JavaRDD<String> handleImbalance(JavaRDD<String> lines) {
	Function<String, Boolean> isTrue = line -> {
		String[] columns = line.split(",");
		return Boolean.parseBoolean(columns[columns.length - 1]);
	};

	JavaRDD<String> trueLines = lines.filter(isTrue);
	JavaRDD<String> falseLines =
			lines.filter(line -> !isTrue.call(line));
	long numOfLines = lines.count();
	long numOfTrueLines = trueLines.count();
	long numOfFalseLines = falseLines.count();
	logger.info("Stats (raw): {}, {} ({}) false, {} ({}) true.",
			numOfLines, numOfFalseLines,
			prettyRatio(numOfFalseLines, numOfLines),
			numOfTrueLines,
			prettyRatio(numOfTrueLines, numOfLines));

	JavaRDD<String> sampledFalseLines =
			falseLines.sample(false, 0.9, Randomness.getSeed());
	JavaRDD<String> balancedLines =
			trueLines.union(sampledFalseLines);
	long numOfSampledFalseLines = sampledFalseLines.count();
	long numOfBalancedLines = balancedLines.count();
	logger.info("Stats (sampled): {}, {} ({}) false, {} ({}) true.",
			numOfBalancedLines, numOfSampledFalseLines,
			prettyRatio(numOfSampledFalseLines, numOfBalancedLines),
			numOfTrueLines,
			prettyRatio(numOfTrueLines, numOfBalancedLines));

	return balancedLines;
};
\end{minted}
\caption{Handling data imbalance with undersampling.}
\label{lst:csvdatatransformer-handleimbalance}
\end{listing}

Without going into too much detail regarding Spark's internals, the basic abstraction it uses are \glspl{rdd}. An \gls{rdd} is an immutable and partitioned collection of records that offers operations such as \inline{map} and \inline{reduce}. \glspl{rdd} serve to increase the fault tolerance in distributed computing. Each distributed node in Spark works on a particular partition of an \gls{rdd} and performs the given sequence of operations, called lineage. If a node crashes, another node can easily jump in and perform the same lineage again without having to deal with corrupt data. Listing~\ref{lst:csvdatatransformer-handleimbalance} shows the \inline{handleImbalance} method which receives the raw \gls{csv} lines as a \inline{JavaRDD}. The method starts with a split into \inline{true} and \inline{false} lines, i.e. it separates the feature vectors that constitute a correct target component from those that do not. Aside from various logging statements, the actual sampling can be found in line~21. The implementation uses \inline{JavaRDD#sample(boolean, double, long)} to keep only \SI{90}{\percent} of the \inline{false} lines, which are chosen randomly using the given seed. Although this only slightly increases the probability distribution of the \inline{true} labels, section~ \ref{subsec:model-accuracy} will show that this is enough to make the \gls{ann} learn. Afterwards, the balanced lines as well as the previously defined transform process are deployed to Spark via the DataVec Spark \gls{api} and finally merged into a single \gls{csv} file again~(see appendices~\ref{lst:raw-csv}, \ref{lst:transformed-csv}, and \ref{lst:transform-datavec-spark} for examples of the raw and the transformed \gls{csv} as well as the deployment on local Spark).

\section{Network Construction and Training}
\label{sec:ann-construction-training}

The construction and training of the \gls{ann} is done by \inline{MonkeyTamer}. Just as \inline{CsvDataTransformer}, the class offers a \inline{main} method as well as a public class method named \inline{trainAndEval}, which both expect the absolute path to the transformed \gls{csv} file; again, the former as a program argument and the latter as a \inline{java.nio.file.Path}. The first thing it does is preparing the data via the \inline{prepareData} method~(see listing~\ref{lst:monkeytamer-preparedata}). The transformed \gls{csv} file is loaded through several DataVec helper classes and then shuffled before it is split into a training~(\SI{70}{\percent}) and a test~(\SI{30}{\percent}) set. Afterwards, both sets are normalized such that they have a mean of 0 and a standard derivation of 1. The corresponding normalizer is persisted in order to be able to reuse it later.

\begin{listing}[h]
\begin{minted}{java}
private static SplitTestAndTrain prepareData(int batchSize, Path csv)
		throws IOException, InterruptedException {
	double percTrain = 0.70;
	logger.info("Preparing data (batch size: {}, training: {}%).",
			batchSize, String.format("%.2f", percTrain * 100.0));
	RecordReader reader = new CSVRecordReader();
	reader.initialize(new FileSplit(csv.toFile()));
	DataSetIterator iter = new RecordReaderDataSetIterator(
			reader, batchSize, LABEL_INDEX, NUMBER_OF_CLASSES);
	DataSet allData = iter.next();
	allData.shuffle();
	SplitTestAndTrain testAndTrain =
			allData.splitTestAndTrain(percTrain);		
	normalize(testAndTrain.getTrain(), testAndTrain.getTest());
	return testAndTrain;
}
\end{minted}
\caption{Data preparation before training.}
\label{lst:monkeytamer-preparedata}
\end{listing}

\begin{listing}[ht!]
\begin{minted}{java}
MultiLayerNetwork initialModel = initModel();
initialModel.setListeners(new StatsListener(
		new RemoteUIStatsStorageRouter("http://localhost:9000")));

logger.info("Training model.");
EarlyStoppingConfiguration<MultiLayerNetwork> earlyStoppingConfig =
		new EarlyStoppingConfiguration.Builder<MultiLayerNetwork>()
				.epochTerminationConditions(
						new MaxEpochsTerminationCondition(400))
				.evaluateEveryNEpochs(1)
				.iterationTerminationConditions(
						new MaxTimeIterationTerminationCondition(1,
								TimeUnit.MINUTES))
				.scoreCalculator(
						new DataSetLossCalculator(
								new ListDataSetIterator(
										testData.asList()), true))
				.build();
EarlyStoppingTrainer trainer = new EarlyStoppingTrainer(
		earlyStoppingConfig, initialModel,
		new ListDataSetIterator(trainingData.asList()));
EarlyStoppingResult<MultiLayerNetwork> result = trainer.fit();
\end{minted}
\caption{Training of the ANN with early stopping.}
\label{lst:monkeytamer-earlystopping}
\end{listing}

\begin{listing}
\begin{minted}{java}
private static MultiLayerNetwork initModel() {
	MultiLayerConfiguration configuration =
			new NeuralNetConfiguration.Builder()
			.seed(Randomness.getSeed())
			.iterations(1)
			.activation(Activation.SIGMOID)
			.weightInit(WeightInit.XAVIER)
			.optimizationAlgo(
					OptimizationAlgorithm.STOCHASTIC_GRADIENT_DESCENT)
			.updater(Updater.RMSPROP)
			.learningRate(1e-2)
			.regularization(true)
			.l2(1e-3)
			.list()
			.backprop(true)
			.pretrain(false)
			.layer(0, new DenseLayer.Builder()
					.nIn(5)
					.nOut(5)
					.build())
			.layer(1, new DenseLayer.Builder()
					.nIn(5)
					.nOut(3)
					.build())
			.layer(2, new OutputLayer.Builder()
					.activation(Activation.SOFTMAX)
					.lossFunction(new LossBinaryXENT(
							Nd4j.create(new double[]{0.8, 1.2})))
					.nIn(3)
					.nOut(2)
					.build())
			.build();
	MultiLayerNetwork model = new MultiLayerNetwork(configuration);

	logger.info("Initializing network with configuration:\n{}",
			configuration);
	model.init();

	return model;
}
\end{minted}
\caption{Construction and initialization of the network.}
\label{lst:monkeytamer-initmodel}
\end{listing}

Listing~\ref{lst:monkeytamer-initmodel} shows how the \gls{ann} is constructed and initialized. The entire configuration corresponds to the discussed network topology and its hyperparameters in section~\ref{subsec:network-topology} and \ref{subsec:hyperparameters}. Again \dlfj{} uses the builder pattern\footnote{\url{https://en.wikipedia.org/wiki/Builder_pattern}.} for this purpose, which allows the configuration to be expressed in a relatively concise way. After the network is constructed, the \inline{trainAndEval} method starts with the actual training of which an excerpt can be found in listing~\ref{lst:monkeytamer-earlystopping}. First of all, a \inline{StatsListener} is registered on the model to collect system and model information during training. In particular, a \inline{RemoteUIStatsStorageRouter} is used, which offers a web-based \gls{gui} to visualize and monitor the training progress in real time. The next step, line~6, sets up the \inline{EarlyStoppingConfiguration} to specify the options for performing the training with early stopping. As can be seen, the training stops after 400~epochs or 1~minute, where the model is evaluated in every epoch. Since there is only little data available, it is sufficient to abort the training process after these stopping conditions. This is also why the training is executed locally as indicated by the address of the remote listener before. In order to start the training of the model, an \inline{EarlyStoppingTrainer} is needed; its \inline{fit} method conducts the early stopping training and returns an \inline{EarlyStoppingResult}, containing information such as the termination reason, the score of the model, and, of course, the best model itself. The not illustrated part of the \inline{trainAndEval} method closes with persisting this model so that it can be reused like the normalizer.

The results of the training, including details on the training process provided by the remote listener, can be found in section~\ref{subsec:model-accuracy}. Section~\ref{sec:evaluation} also describes how the~(two-fold) evaluation scenario is set up and particularly how the training data is obtained~(i.e. which \gls{sut} and which tests are used).

\section{Integration with ReTest}
\label{sec:retest-integration}

To now integrate the previously trained model in ReTest, \inline{MonkeyExecutor} is adapted so that the existing system can use the acquired knowledge while creating the initial population of the \gls{ga}. As mentioned in section \ref{sec:overall-concept}, the \inline{getNextAction} method uses a cascade of \inline{if} statements to choose the next action. \inline{MonkeyModel} respectively its implementation, \inline{MonkeyModelImpl}, is inserted between the second and the third rule, which leads to the following set of rules in descending order:
\begin{enumerate}
\item Terminate the \gls{sut} if the current state is an exit state.
\item Follow the current road map~(see below).
\item \emph{Execute a predicted action.}
\item Execute an unexplored action.
\item Create and follow a road map to a state with unexplored actions.
\item Execute a random action.
\end{enumerate}
The new third rule is coupled to the \inline{MONKEY_MODEL_USAGE_PROBABILITY} property, which determines the probability with which \inline{MonkeyModelImpl} is used. It is defined within the \inline{MonkeyModel} interface and accepts integer values between 0--100 that are mapped to the corresponding percentage disclosure. The property enables the user to define to which degree the \gls{ann} is used as he might still want to focus on pure monkey testing for reliability reasons, rather than having a portion of human behavior in the generated tests.

\begin{listing}[h]
\begin{minted}{java}
@Override
public Action getNextAction(Action previousAction,
		NormalState fromState) {
	if (previousAction == null) {
		logger.debug("Previous action is null, returning null.");
		return null;
	}

	ComponentDescriptor previousTarget =
			previousAction.getTargetComponentDescriptor();
	List<ComponentDescriptor> flattenedPossibleTargets =
			ComponentDescriptors.flattenAllComponents(
					new ArrayList<>(fromState.getWindowDescriptors()));
	List<ComponentDescriptor> sortedPredictedTargets =
			getSortedPredictedTargets(
					previousTarget, flattenedPossibleTargets);

	return createActionFor(sortedPredictedTargets, fromState);
}
\end{minted}
\caption{Implemented \inline{MonkeyModel} interface.}
\label{lst:monkeymodelimpl-getnextaction}
\end{listing}

Listing~\ref{lst:monkeymodelimpl-getnextaction} shows the implemented \inline{MonkeyModel} interface that is used to retrieve the predicted target in \inline{MonkeyExecutor}. The method first checks if the given previous~(maybe first) action is \inline{null} to optionally return \inline{null} as well---which would make~\inline{MonkeyExecutor} execute an unexplored action if feasible. Just like during the feature extraction, the next steps extract the three core domain objects. The main difference here is that there is no label as this is what the model is supposed to predict. After the possible target components are flattened~(i.e. the nested component hierarchy is transformed into a \enquote{flat} list), they are handed over to the \inline{getSortedPredictedTargets} method together with the previous target component. As shown in listing~\ref{lst:monkeymodelimpl-getsortedpredictedtargets}, the overloaded method defines a two-stage process. The first method, which receives a \inline{ComponentDescriptor} as its first parameter, starts with converting the possible target components to a feature matrix. A feature matrix is simply a matrix of feature vectors, i.e. each row of that matrix encloses a feature vector. The \gls{ann} needs its input to be in the same format that was used during training. Hence, the conversion must execute the extraction and transformation again---including the normalization of these values with the aid of the previously persisted normalizer---but programmatically during runtime~(see appendix~\ref{lst:convert-feature-matrix} for the corresponding implementation). The return type, \inline{INDArray}, is also part of \dlfj{}\footnote{\url{https://nd4j.org/}.} and represents an $n$-dimensional array that is used to store the feature matrix.

Invoking the \inline{output} method in line~6 retrieves the predictions from the trained network, which also returns an \inline{INDArray}. But instead of a matrix of feature vectors, it is now a matrix of probabilities for each label. That is, a feature vector $(x_{i, 0} \; x_{i, 1} \; x_{i, 2} \; x_{i, 3} \; x_{i, 4})$ in row $i$, where $x_0$--$x_4$ are the numerical input values for the particular features enabled, preferred type, focus distance and so forth, becomes an output vector $(o_{i, 0} \; o_{i, 1})$, in which $o_{i, 0}$ constitutes the probability of the label \inline{false} and $o_{i, 1}$ of the label \inline{true}:
\begin{equation*}
\begin{pmatrix}
x_{0, 0} & x_{0, 1} & x_{0, 2} & x_{0, 3} & x_{0, 4} \\
x_{1, 0} & x_{1, 1} & x_{1, 2} & x_{1, 3} & x_{1, 4} \\
x_{2, 0} & x_{2, 1} & x_{2, 2} & x_{2, 3} & x_{2, 4} \\
x_{3, 0} & x_{3, 1} & x_{3, 2} & x_{3, 3} & x_{3, 4} \\
\vdots   & \vdots   & \vdots   & \vdots   & \vdots
\end{pmatrix}
\Rightarrow
\begin{pmatrix}
o_{0, 0} & o_{0, 1} \\
o_{1, 0} & o_{1, 1} \\
o_{2, 0} & o_{2, 1} \\
o_{3, 0} & o_{3, 1} \\
\vdots   & \vdots
\end{pmatrix}
\end{equation*}
The corresponding result \inline{predictions} is then passed to the second method.~(As the missing access level modifier indicates, tests are created for this method, too, as it represents a critical part of the prototype.) Since one is only interested in the probabilities for the \inline{true} label, the method creates another \inline{INDArray} from \inline{predictions} which now conducts a single-column array that contains the guessed probability for each possible target component. In order to sort these values from best to worst, \inline{flattenedPossibleTargets} must be sorted accordingly since the indices of both belong to each other. Line~18 first creates a \inline{Pair} for every index which consists of the index itself as well as the corresponding probability. It then sorts the values in descending order with respect to the probabilities, which is followed by a mapping from the indices to the respective possible target component. Thus, the returned list represents the predicted possible target components, starting with the best recommended component at index~0.

\begin{listing}
\begin{minted}{java}
private List<ComponentDescriptor> getSortedPredictedTargets(
		ComponentDescriptor previousTarget,
		List<ComponentDescriptor> flattenedPossibleTargets) {
	INDArray featureMatrix = convertToFeatureMatrix(previousTarget,
			flattenedPossibleTargets);
	INDArray predictions = model.output(featureMatrix);

	return getSortedPredictedTargets(predictions,
			flattenedPossibleTargets);
}

static List<ComponentDescriptor> getSortedPredictedTargets(
		INDArray predictions,
		List<ComponentDescriptor> flattenedPossibleTargets) {
	INDArray truePredictions = predictions.getColumn(1);

	return IntStream.range(0, truePredictions.size(0))
			.mapToObj(index -> Pair.of(
					index, truePredictions.getDouble(index)))
			.sorted(Comparator.<Pair<Integer, Double>> comparingDouble(
					Pair::getRight).reversed())
			.map(pair -> flattenedPossibleTargets.get(pair.getLeft()))
			.collect(Collectors.toList());
}
\end{minted}
\caption{Sorting of the predicted correct target components.}
\label{lst:monkeymodelimpl-getsortedpredictedtargets}
\end{listing}

The final task of the \inline{getNextAction} method is to generate an action for the predicted correct target components, which is done within the \inline{createActionFor} method. The state graph offers two methods to retrieve the currently available unexplored and random actions; both are invoked with the given state and if the unexplored actions are not empty, they are given precedence for two reasons:
\begin{enumerate}
\item To respect the set of rules defined by \inline{MonkeyExecutor}, which prefers unexplored actions over random actions.
\item To address user story~\#\ref{story:monkey-performance}~(monkey testing performance) by still focusing on exploration and, therefore, implicitly optimizing towards branch coverage.
\end{enumerate}
If actions are available, line~16 starts looping over the \inline{sortedPredictedTargets}, going from best to worst. It then filters the available actions based on a) wether they address the predicted correct target component and b) if they are not contained in \inline{previousTargets}. The latter is an \inline{EvictingQueue} from the Google Guava library\footnote{\url{https://github.com/google/guava/}.}; the data structure represents a queue that automatically evicts elements from its head when new elements are added although it is full, similar to a circular buffer. This avoids that the model keeps recommending one or more \gls{gui} components over and over again. Consequently, since the size is set to 10, the queue ensures that the 10 previous target components are distinct. \inline{findAny()} then tries to select an action from the resulting stream, which returns an \inline{Optional}. If the value is present, it is added to the \inline{previousTargets} queue and then returned. If the method returns \inline{null}---either because they are no unexplored nor random actions or the value of the \inline{Optional} is absent---\inline{MonkeyExecutor} skips to the next rule to select an unexplored actions for a random target component. This behavior also promotes diversity for the evolution of the \gls{ga} since there is still some amount of randomness involved. As mentioned before, the balance between predicted and random actions respectively their underlying components can be explicitly controlled via the \inline{MONKEY_MODEL_USAGE_PROBABILITY} property.

\begin{listing}
\begin{minted}{java}
Action createActionFor(
		List<ComponentDescriptor> sortedPredictedTargets,
		NormalState fromState) {
	List<Action> unexploredActions =
			stateGraph.getStateNeverExploredActions(fromState);
	List<Action> randomActions =
			stateGraph.getRandomActions(fromState);
	List<Action> actions = !unexploredActions.isEmpty()
			? unexploredActions : randomActions;

	if (actions.isEmpty()) {
		logger.debug("No action available for given state.");
		return null;
	}

	for (ComponentDescriptor predicted : sortedPredictedTargets) {
		Optional<Action> createdAction = actions.stream()
				.filter(action -> ComponentDescriptors.pathEquals(
						action.getTargetComponentDescriptor(),
						predicted))
				.filter(action -> !previousTargets.contains(
						action.getTargetComponentDescriptor()
								.getIdentificationCriteria()
								.getPathTyped()))
				.findAny();

		if (createdAction.isPresent()) {
			previousTargets.add(predicted.getIdentificationCriteria()
					.getPathTyped());
			return createdAction.get();
		}
	}

	logger.debug("No action available for predicted target.");

	return null;
}
\end{minted}
\caption{Action creation for the predicted correct target components.}
\label{lst:monkeymodelimpl-createactionfor}
\end{listing}

\section{Evaluation}
\label{sec:evaluation}

The evaluation of the approach is two-fold: The first part, section~\ref{subsec:model-accuracy}, focuses on the evaluation of the model in terms of its performance on the test set. The second part in section~\ref{subsec:test-generation} then evaluates the resulting model integrated with ReTest to see how it affects the test generation mechanism. Hence, part two is based on part one since it uses the same model. Just as in the previous chapters, ReTest's demo application serves as the \gls{sut}. In particular, it is used to create tests that can be leveraged for feature extraction and it is also used to evaluate the test generation mechanism with and without the \gls{ann}.

\subsection{Model Accuracy}
\label{subsec:model-accuracy}

In order to evaluate the accuracy of the \gls{ann}, it is necessary to first create some tests that can be used to extract features. Therefore, two suites have been created that aim to test the most important parts of the \gls{sut}. The \inline{address-book} suite~(see appendix~\ref{lst:address-book-feature-extraction}) consists of 3 tests that contain $12 + 6 + 8 = 26$ actions in total. The \inline{calculator} suite~(see appendix~\ref{lst:calculator-feature-extraction}) is composed of 4 tests with $8 + 8 + 7 + 7 = 30$ actions. \inline{CsvDataExtractor} was able to extract \num{4629} feature vectors of which $26 + 30 = 56$~(\SI{1.21}{\percent}) are labeled \inline{true}, hence, they represent correct target components. Due to undersampling, this is reduced to \num{4153} examples, which slightly increases the proportion of those labeled \inline{true} to \SI{1.35}{\percent}.

\begin{table}[h]
\small
\centering
\begin{tabular}{l l l l} \toprule
Accuracy &
Precision &
Recall &
F1 \\ \midrule
\SI{82.05}{\percent} &
\SI{76.92}{\percent} &
\SI{81.38}{\percent} &
\SI{79.09}{\percent} \\ \bottomrule
\end{tabular}
\captionbelow{Model score overview.}
\label{tab:model-score}
\end{table}

After applying the aforementioned data transformations, the \gls{ann} was able to reach an accuracy\footnote{\url{https://en.wikipedia.org/wiki/Accuracy_and_precision}.} of \SI{82}{\percent} as illustrated in table~\ref{tab:model-score}. The accuracy metric is the ratio of all correctly predicted labels with respect to the total number of test examples. Precision and recall\footnote{\url{https://en.wikipedia.org/wiki/Precision_and_recall}.} as well as the F1 score\footnote{\url{https://en.wikipedia.org/wiki/F1_score}}, including the accuracy itself, are calculated based on the following categories:
\begin{description}
\item[True positives (TP)] Correctly assigned 1 respectively \inline{true}.
\item[False positives (FP)] Wrongly assigned 1 respectively \inline{true}.
\item[True negatives (TN)] Correctly assigned 0 respectively \inline{false}.
\item[False negatives (FN)] Wrongly assigned 0 respectively \inline{false}.
\end{description}
The performance metrics are then defined as follows:
\begingroup
\addtolength{\jot}{1em}
\begin{align*}
\text{Accuracy} &= \frac{TP + TN}{TP + FP + TN + FN} \\
\text{Precision} &= \frac{TP}{TP + FP} \\
\text{Recall} &= \frac{TP}{TP + FN} \\
\text{F1 score} &= 2 \cdot \frac{\text{Precision} \cdot \text{Recall}}{\text{Precision} + \text{Recall}}
\end{align*}
\endgroup
Consequently, precision denotes the proportion of correctly assigned \inline{true} labels over all examples that haven been labeled \inline{true}, whereas recall represents the ratio of correctly assigned \inline{true} labels within all examples that should be \inline{true}. The F1 score is simply the weighted average of these two metrics.

An example of a corresponding per-batch confusion matrix\footnote{\url{https://en.wikipedia.org/wiki/Confusion_matrix}.} can be found in table~\ref{tab:confusion-matrix}. The diagonal of the table shows all correct predictions~($TP + TN$), whereas the values outside the diagonal are those examples that have been wrongly classified~($FP + FN$). The shown confusion matrix consists of $128 \cdot \SI{30}{\percent} \approx 39$ examples, where 128 refers to the minibatch size and \SI{30}{\percent} is the test set size. It illustrates that shuffling the data may lead to minibatches that do not reflect the actual label distribution, because \inline{true} labels are usually less common.

\begin{table}[h]
\small
\centering
\begin{tabular}{l r r} \toprule
&
0 &
1 \\ \midrule
0 labeled as \ldots &
8 &
2 \\
1 labeled as \ldots &
5 &
24 \\ \bottomrule
\end{tabular}
\captionbelow{Per-batch confusion matrix example.}
\label{tab:confusion-matrix}
\end{table}

\begin{figure}[h]
\centering
\includegraphics[scale=0.55]{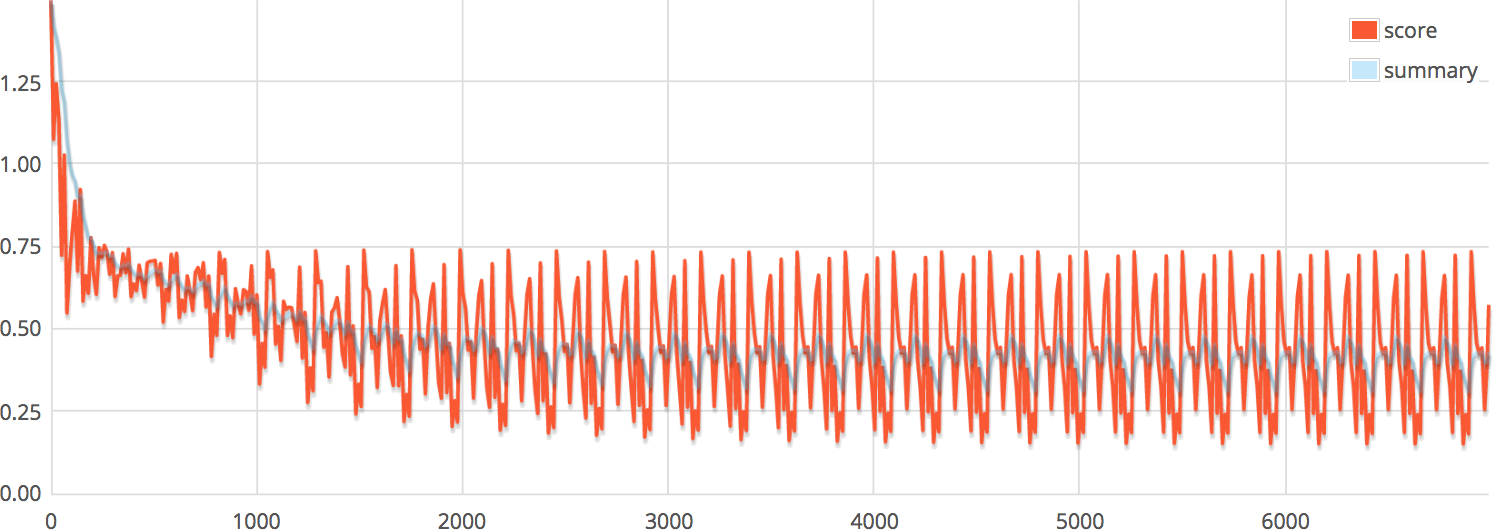}
\captionbelow{Loss vs. iteration.}
\label{fig:loss-vs-score}
\end{figure}

The \dlfj{} remote listener visualizes the progress of the model during the training process in terms of loss~($y$-axis) compared to iterations~($x$-axis) as shown in figure~\ref{fig:loss-vs-score}. The result basically looks reasonable, but two things should be noted:
\begin{enumerate}
\item The speed of decay seems to be a bit low, which often indicates a too small learning rate. But adjusting the hyperparameter did not lead to better results.
\item Since the loss is quite noisy, one typically tries to increase the minibatch size. However, enlarging the minibatches actually created worse results.
\end{enumerate}
In addition, appendix~\ref{fig:parameter-ratios} shows the parameter update ratio by layer~($\log_{10}$) vs. iteration. As a rule of thumb, it should be $1:\num{1000} = \num{0.001} = \log_{10}-3$~\cite{sky17b}, which was achieved almost exactly. According to the authors, the standard deviations of the layer activations---shown in appendix~\ref{fig:std-dev-activations}---should range from \SIrange{0.5}{2.0}{}, i.e. $[\log_{10}0.5,\; \log_{10}2.0] \approx [-0.3,\; 0.3]$. Hence, this is appropriate, too.

Due to the limited amount of data, both training and evaluation were performed locally on an Apple MacBook Pro~(Retina \SI{13}{\inch}, early 2015) with a \SI{2.7}{\giga\hertz} Intel Core i5 CPU and \SI{8}{\giga\byte} of RAM, using macOS Sierra~(10.12.6), Oracle Java~SE 8u144, and \dlfj{} 0.8.0. All 400 epochs have been executed in less than a minute.

\subsection{Test Generation}
\label{subsec:test-generation}

To benchmark the performance of the \gls{ann} integrated with ReTest, the scenario in listing~\ref{lst:eval-script} has been scripted and executed , both without and with the trained model~(i. e. \inline{MONKEY_MODEL_USAGE_PROBABILITY} set to 0 and 100). The evaluation ran on a Windows~7~(SP1) virtual machine with a \SI{3.6}{\giga\hertz} dual core CPU and \SI{8}{\giga\byte} of RAM, again using the Oracle Java~SE 8u144 and \dlfj{} 0.8.0. Besides various directory structure-specific commands, the script invokes the test generation mechanism 15 times via ReTest's Ant\footnote{\url{http://ant.apache.org/}.} interface, where each run tries to achieve a branch coverage of \SI{50}{\percent} within 2~minutes.~(This may sound a lot, but the conditions are adequate since the \gls{sut} is quite small.) Each iteration creates a separate log file and removes the persisted generation so that the next iteration cannot benefit from it. After all 15~iterations have been executed, all generated test suites are replayed to also create a corresponding \gls{tap} report. Similar to \cite{ep16}, the evaluation addresses the following questions:
\begin{enumerate}
\item How effective is the approach in terms of branch coverage?
\item How efficient is the implementation when it comes to execution time?
\item How did the characteristics of the generated tests evolve?
\end{enumerate}

To answer the first question, table~\ref{tab:branch-coverage} shows the branch coverage of all generated test suites without~(column \enquote{before}) and with~(column \enquote{after}) using the trained model. As can bee seen, generating tests without the \gls{ann} is slightly better, but only \SI{0.97}{\percent} on average. However, this statistically insignificant difference is negligible. Moreover, the deficit is also comprehensible: Human testers usually create test cases in order to cover specific scenarios, not to increase branch coverage in the first place. For instance, in case of the given \gls{sut}, such a scenario might be to login, add a new contact to the address book, and save it~(see \inline{add-contact} in appendix~\ref{lst:address-book-feature-extraction}). Because all tests that have been used to train the model follow this principle, the \gls{ann} is influenced accordingly.

\begin{listing}[h]
\begin{minted}{bat}
cd ..

echo Starting evaluation ...

echo Generating execsuites ...
for /l %%i in (1, 1, 15) do (
  echo Running generation %%i ...
  ant generate > .\evaluation\logs\generated-%%i.log
  rmdir .\generations\ /s /q
)

echo Replaying execsuites ...
ant replay > .\evaluation\logs\replay.log

echo Finished evaluation.
\end{minted}
\caption{Batch script for test generation evaluation.}
\label{lst:eval-script}
\end{listing}

\begin{table}[ht]
\small
\centering
\begin{tabular}{r r r} \toprule
Test suite &
Before~(in \si{\percent}) &
After~(in \si{\percent}) \\ \midrule
1 &
47.37 &
46.75 \\
2 &
49.85 &
47.06 \\
3 &
47.99 &
47.37 \\
4 &
47.06 &
47.37 \\
5 &
44.58 &
46.75 \\
6 &
47.37 &
46.44 \\
7 &
46.44 &
46.44 \\
8 &
48.92 &
47.68 \\
9 &
47.68 &
46.13 \\
10 &
49.85 &
47.06 \\
11 &
49.54 &
46.44 \\
12 &
48.61 &
47.06 \\
13 &
47.99 &
46.44 \\
14 &
47.68 &
47.37 \\
15 &
46.44 &
46.44 \\\addlinespace
Avg. &
47.82 &
46.85 \\ \bottomrule
\end{tabular}
\captionbelow{Branch coverage results.}
\label{tab:branch-coverage}
\end{table}

In terms of efficiency, the values vary a lot. The measurements from \num{1807} invocations of \inline{MonkeyModelImpl#getNextAction(Action, NormalState)} revealed that the delay ranges from \SIrange{4}{5847}{\ms}. While \SI{4}{\ms} is very good, the latter is unacceptable. Nonetheless, only 149 predictions took more than \SI{1000}{\ms}; 760 needed less than \SI{100}{\ms}, the remaining 900 invocations took between \SI{100}{\ms} and \SI{1000}{\ms}. Considering the fact that this only happens once per \gls{gui} action during the population initialization of the \gls{ga}, the implementation appears to be relatively efficient. But in the case of a use in production, it is first necessary to investigate why the severe delays occurred and how they can be avoided.

Since the main objective of this thesis is to reduce the gap between manually created and automatically generated regression tests, question number three can be considered to be the most important one. In order to answer it, the resulting \gls{tap} reports from the generated suites have been compared semantically to the reports of the manually created ones, which were also used to train the \gls{ann}. This is done exemplary for the fittest---in terms of the achieved branch coverage---generated suites in both cases. Before the \gls{ann} was integrated, suite~2 and 10~(\SI{49.85}{\percent}) yielded the best results, whereas it was suite~8~(\SI{47.68}{\percent}) after the integration. For simplification, the comparison will refer to suite~2 before the integration as $B_2$ and to suite~10 after the integration as $A_{10}$. $B_2$ consists of 15 tests that contain 379 actions, $A_{10}$ also has 15 tests but only $249$ actions. That is, $A_{10}$ achieved almost the same coverage~(\SI{2.17}{\percent} less) with 130 actions less than $B_2$. Consequently, the average test case size is $|C|_{\text{avg}} = \num[parse-numbers=false]{25.2\overline{6}}$ for $B_2$, whereas $A_{10}$ only has $|C|_{\text{avg}} = \num{16.60}$. As mentioned in section~\ref{subsec:retest-ga}, short tests usually require less execution time and less maintenance effort. 13 out of 15 tests of $B_2$ logged into the \gls{sut} after two attempts. In case of $A_{10}$, only 12 tests performed a login, which took three attempts. However, $B_2$ did only try how the \gls{sut} behaves without credentials, whereas $A_{10}$ tried both no and invalid credentials. When it comes to human behavior, $A_{10}$ clearly outperforms $B_2$. For example, $B_2$ never added, deleted, or edited an entry in the address book tab, but $A_{10}$ did so multiple times. Most importantly, $A_{10}$ was able to fill out forms such as the one for new addresses almost completely. $B_2$ never executed more than one text enter action in a row after the login dialog; $A_{10}$ entered text to coherent text fields up to five times consecutively as can be seen in listing~\ref{lst:test-example-after}.

\begin{listing}[h]
\begin{minted}[breaklines=true]{text}
ok 7 generated
    1..16
    ok 1 Entering text 'Max' into JTextField Benutzername
    ok 2 Entering text 'ReTest' into JPasswordField Passwort
    ok 3 Click on JButton [Login]
    ok 4 Click on BasicInternalFrameTitlePane$NoFocusButton [InternalFrameTitlePane.maximizeButton]
    ok 5 Click on BasicInternalFrameTitlePane$NoFocusButton [InternalFrameTitlePane.closeButton]
    ok 6 Entering text 'ainategrump' into JTextField
    ok 7 Click on JButton [Add address]
    ok 8 Entering text 'yre' into JTextField
    ok 9 Click on JButton [Delete address]
    ok 10 Entering text 'AddressbookPanel.labelCaption' into JTextField
    ok 11 Entering text 'Postal code' into JTextField
    ok 12 Entering text 'ackeialroma' into JTextField
    ok 13 Entering text '29' into JTextField
    ok 14 Entering text 'InternalFrameTitlePane.maximizeButton' into JTextField
    ok 15 Click on JButton [Edit entry]
    ok 16 Click on JButton [Close]
\end{minted}
\caption{Generated test case example after the integration.}
\label{lst:test-example-after}
\end{listing}

Yet, the \gls{tap} report of this particular test case also illustrates the limitations of the approach. Although all of the three aformentioned address book actions have been executed, they are not executed together with \emph{meaningful} text enter actions which address corresponding fields such as \enquote{First name} or \enquote{Last name}~(see appendix~\ref{app:demo-address-book}). The reasons for this have already been discussed by Ermuth and Pradel as they face a similar problem:
\begin{displaycquote}[89]{ep16}
First, to be able to cover a scenario, the test generator must reach a particular state, which it fails to do for some scenarios \textelp{} Second, even when the program is in the specific state, the test generator may not trigger the \enquote{right} macro event that covers the scenario. For example, there may be other applicable macro events that the test generator triggers, or it may also select an event randomly.
\end{displaycquote}
In order to cover a scenario a human tester would consider meaningful, the test generator must follow a certain sequence of actions that is reasonable in the context of the given \gls{sut}. Although the state graph helps doing so, due to the domination of the \gls{ga}---which favors unexplored over meaningful actions---it is rather difficult to create sequences that cover such scenarios if they do not increase the coverage significantly.

%% file: chapters/6-conclusion.tex


\chapter{Conclusion}
\label{ch:conclusion}

This final chapter summarizes and reflects the findings of this thesis in section~\ref{sec:summary}, which is followed by a discussion of possible future work that is based on the acquired results in section~\ref{sec:future-work}.

\section{Summary}
\label{sec:summary}

This thesis investigated how ReTest's branch coverage-optimizing \gls{ga} could be improved with the aid of \gls{ml} to reduce the gap between manually created and automatically generated regression tests. In doing so, a simple \gls{ann} was designed and implemented with \dlfj{} to rank \gls{gui} actions respectively their underlying \gls{gui} components at runtime. The identified and extracted features, for which a dedicated \gls{etl} pipeline was implemented as well, led to an accuracy of \SI{82}{\percent} based on the knowledge obtained from existing tests. On top of ReTest, the prototype forms an \gls{mlec} algorithm that improves the initial population of the corresponding \gls{ga}. The presented methods describe a general framework that can be easily used to guide any monkey testing mechanism with \gls{ml}.

The evaluation of the approach integrated with ReTest showed that the test generator was able to maintain its branch-coverage performance while reducing the required amount of \gls{gui} actions to less than two-thirds without a significant increase in terms of execution time. Most importantly, the characteristics of the generated tests evolved such that they created a more human-like usage of the \gls{sut} respectively its \gls{gui}. However, since the proposed prototype was only evaluated with ReTest's demo \gls{sut}, the limited amount of data may be a serious threat to validity. A key issue is that the developed \gls{ann} generalizes over the entire \gls{gui}. Consequently, if the \gls{sut} is composed of many different looking windows, this generalization might become infeasible. On the one hand, it can be argued that a large amount of heterogeneous windows indicates design flaws. On the other hand, especially legacy systems in business context are required to present data in many different ways: forms, tables, charts and so forth. Although the approach denotes a reasonable step towards automated functional testing, it cannot achieve the capabilities of manually created tests. Nonetheless, compared to~(dumb and smart) monkey testing respectively a random initial population, the developed system is already considerably better. In comparison with related approaches such as \cite{arb17}, the training data does not require manual intervention since the labels are generated automatically. Moreover, because the data consists of text only, it is memory efficient and leads to short training times. Other systems in the field of \gls{sbse} respectively \gls{sbst}, for instance, \cite{ep16} or \cite{mhj16}, typically only apply either \gls{ec} or \gls{ml}. The author of this thesis is not aware of any solution that employs an \gls{mlec} algorithm in the context of \gls{gui}-based regression testing.

Finally, each of the formulated user stories have been addressed by the prototypical implementation. Training data can be extracted from existing tests~(user story~\#\ref{story:data-extraction}) and used to train and evaluate the \gls{ann}~(user story~\#\ref{story:training-evaluation}). Also, user story~\#\ref{story:enhanced-monkey}---enhancing the generated tests in terms of human behavior---has been fulfilled without affecting the overall performance~(user story~\#\ref{story:monkey-performance}) as shown in section~\ref{subsec:test-generation}. With \dlfj{}, both user story~\#\ref{story:open-source} and~\#\ref{story:java-compatibility}~(which demanded a prototype that uses open-source and Java-compatible libraries) were implemented as well. Critical functionalities have been backed by unit and integration tests to create a robust and extensible proof of concept, which also addresses user story~\#\ref{story:robustness-extensibility}.

\section{Future Work}
\label{sec:future-work}

Overall, the proposed approach yields promising results that offer several connecting factors for possible extensions and deeper investigations. For example, a large-scale evaluation could verify the results for complex \glspl{sut} with many different windows. The question arises whether the \gls{ann} can cope with this situation simply by adding more data~(i.e. more tests that can be used for feature extraction) or if it is not able to learn at all. Another way to overcome this issue could be an approach like \cite{silp16}. Similar to the \gls{ann}-guided game tree search used by Silver et al., the model could use a tree of \gls{sut} states~(e.g. a state graph) to find reasonable actions executed by human testers or actual users. The simple configuration of the \gls{ann} that was used in the present thesis generally leaves room for various optimizations.

Another potential research direction could be towards feature engineering. Today, \glspl{ann} are capable of processing many more features than those selected in this thesis. However, selecting appropriate features is a difficult topic and requires deep knowledge of the problem domain. Therefore, it is worthwhile to extend respectively adapt the given set of selected features.

Since the approach is currently only implemented for Swing-based \glspl{gui}, it could be ported to other technologies such as the Android platform to be used together with the UI/Application Exerciser Monkey. In the case of web applications, test generators often struggle with identifying reasonable target components. A \inline{<div>} attached with a certain CSS class or a special JavaScript event handler may only be feasible for a particular category of \glspl{sut} or even only for a single \gls{sut}. The component-based approach of this thesis could learn which types make sense and which do not. Within this context, a study could be conducted on how a system that uses \gls{gui} actions instead of \gls{gui} components for learning might look like and if it is superior to an approach that is based on the latter. From an \gls{mlec} perspective, it would also be interesting to see what other parts of an \gls{ec}~(e.g. fitness evaluation and selection or the adaption of the algorithm itself) can be improved with the aid of \gls{ml} when it comes to test generation and how these techniques can help to address open problems and challenges especially in the field of \gls{sbst}~\cite{hjz15}.

%% file: chapters/7-appendix.tex


\appendix
\renewcommand\thesection{\Alph{section}}
\counterwithin{figure}{section}
\counterwithin{listing}{section}
\addchap{Appendix}

\section{ReTest's Demo SUT}

\begin{figure}[ht]
\centering
\includegraphics[width=\textwidth]{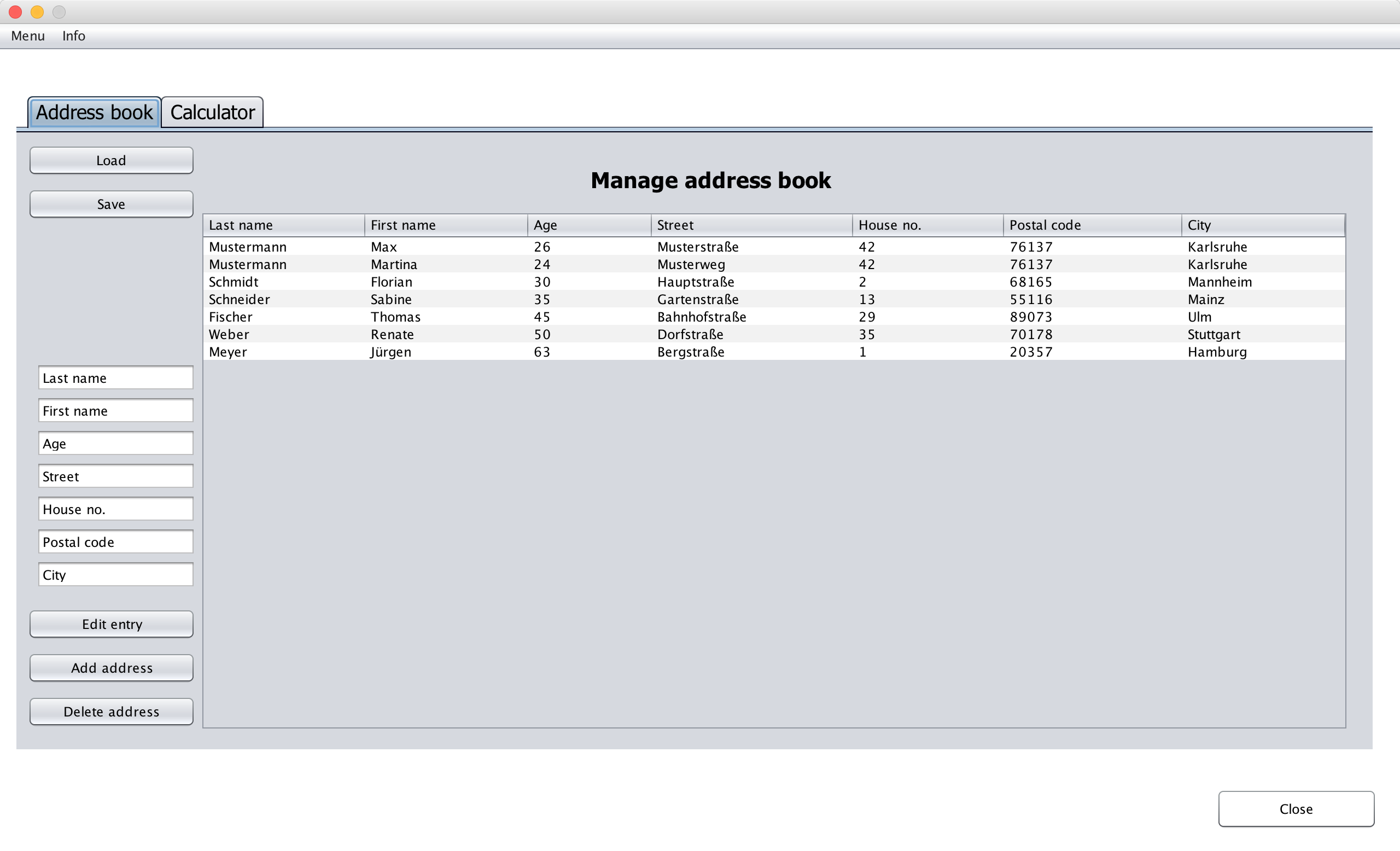}
\captionbelow{Address book tab of ReTest's demo SUT.}
\label{app:demo-address-book}
\end{figure}

\clearpage

\begin{figure}
\centering
\includegraphics[width=\textwidth]{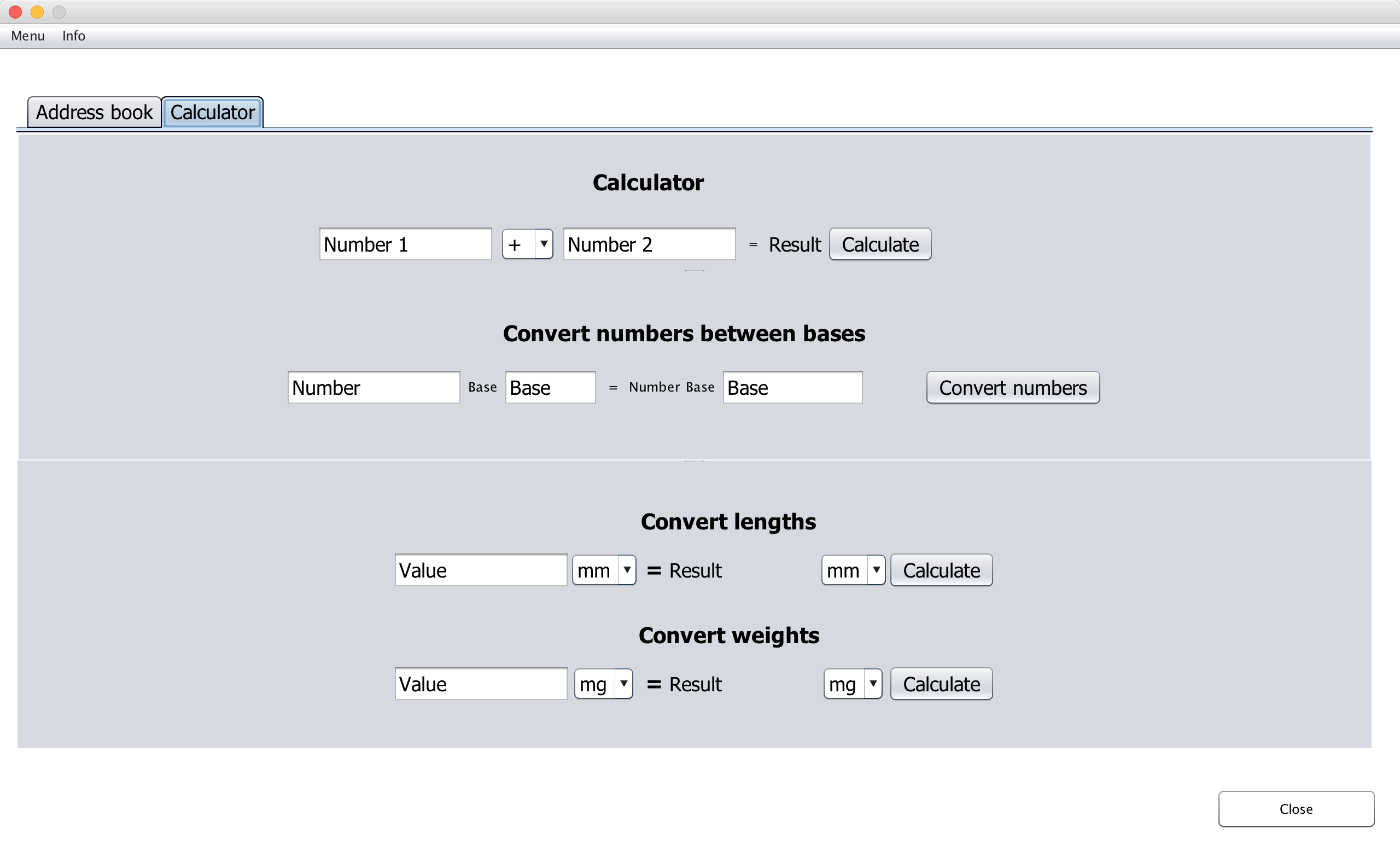}
\captionbelow{Calculator tab of ReTest's demo SUT.}
\label{app:demo-calculator}
\end{figure}

\section{Supplementary Listings}
\label{app:supplementary-listings}

\begin{listing}[h]
\begin{minted}{java}
public static boolean isEnabled(ComponentDescriptor cd) {
	Object enabled = cd.getStateCriteria().get(StateCriteria.ENABLED);
	return enabled instanceof Boolean ? (Boolean) enabled : true;
}
\end{minted}
\caption{Enabled feature extraction.}
\label{lst:extract-enabled-feature}
\end{listing}

\begin{listing}[hb!]
\begin{minted}{java}
public static boolean isPreferredType(ComponentDescriptor cd) {
	String type = cd.getIdentificationCriteria().getType();
	return ReflectionUtils.instanceOf(type, AbstractButton.class)
			|| ReflectionUtils.instanceOf(type, JTextComponent.class);
}
\end{minted}
\caption{Preferred type feature extraction.}
\label{lst:extract-preferredtype-feature}
\end{listing}

\begin{listing}
\begin{minted}{java}
public static int getPathDistance(ComponentDescriptor from,
		ComponentDescriptor to) {
	String fromPath = from.getIdentificationCriteria().getPath();
	String toPath = to.getIdentificationCriteria().getPath();
	String[] fromPathElements = fromPath.split(Path.PATH_SEPARATOR);
	String[] toPathElements = toPath.split(Path.PATH_SEPARATOR);
	int minLength =
			Math.min(fromPathElements.length, toPathElements.length);
	int maxLength =
			Math.max(fromPathElements.length, toPathElements.length);
	
	for (int commonPrefix = 0; commonPrefix < minLength; commonPrefix++) {
		if (ObjectUtils.notEqual(fromPathElements[commonPrefix],
				toPathElements[commonPrefix])) {
			return maxLength - commonPrefix;
		}
	}

	return maxLength - minLength;
}
\end{minted}
\caption{Path distance feature extraction.}
\label{lst:extract-pathdistance-feature}
\end{listing}

\begin{listing}
\begin{minted}{java}
public static double getPointDistance(ComponentDescriptor from,
		ComponentDescriptor to) {
	Point fromLocation =
			from.getIdentificationCriteria().getOutline().getLocation();
	Point toLocation =
			to.getIdentificationCriteria().getOutline().getLocation();
			
	return fromLocation.distance(toLocation);
}
\end{minted}
\caption{Point distance feature extraction.}
\label{lst:extract-pointdistance-feature}
\end{listing}

\begin{listing}
\begin{minted}{text}
JButton [Login],true,true,-1,1,44.0,true
JButton [Beenden],true,true,-3,1,91.30169768410661,false
JLabel [           ],true,false,-10,1,116.77756633874505,false
JPanel,true,false,0,1,128.14444974324874,false
JPanel,true,false,0,1,107.64757312638311,false
JPanel,true,false,-10,2,107.64757312638311,false
JLayeredPane,true,false,-10,3,107.64757312638311,false
JRootPane,true,false,-10,4,107.64757312638311,false
LoginDialog [Login],true,false,-10,5,111.50336317797773,false
JButton [Schließen],true,true,0,2,1183.3110326537144,false
Tab [Adressbuch],true,false,-10,2,77.17512552629896,true
Tab [Rechner],true,false,-10,2,55.86591089385369,false
JLabel [Adressbuch verwalten],true,false,-10,4,452.58479868418027,false
JButton [Löschen],true,true,-1,4,511.8915900852445,false
\end{minted}
\caption{Raw CSV file example.}
\label{lst:raw-csv}
\end{listing}

\begin{listing}
\begin{minted}{text}
1,1,-1,1,44.0,1
1,1,-3,1,91.30169768410661,0
1,0,-10,1,116.77756633874505,0
1,0,0,1,128.14444974324874,0
1,0,0,1,107.64757312638311,0
1,0,-10,2,107.64757312638311,0
1,0,-10,3,107.64757312638311,0
1,0,-10,4,107.64757312638311,0
1,0,-10,5,111.50336317797773,0
1,1,0,2,1183.3110326537144,0
1,0,-10,2,77.17512552629896,1
1,0,-10,2,55.86591089385369,0
1,0,-10,4,452.58479868418027,0
1,1,-1,4,511.8915900852445,0
\end{minted}
\caption{Transformed CSV file example.}
\label{lst:transformed-csv}
\end{listing}

\begin{listing}
\begin{minted}{java}
private static void runOnSpark(Path input,
		TransformProcess transformProcess, Path output) {
	SparkConf sparkConf = new SparkConf();
	sparkConf.setMaster("local[*]");
	sparkConf.setAppName("Storm Reports Record Reader Transform");

	try (JavaSparkContext sc = new JavaSparkContext(sparkConf)) {
		JavaRDD<String> lines = sc.textFile(input.toString());
		JavaRDD<String> balancedLines = handleImbalance(lines);
		
		JavaRDD<List<Writable>> stormReports = balancedLines.map(
				new StringToWritablesFunction(new CSVRecordReader()));
		JavaRDD<List<Writable>> processed = SparkTransformExecutor
				.execute(stormReports, transformProcess);
				
		JavaRDD<String> toSave = processed.map(
				new WritablesToStringFunction(
						CSVRecordReader.DEFAULT_DELIMITER));
		toSave.coalesce(1).saveAsTextFile(output.toString());
	}
}
\end{minted}
\caption{Data transformation via DataVec on local Spark.}
\label{lst:transform-datavec-spark}
\end{listing}

\begin{listing}
\begin{minted}{java}
private INDArray convertToFeatureMatrix(
		ComponentDescriptor previousTarget,
		List<ComponentDescriptor> flattenedPossibleTargets) {
	List<FeatureVector> vectors = flattenedPossibleTargets.stream()
			.map(possibleTarget -> FeatureVector.of(
					previousTarget, possibleTarget, environment))
			.collect(Collectors.toList());
	INDArray featureMatrix = convertToFeatureMatrix(vectors);
	normalizer.transform(featureMatrix);
	return featureMatrix;
}

static INDArray convertToFeatureMatrix(List<FeatureVector> vectors) {
	return Nd4j.create(vectors.stream()
			.map(vector -> new double[]{vector.enabled ? 1.0 : 0.0,
					vector.preferredType ? 1.0 : 0.0,
					vector.focusDistance,
					vector.pathDistance,
					vector.pointDistance})
			.toArray(double[][]::new));
}
\end{minted}
\caption{Conversion of possible target components to a feature matrix.}
\label{lst:convert-feature-matrix}
\end{listing}

\section{Evaluation Details}
\label{app:evaluation-details}

\begin{listing}[h!]
\begin{minted}[breaklines=true]{text}
1..1
ok 1 address-book
    1..3
    ok 1 address-book/add-contact
        1..12
        ok 1 Entering text 'Max' into JTextField Benutzername
        ok 2 Entering text 'ReTest' into JPasswordField Passwort
        ok 3 Click on JButton [Login]
        ok 4 Click on Tab [Adressbuch]
        ok 5 Entering text 'John' into JTextField
        ok 6 Entering text 'Doe' into JTextField
        ok 7 Entering text '42' into JTextField
        ok 8 Entering text 'Musterstraße' into JTextField
        ok 9 Entering text '13' into JTextField
        ok 10 Entering text '12345' into JTextField
        ok 11 Entering text 'Musterstadt' into JTextField
        ok 12 Click on JButton [Hinzufügen]
    ok 2 address-book/delete-contact
        1..6
        ok 1 Entering text 'Max' into JTextField Benutzername
        ok 2 Entering text 'ReTest' into JPasswordField Passwort
        ok 3 Click on JButton [Login]
        ok 4 Click on Tab [Adressbuch]
        ok 5 Click on TableCell [Mustermann] (1/1) of JTable[Adressbuch]
        ok 6 Click on JButton [Löschen]
    ok 3 address-book/update-contact
        1..8
        ok 1 Entering text 'Max' into JTextField Benutzername
        ok 2 Entering text 'ReTest' into JPasswordField Passwort
        ok 3 Click on JButton [Login]
        ok 4 Click on Tab [Adressbuch]
        ok 5 Click on TableCell [Schmidt] (1/2) of JTable[Adressbuch]
        ok 6 Entering text 'Anderer Weg' into JTextField
        ok 7 Entering text '21' into JTextField
        ok 8 Click on JButton [Bearbeiten]
\end{minted}
\caption{Address book suite for feature extraction as TAP report.}
\label{lst:address-book-feature-extraction}
\end{listing}

\begin{listing}
\begin{minted}[breaklines=true]{text}
ok 1 calculator
    1..4
    ok 1 calculator/multiply
        1..8
        ok 1 Entering text 'Max' into JTextField Benutzername
        ok 2 Entering text 'ReTest' into JPasswordField Passwort
        ok 3 Click on JButton [Login]
        ok 4 Click on Tab [Rechner]
        ok 5 Entering text '1' into JTextField
        ok 6 Select [*] on JComboBox
        ok 7 Entering text '2' into JTextField
        ok 8 Click on JButton [Berechnen]
    ok 2 calculator/convert-base
        1..8
        ok 1 Entering text 'Max' into JTextField Benutzername
        ok 2 Entering text 'ReTest' into JPasswordField Passwort
        ok 3 Click on JButton [Login]
        ok 4 Click on Tab [Rechner]
        ok 5 Entering text '1011' into JTextField
        ok 6 Entering text '2' into JTextField zur Basis
        ok 7 Entering text '10' into JTextField zur Basis
        ok 8 Click on JButton [Zahl umrechnen]
    ok 3 calculator/convert-length
        1..7
        ok 1 Entering text 'Max' into JTextField Benutzername
        ok 2 Entering text 'ReTest' into JPasswordField Passwort
        ok 3 Click on JButton [Login]
        ok 4 Click on Tab [Rechner]
        ok 5 Entering text '1000' into JTextField
        ok 6 Select [m] on JComboBox
        ok 7 Click on JButton [Berechnen]
    ok 4 calculator/convert-weight
        1..7
        ok 1 Entering text 'Max' into JTextField Benutzername
        ok 2 Entering text 'ReTest' into JPasswordField Passwort
        ok 3 Click on JButton [Login]
        ok 4 Click on Tab [Rechner]
        ok 5 Entering text '1000' into JTextField
        ok 6 Select [kg] on JComboBox Ergebnis
        ok 7 Click on JButton [Berechnen]
\end{minted}
\caption{Calculator suite for feature extraction as TAP report.}
\label{lst:calculator-feature-extraction}
\end{listing}

\begin{figure}
\centering
\includegraphics[scale=0.55]{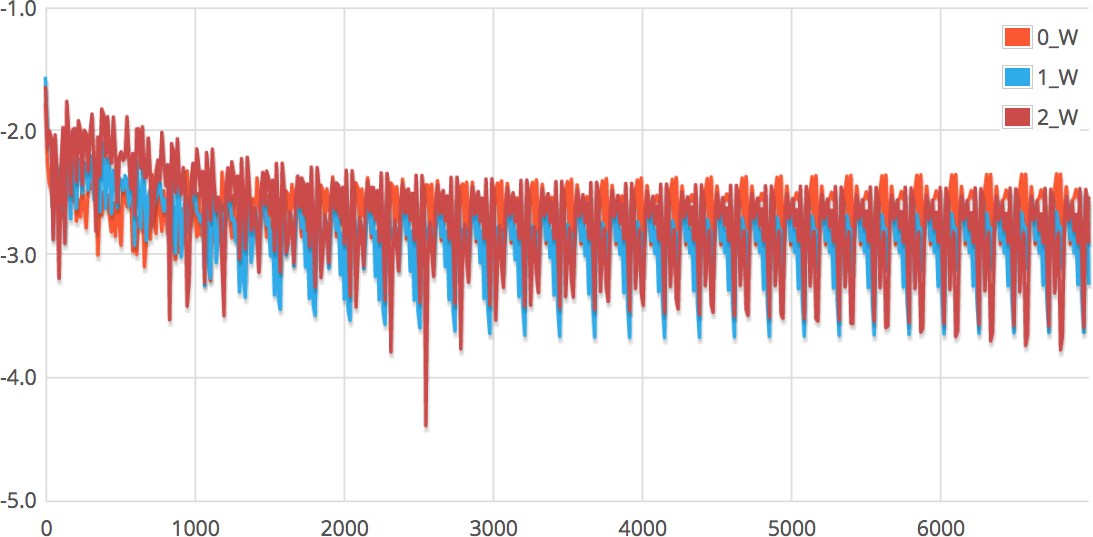}
\captionbelow{Parameter update ratio by layer~($\log_{10}$) vs. iteration.}
\label{fig:parameter-ratios}
\end{figure}

\begin{figure}
\centering
\includegraphics[scale=0.55]{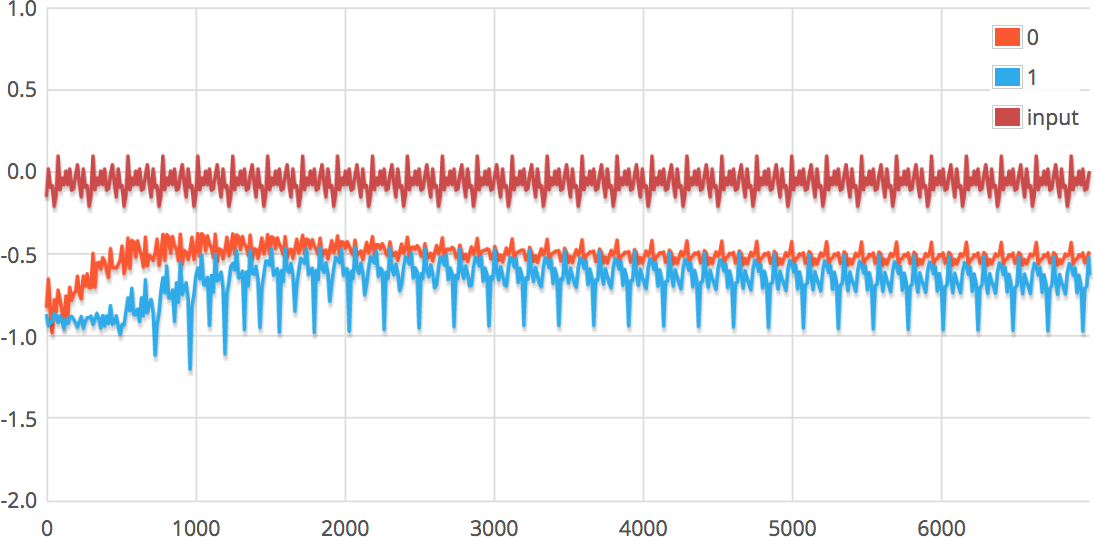}
\captionbelow{Activation standard deviations~($\log_{10}$) vs. iteration.}
\label{fig:std-dev-activations}
\end{figure}